\documentclass{article}
\usepackage[utf8]{inputenc}
\usepackage[table,xcdraw]{xcolor}
\usepackage{amsmath}
\usepackage{amssymb}
\usepackage{mathabx}
\usepackage{amsthm}
\usepackage{bm,bbm}
\usepackage[framemethod=tikz]{mdframed}
\usepackage{graphicx}\usepackage{pgfplots,float,makecell}\usetikzlibrary{arrows}
\usepackage{tikz}
\usetikzlibrary{shapes.geometric}
\usetikzlibrary{positioning}
\usepackage{subfigure}

\usepackage[colorlinks=true,breaklinks=true,bookmarks=true,urlcolor=blue,
     citecolor=blue,linkcolor=blue,bookmarksopen=false,draft=false]{hyperref}

\def\qed{\hfill \vrule height 7pt width 7pt depth 0pt\medskip}
\newcommand{\ds}{\displaystyle}

\newcommand{\ba}{\begin{array}}
\newcommand{\ea}{\end{array}}

\newcommand{\1}{\mathbbm{1}}

\newcommand{\be}{\begin{equation}}
\newcommand{\ee}{\end{equation}}

\newcommand{\mc}{\mathcal}

\DeclareMathOperator{\argmax}{arg\,max}
\DeclareMathOperator{\argmin}{arg\,min}
\newcommand{\ov}{\overline}

\newcommand{\tcb}{\textcolor{black}}

\newcommand{\E}{\mathbb{E}}

\newcommand{\R}{\mathbb{R}}
\newcommand{\N}{\mathbb{N}}

\newcommand{\se}{\text{ if }}

\DeclareMathOperator{\sgn}{sgn}
\DeclareMathOperator{\nash}{\mc X^{*}}
\DeclareMathOperator{\recnash}{\mc X^{\circ}}
\DeclareMathOperator{\nashZ}{\mc X^{Z}}
\DeclareMathOperator{\strictnash}{\mc X^{\bullet}}
 
\renewcommand{\deg}{d}
\newcommand{\degprof}{\mathbf{d}}

\def\N{\mathbb{N}}
\def\E{\mathbb{E}}
\def\R{\mathbb{R}}

\newtheorem{theorem}{Theorem}
\newtheorem{proposition}{Proposition}
\newtheorem{lemma}{Lemma}
\newtheorem{corollary}{Corollary}
\newtheorem{remark}{Remark}

\newtheorem{example}{Example}

\title{On a Network Centrality Maximization Game}

\author{Costanza Catalano\small{$^1$}, \large Maria Castaldo\small{$^2$}, \\Giacomo Como\small{$^3$}\thanks{Giacomo Como is also with the Department of Automatic Control, Lund University, BOX 118, SE-22100, Lund, Sweden.} , \large Fabio Fagnani\small{$^3$}}

\date{\small $^1$Department of Mathematics and Computer Science, Universit\`{a} degli Studi di Firenze, Viale Morgagni 67/a, 50134, Firenze, Italy.\\ costanza.catalano@unifi.it\\
$^2$Univ. Grenoble Alpes, CNRS, Inria, Grenoble INP, GIPSA-lab, \\11 rue des Math\'ematiques, F-38000, Grenoble, France.\\ maria.castaldo@grenoble-inp.fr\\
$^3$Department of Mathematical Sciences, Politecnico di Torino,\\ Corso Duca degli Abruzzi 24, 10129, Torino, Italy. \\ giacomo.como@polito.it, fabio.fagnani@polito.it} 

\begin{document}

\maketitle
%%%%%%%%%%%%%%%%

%\setcounter{page}{1}
%\DOI{10.1287/xxxx.0000.0000}%

\begin{abstract}
	We study a network formation game where $n$ players, identified with the nodes of a directed graph to be formed, choose where to wire their outgoing links in order to maximize their PageRank centrality. Specifically, the action of every player $i$ consists in the wiring of a predetermined number $\deg_i$ of directed out-links, and her utility is her own PageRank centrality in the network resulting from the actions of all players. We show that this is a potential game and that the best response correspondence always exhibits a local structure in that it is never convenient for a node $i$ to link to other nodes that are at incoming distance more than $\deg_i $ from her. We then study the equilibria of this game determining necessary conditions for a graph to be a (strict, recurrent) Nash equilibrium. Moreover, in the homogeneous case, where players all have the same number $\deg$ of out-links, we characterize the structure of the potential maximizing equilibria and, in the special cases
$ \deg=1 $ and $ \deg=2 $, we provide a complete classification of the set of (strict, recurrent)  Nash equilibria. Our analysis shows in particular that the considered formation mechanism leads to the emergence of undirected and disconnected or loosely connected networks. 
\end{abstract}

\maketitle

\section{Introduction}
The notion of \emph{centrality}  is ubiquitous in network science and engineering:  
it provides a measure of the relative relevance of nodes in a network and finds a wide range of applications \cite{Newman:03,JacksonBook2008,Easley.Kleinberg:2010}. 
Various definitions of network centrality have appeared in the literature \cite{Boldi2014}, many of which tailored to specific applications. In this paper, we focus on the so-called \emph{PageRank centrality} \cite{SB-LP:98}, which is closely related to the notions introduced by Katz and Bonacich \cite{katz:53,PB:87,Friedkin:91}. 

Understanding how the PageRank centrality measure can be efficiently computed and how it can be modified by perturbing the network has been identified as key problem and has received significant attention in the recent literature  \cite{Ishii.Tempo:2014,Como.Fagnani:2015}. The effect on the centrality caused by adding or deleting links in the network is not obvious. While it can be shown  \cite{Chien2004} that the addition of a directed link always increases the PageRank centrality of its head node, it is less clear how it affects the centrality of the other nodes in the network. In applications such as citation networks, the World Wide Web, or (on-line) social networks, each node can decide where to direct its out-links, and there is an interest in gaining visibility (that is, to increase its own centrality in the network). A natural question is how such choices modify its centrality and what is the rewiring that can possibly optimize it. 
A first analysis in this sense can be found in \cite{Avrachenkov06,dekerchove08}, while \cite{Jungers10} explores computation time issues of these problems. 
\tcb{In \cite{Corbo06} and \cite{Belhaj16}, optimal network intervention problems are studied where an external planner seeks to maximize the sum of Bonacich centralities with either a budget on the links to be created or a cost in the objective function.}

In this paper, we study a network formation model where the nodes of a directed graph to be formed are left free to choose where to place their out-links, with the constraint that each node has a fixed number of links at disposal (and it has to place them all).  We cast the problem into a game-theoretic setting where utilities of the nodes are exactly their PageRank centralities. We first show that this class of network formation games are potential games. We then study the structure of the best response correspondence and the structure of the equilibria of such games. Our analysis highlights two effects that network formation mechanisms purely based on centrality maximization have on the emerging structure:
\begin{itemize}
\item a preference for local interactions that induces equilibria with a large number of undirected links and short cycles;
\item a large number of isolated connected components and a small hierarchical depth.  
\end{itemize}
In particular,  in the homogeneous case, where players all have the same number $\deg$ of out-links, our results provide a characterization of the potential maximizing equilibria, and a complete classification of the Nash equilibria for the two special cases when $\deg=1$ and $\deg=2$, respectively.

A similar  network formation game is considered in \cite{Hopcroft_2008} with the important difference that each player is left free to choose the number of her out-links (that can also be zero). In this framework, the best action of a node is always to link back to nodes in its in-neighborhood. Using this fact, the authors of \cite{Hopcroft_2008} prove that all Nash equilibria have an undirected graph as a core, with possibly a set of nodes linking to the core and having no in-links. The work \cite{Chen2009} later proved that these Nash equilibria explicitly depend on the discount factor of the PageRank centrality, answering a question left open by \cite{Hopcroft_2008}. Although certain qualitative features of the Nash equilibria are similar in the two models, none of the results in \cite{Hopcroft_2008} can be extended to our model and various counterexamples are presented in this paper. The recent work \cite{scarsini} is also closely related to ours: there, the authors prove the existence of Nash equilibria for a generalized version of our game. However, their proof is non-constructive and our classification of Nash equilibria cannot be derived from their results.

The problem considered in this paper is an instance of a \emph{network formation game}, where players are identified with nodes and their actions determine the underlying network. {\color{black} The related literature is vast and goes back to the seminal paper \cite{JacksonW96} proposing a model where the utility of a node consists in a discounted sum of distances from all other nodes and of a cost determined by the number of out-links maintained by the node. In their model, undirected links are considered  and for creating a link both incident nodes must agree, while each node can unilaterally prune a link. For this and similar models, the authors  of \cite{JacksonW96} introduce and study a related concept of pairwise stability and of social efficiency. Other works, including \cite{Bala00,Fabrikant03,Demaine07},  have reconsidered the model in  \cite{JacksonW96} and studied network formation games where nodes can autonomously create and delete links. These works focus on the structure of Nash equilibria and the analysis of the Price of Anarchy, while their contributions differ for the way links are considered, either directed or undirected, and in the form of the utility functions. 

The works \cite{Laoutaris08} and \cite{Ehsani11} consider bounded budget models that are closer in spirit to our framework. In these works, links are not associated to a cost but rather the number of out-links that a node can have is bounded or exactly specified a-priori. Two different games are investigated where the nodes' costs are either the sum  or the maximum of all distances from the other nodes. In \cite{Laoutaris08}, the emerging graph is considered directed, and existence of Nash equilibria under uniform budget is proven as well as estimations on Price of Anarchy are derived. In \cite{Ehsani11}, link ownership belongs to just one node, but its effect is bidirectional: the authors prove existence and connectedness of Nash equilibria and establish bounds for the Price of Anarchy. Somewhat related is the model considered in \cite{Alon10} where the set of moves that a player can make is restricted to swapping one incident link at a time: here, links are considered as undirected and a broader notion of swap equilibrium is introduced and studied.  It is also worth citing the work \cite{Konig.ea:2014}, where a game-theoretic dynamic model of network evolution is proposed and the emergence of so-called \emph{nested split graphs} is studied.}

{\color{black}We remark some striking differences between the model studied in this paper and the models proposed in the literature on network formation games we just reviewed. Since the PageRank centrality is a normalized measure (i.e., the sum of all centralities is always equal to $1$), our model results in a constant-sum game. Yet, as mentioned earlier, we prove that our game is potential: more precisely, the closely related game where the utilities are the $\log$ of the PageRank centralities is an exact potential game, so that ours is an ordinal potential game. This is in contrast with the previously cited network formation games based on distances and has relevant consequences both in terms of tools used in the analysis and in the type of emerging network structures. E.g., the bounded budget model studied in \cite{Ehsani11}, which is one of the closest to our model, yields Nash equilibria that are connected, while, as we will see, in our model the lack of connectivity of equilibria is the norm. }

The remainder of this paper is organized as follows. Section \ref{sec:graph} introduces the required graph theoretical and network centrality background. 
Section \ref{sec:CMG-sub} provides a formal definition of the centrality maximization games studied in this paper, along with some of their fundamental properties (including the fact that they are potential games. In Section \ref{sec:mainResults}, we develop an equilibrium analysis of the centrality maximization games for general out-degree profiles. Our main results there are Theorem \ref{thm:best_response} showing how rational choices yield local connections, Theorem \ref{thm:condensation_graph_generalm} describing the structure of a Nash equilibrium in terms of its connected components, and Theorem \ref{theo:betato1} providing an alternative characterization of potential-maximizing equilibria for large enough values of the discount factor. We also have a number of secondary results such as Corollary \ref{prop:degrees} providing a lower bound on the number of undirected links and $3$-cycles in Nash equilibria. In Section \ref{sec:homogeneous}, we focus on homogeneous out-degree profiles: first we prove a result on the structure of potential-maximizing equilibria (Theorem \ref{theo:d=k})
and then derive a complete classification of Nash equilibria when nodes have all out-degree equal to $1$ (Theorem \ref{thm:nash_m=1}) and all out-degree equal to $2$ (Theorem \ref{thm:trapping_setsM2}). 
Section \ref{sec:num_results} presents numerical experiments on the role of the game parameters in the network formation.
Finally, Section \ref{conclusions} concludes with a summary and the description of some open problems. 

{A preliminary and incomplete version of our work was presented at the 21st IFAC World Conference \cite{Castaldo.ea:2020}. The results reported in \cite{Castaldo.ea:2020} are limited to the homogeneous out-degree profiles and correspond to a subset of those presented in Section \ref{sec:homogeneous}. More specifically, \cite[Theorem 6]{Castaldo.ea:2020} corresponds to points (i) and (iii) of Theorem \ref{thm:nash_m=1}, while \cite[Theorem 8]{Castaldo.ea:2020} corresponds Theorem \ref{thm:trapping_setsM2}(i).} On the other hand,  \cite{Castaldo.ea:2020} did not contain any of the other results reported in this paper.  

\tcb{Throughout the paper, we shall use the following notational convention. The indicator function of a set $\mc A$ is denoted by $\1_{\mc A}$, so that $\1_{\mc A}(a)=1$ if $a\in\mc A$ and $\1_{\mc A}(a)=0$ if $a\notin\mc A$. The all-one vector is denoted by $\mathbf1$, while $\delta^i$ for the vector with all zero entries except for entry $i$ which is equal to $1$. For two real-valued functions $f$ and $g$, the asymptotic notation $f\sim g$ means that $\lim f/g=1$, whereas $f\asymp g$ means that $0<\liminf f/g\le\limsup f/g<+\infty$. 
}

\section{Graph-Theoretic Notions and PageRank Centrality} \label{sec:graph}
In this section, we review some graph theoretic concept and introduce the notion of PageRank centrality in a network. 

\begin{figure}\begin{center}
\subfigure[\ $K_8$]{\includegraphics[height=3.5cm]{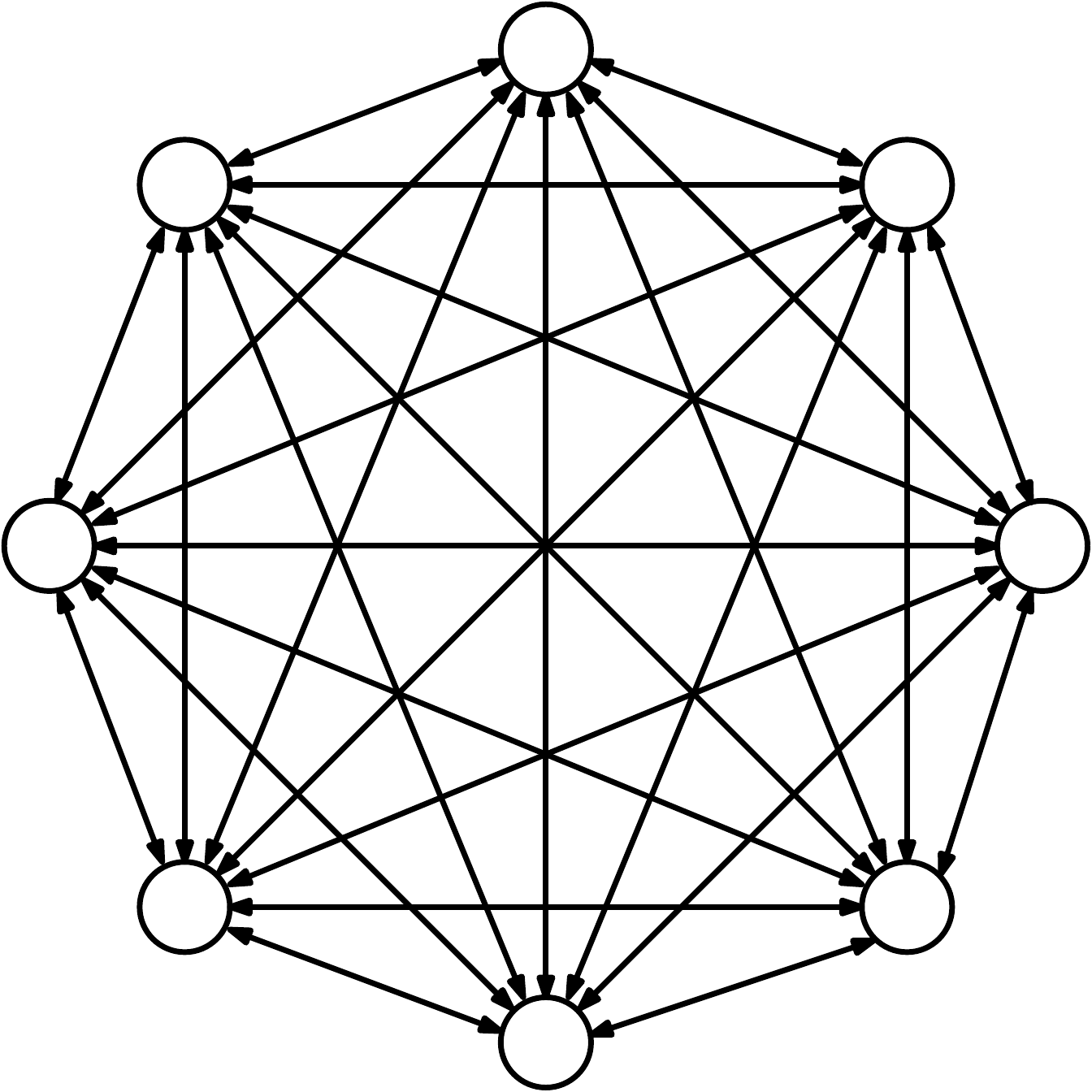}}\hspace{0.2cm}
\subfigure[\ $S_8$]{\includegraphics[height=3.5cm]{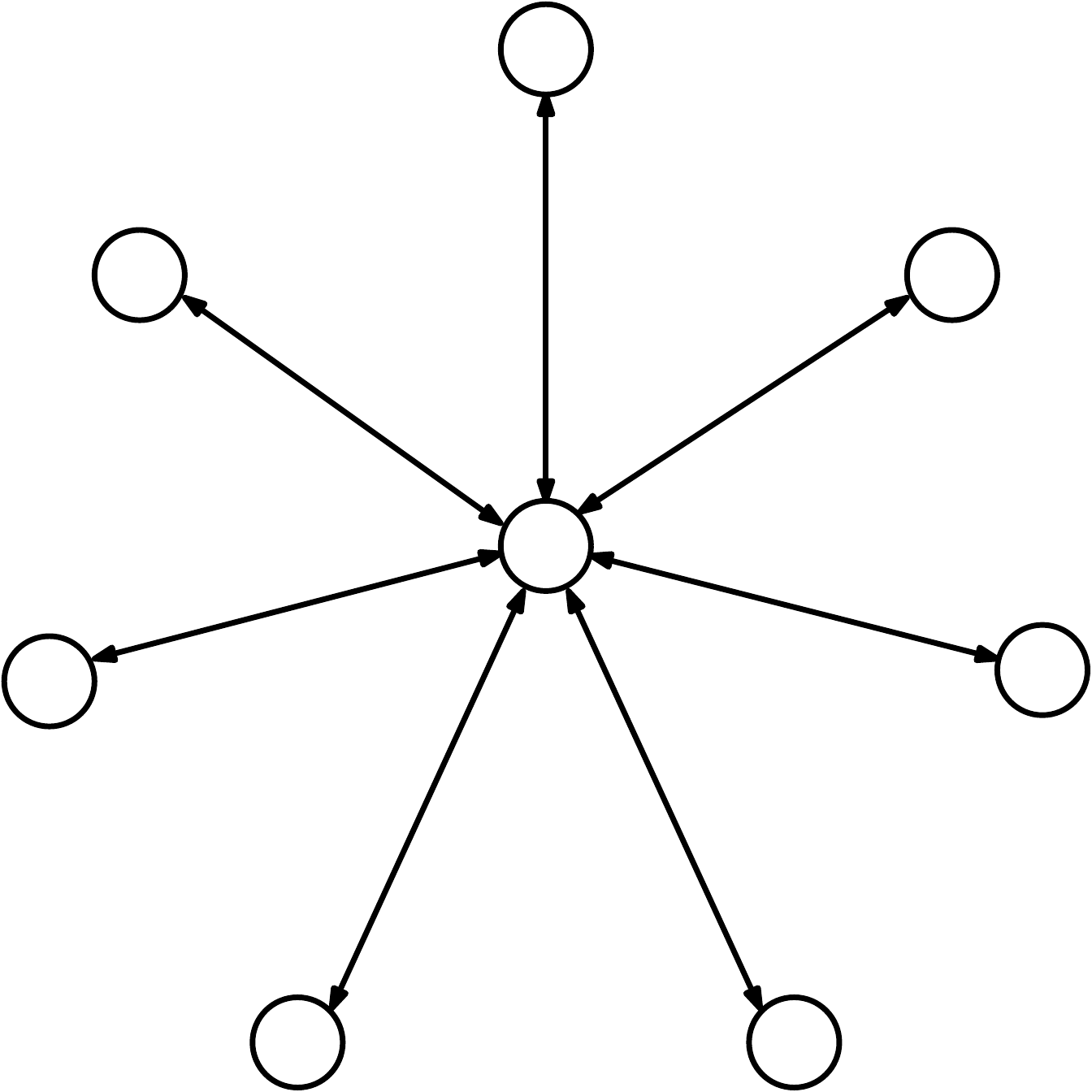}}\hspace{0.2cm}
\subfigure[\ $R_8$]{\includegraphics[height=3.5cm]{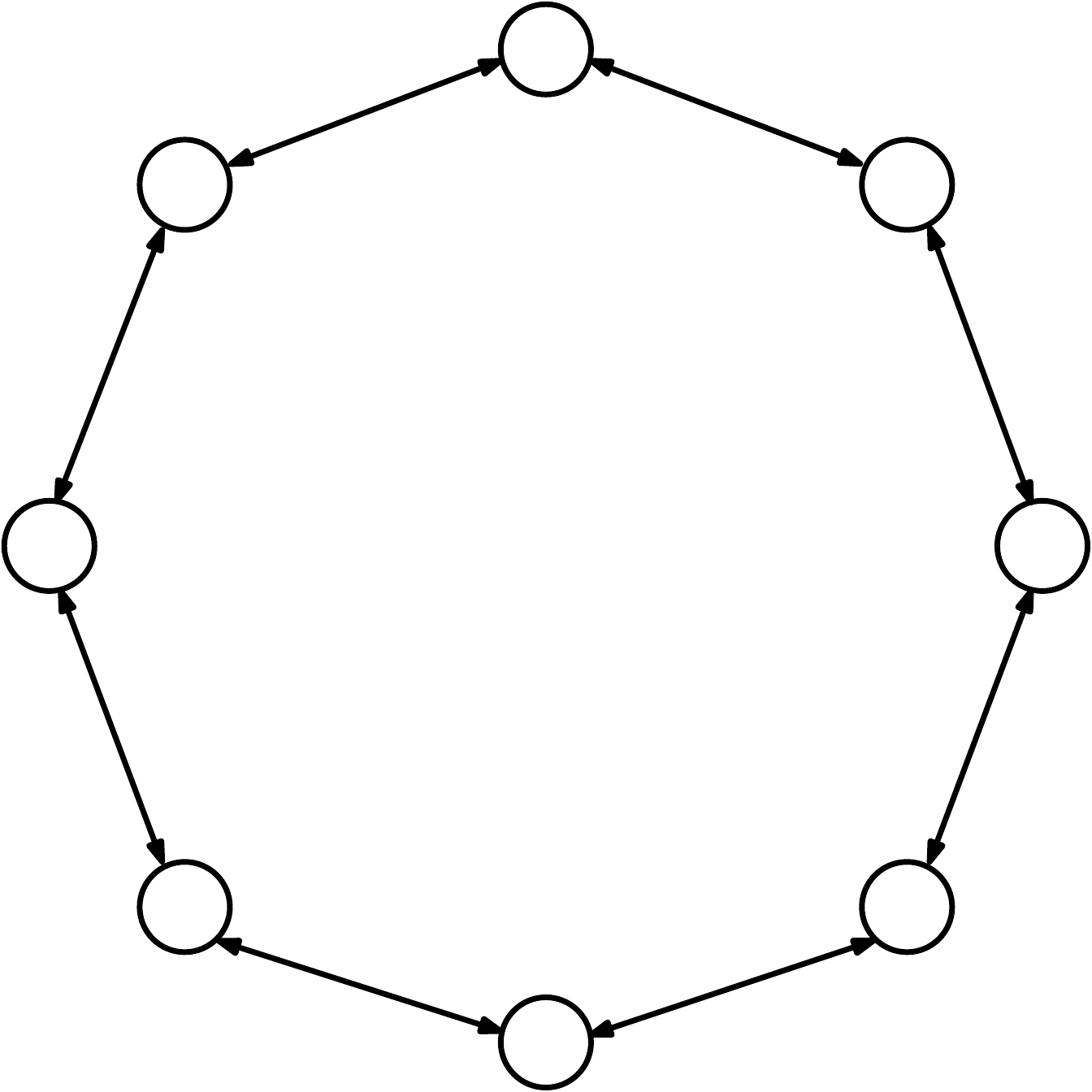}}
\end{center}
\tcb{\caption{\label{fig:complete-star-ring}The complete, star, and ring graphs with $n=8$ nodes.  }}
\end{figure}

We shall consider finite directed graphs $\mc G=(\mc V, \mc E)$ with node set $\mc V$ and link set $\mc E\subseteq \mc V\times\mc V$, containing no self-loops, i.e., such that $(i,i)\not\in\mc E$ for any $i$ in $\mc V$. The \emph{out-neighborhood} and the \emph{out-degree} of a node $i$ in $\mc V$ will be denoted by $\mc N_i=\{j\in\mc G:\,(i,j)\in\mc E\}$ and $\deg_i=|\mc N_i|$,  respectively. The vector $\degprof=(\deg_i)_{i\in\mc V}$ will be referred to as the \emph{out-degree profile}. We shall always consider cases where there are no sink nodes, i.e., $\deg_i\ge1$ for every $i$ in $\mc V$. An \emph{undirected link} in $\mc G$ is a pair $\{i,j\}$ such that $(i,j)$ and $(j,i)$ are both in $\mc E$. A graph $\mc G=(\mc V, \mc E)$ is referred to as \emph{undirected} whenever $(i,j)\in\mc E$ implies that $(j,i)\in\mc E$. Standard examples of undirected graphs with $n$ nodes include the complete graph $K_n$ (i.e., the graph $\mc G=(\mc V,\mc E)$ where $\mc E=\{(i,j):i\ne j\in\mc V\}$), the star graph $S_n$ (i.e., the graph $\mc G=(\mc V,\mc E)$ with $\mc V=\{1,2,\ldots,n\}$ and $\mc E=\{(1,i):\,2\le i\le n\}\cup\{(i,1):\,2\le i\le n\}$), and the ring graph $R_n$ (i.e., the graph $\mc G=(\mc V,\mc E)$ with $\mc V=\{1,2,\ldots,n\}$ and $\mc E=\{(i,j):\,|i-j|=1\text{ mod }n\}$), all displayed in Figure \ref{fig:complete-star-ring}. 

An automorphism of a graph $\mc G=(\mc V,\mc E) $ is a one-to-one map $f:\mc V\to \mc V$ such that $ (i,j)\in \mc E $ if and only if $ (f(i),f(j))\in \mc E $.
Two graphs $\mc G=(\mc V,\mc E) $ and $\mc G=(\mc V',\mc E') $ are referred to as \emph{isomorphic} if they simply differ by a relabeling of the vertices, i.e., if there exists a one-to-one map $ f:\mc V\to \mc V'$ such that $ (i,j)\in \mc E $ if and only if $ (f(i),f(j))\in \mc E' $. The subgraph induced by a subset of nodes $\mc W\subseteq \mc V$ is  $\mc G_{\mc W}=(\mc W, \mc E\cap(\mc W\times\mc W))$.  A $k$-clique is a subgraph that is isomorphic to $K_k$.
A length-$l$ \emph{walk} in $\mc G$ from $i$ to $j$ is an $(l+1)$-tuple of nodes $(r_0, r_1,\dots ,r_l)$ such that $r_0=i$, $r_l=j$, and $(r_{h-1},r_{h})\in\mc E$ for every $1\le h\le l$. A \emph{path} is a walk $(r_0, r_1,\dots ,r_l)$ such that $r_h\ne r_{k}$ for every $0\le h<k\le l$, except for possibly $r_0=r_l$, in which case the path is referred to as \emph{closed}. A closed path of length $l\ge3$ is called a \emph{cycle}. For two distinct nodes $i$ and $j$ in $\mc V$, we say that a subset $\mc C\subseteq \mc V\setminus\{i,j\}$ \emph{cuts} (or is a \emph{cut-set}) between $i$ and $j$ if every path from $i$ to $j$ in $\mc G$ has at least a node in $\mc C$. Given a node $i$ in $\mc V$, we denote by $\mc N^{-l}_i$ the set of nodes $j$  for which there exists a walk of length at most $l$ from $j$ to $i$ in the graph $\mc G$. Let also $\mc N^{-\infty}_i=\bigcup_{l\ge 0}\mc N^{-l}_i$ be the set of all nodes from which $i$ is reachable in $\mc G$. A graph $\mc G=(\mc V,\mc E)$ is \emph{strongly connected} if $\mc N^{-\infty}_i=\mc V$ for every node $i$ in $\mc V$. Every graph $\mc G$ can be decomposed into its \emph{connected components}, i.e., its maximal strongly connected subgraphs. A connected component $\mc G_{\mc S}$ of $\mc G$ is referred to as: a \emph{sink} if there is no link in $\mc G$ from $\mc S$ to $\mc V\setminus \mc S$; a \emph{source} if $\mc G_{\mc S}$ is not a sink and there is no link in $\mc G$ from $\mc V\setminus \mc S$ to $\mc S$. Notice that, with this convention, isolated connected components are to be considered as sinks and not as sources.

To a graph $\mc G=(\mc V,\mc E)$ we associate its normalized adjacency matrix $R$ in $\R^{\mc V\times\mc V}$ with entries 
\be\label{adjacency} R_{ij}=\left\{\ba{lcl}1/\deg_i&\se&j\in\mc N_i\\0&\se&j\in\mc V\setminus\mc N_i\,.\ea\right.\ee
For a parameter $\beta$ in $(0,1)$ and a probability vector $\eta$ in $\R_+^{\mc V}$, to be referred to respectively as the \emph{discount factor} and the \emph{intrinsic centrality}, the \emph{{PageRank} centrality} vector is then defined as 
\be\label{bon2}\pi=(1-\beta)(I-\beta R^\top)^{-1}\eta=(1-\beta)\sum_{k=0}^{+\infty}\beta^k(R^\top)^k\eta\,.\ee
The fact that $R$ is a row-stochastic matrix, hence with spectral radius $1$, ensures correctness of equation \eqref{bon2} and  also implies that $\pi$ is a probability vector.
Indeed, \eqref{bon2} implies that $\pi=P^\top\pi$ is the unique invariant probability distribution of the  irreducible stochastic matrix \be\label{def:P}P=\beta R+(1-\beta)\mathbf{1}\eta^\top\,.\ee

While the PageRank centrality has often been proposed as a measure of the relative relevance of the nodes in network in an axiomatic way, the following example illustrates how the PageRank centrality can naturally emerge as a natural measure of influence in opinion dynamics models. 
\begin{example}
Consider the Friedkin-Johnsen opinion dynamics \cite{NEF-ECJ:90,Como.Fagnani:2016,proskurnikov:2017tutorial}, whereby nodes $i$ in $\mc V$ are identified with agents in a social network, each holding a scalar opinion value $y_i(t)$ that gets updated as follows: 
$$y_i(t+1)=\beta\sum\nolimits_{j}R_{ij}y_j(t)+(1-\beta)a_i\,,$$
where $R$ is a stochastic matrix, $\beta$ in $(0,1)$ is a discount factor, and $a_i$ in an anchor opinion value for agent $i$ (originally $a_i=y_i(0)$). 
It is then easily verified that 
$$y^*=\lim\limits_{t\to +\infty}y(t) =(1-\beta)(I-\beta Q)^{-1}a\,.$$
For a given vector of weights $\eta$ in $\R_+^{\mc V}$ such that $\sum_i\eta_i=1$, the weighted average equilibrium opinion satisfies
$$\sum\nolimits_i\eta_iy^*_i=\eta^\top(1-\beta)(I-\beta R)^{-1}a=\sum\nolimits_i\pi_ia_i\,,$$
i.e., it coincides with the weighted average of the anchor values weighted by the PageRank centrality vector defined in \eqref{bon2}.  
\end{example}

\tcb{In the sequel, it will prove useful to rely on the following probabilistic interpretation of the \tcb{PageRank} centrality. 
Consider a discrete-time Markov chain $(V_t)_{t\ge0}$ with finite state space $\mc V$, and transition probability matrix $P$ as in \eqref{def:P}, so that, given that its current state $V_{t}=i$, with probability $\beta$ the next state $V_{t+1}$ will be chosen uniformly at random among the $\deg_i$ out-neighbors of $i$, while with probability $(1-\beta)$ the next state $V_{t+1}$ will be sampled from the probability distribution $\eta$. Then, the \tcb{PageRank} centrality vector $\pi=P^\top\pi$ coincides with the unique stationary distribution of $V_t$. Moreover, for two nodes $i$ and $j$ in $\mc V$, let $T_i=\min\{t\ge0:\,V_t=i\}$ be the hitting time in $i$ and let $\tau_j^i=\E[T_i:\,V_0=j]$ be its conditional expected value when starting from node $j$. In other terms, $\tau_j^i$ represents the time it takes, in expectation, for the Markov chain $V_t$ to go from node $j$ to node $i$. 
By \cite[Theorem 1.3.5]{norris_1997}, for every $i$ in $\mc V$, the vector $(\tau^i_j)_{j\in\mc V}$ coincides with the unique solution of the following linear system:  
\be\label{eq:system_tau2}
\tau_i^i=0\,,\qquad  
\tau_j^i
=\displaystyle1+(1-\beta)\sum_{k\in \mc V}\eta_k\tau_k^i+\frac{\beta}{d_j}\sum\limits_{k\in\mc N_j}\tau_k^i\,,\quad\forall j\neq i\,.\ee
Equation \eqref{eq:system_tau2} can in fact be derived by a conditioning argument and the Markov property: the time to reach $i$ when starting from $i$ itself is obviously $0$, whereas, when starting from $j\ne i$, it takes one time step to the Markov chain to make the first move and such a move, with conditional probability $P_{ik}=(1-\beta)\eta_k+\beta\1_{\mc N_j}(k)/\deg_j$,  will lead to a node $k$ from which the expected time to reach node $i$ is $\tau^i_k$.  
A related argument can be used to get the following representation of the Page-Rank centrality.  
}

\begin{lemma}\label{lemma:tau} 
In a graph $\mc G=(\mc V,\mc E)$, the Page-Rank centrality  with discount factor $\beta$  and intrinsic centrality $\eta$  is given by 
\begin{equation}\label{eq:pi}
\pi_i=\frac1{\ds1+(1-\beta)\tau_{\eta}^i+\frac{\beta}{\deg_i}\sum_{j\in\mc N_i}\tau_j^i}\,,\qquad \tau_{\eta}^i=\sum_{j\in \mc V}\eta_j\tau_j^i\,,
\end{equation}
for every node $i$ in $\mc V$.
\end{lemma}
\begin{proof}
Let  $T^{+}_i=\min\{t\geq 1\,|X_t=i\}$ be the return time  in a node $i$ in $\mc V$. 
By conditioning on $X_1$, using the Markov property and equation \eqref{eq:system_tau2}, we have
$$\E[T_i^+\,|\, X_0=i]=
1+\sum_{j\in \mc V}P_{ij}\tau_j^i
=1+(1-\beta)\tau_{\eta}^i+\frac{\beta}{\deg_i}\sum_{j\in\mc N_i}\tau_j^i\,.$$
Combining the above with Kac's formula $\pi_i={1}/{\E[T_i^+\,|\, X_0=i]}$  \cite[Theorem 1.7.7]{norris_1997}, we get \eqref{eq:pi}. 
\qed
\end{proof}

We close this section by observing that on the one hand the expected hitting times $\tau^i_j$  are independent from node $i$'s out-neighborhood $\mc N_i$, 
on the other hand they do depend not only on the structure of the rest of the graph $\mc G$, but also on the discount factor $\beta$ and on the intrinsic centrality $\eta$.

\section{Centrality Maximization Network Formation Games}\label{sec:CMG-sub}In this section, we introduce a class of directed network formation games based on competitive centrality maximization mechanism that will be the object of our study. We then prove some fundamental results on their structure. 

We consider directed network formation games where a finite set $\mc V$ of $n$ nodes choose where to wire a predetermined number of out-links in order to maximize their own PageRank centrality. Specifically, for an integer vector $\degprof$ in $\{1,\ldots,n-1\}^{\mc V}$, a probability vector $\eta$ in $\R_+^{\mc V}$, and a scalar parameter $\beta$ in $(0,1)$, we assume that each node $i$ in $\mc V$ is to choose $\deg_i$ distinct other nodes to link to with the aim of maximizing its own PageRank centrality in the resulting graph $\mc G$ with intrinsic centrality $\eta$ and discount factor $\beta$.  Hence, we model such multi-objective decision problems as a finite game $ \Gamma(\mc V,\beta,\eta, \mathbf{d}) $ where:  
\begin{itemize}
\item the player set is $\mc V$; 
\item the action set $\mc A_i$ of a player $i$ in $\mc V$ coincides with the family of all $\binom{n-1}{\deg_i}$ subsets of $\mc V\setminus\{i\}$ of cardinality $\deg_i$;  
\item the utility $u_i(x)$ of a player $i$ in $\mc V$ in a \emph{configuration}  $x=(x_1,\dots ,x_n)$ in $\mathcal{X}=\prod_{i\in\mc V}\mc A_i $ 
coincides with the $i$-th entry of the PageRank centrality vector 
$$\pi(x)=(1-\beta)(I-\beta R^\top(x))^{-1}\eta$$
where $R(x)$ is the normalized adjacency matrix 
of the graph $\mc G(x)=(\mc V, \mc E(x))$ with node set $\mc V$ and link set
$\mc E(x)=\{(i,j)\;|\; i\in\mc V,\; j\in x_i\}\,.$
\end{itemize}
In the rest of this paper we shall refer to a game as above as the \emph{centrality game} $\Gamma(\mc V,\beta,\eta, \degprof)$. 
We shall use the following standard game-theoretic notions. 
For a configuration $ x=(x_1,\dots ,x_n)$ in $\mc X$ and a player $i$ in $\mc V$, $x_{-i}$ in $\mc X_{-i}=\prod_{k\neq i}\mc A_k$ is the action profile of all players but $i$ and  we write $u_i(x)=u_i(x_i, x_{-i})$ for her utility. 
The \emph{best response set} of player $i$ to an action profile $x_{-i}$ in $\mc X_{-i}$ is 
$$\mathcal{B}_i(x_{-i})= \left\{a\in \mc A_i\,|\, u_i(a, x_{-i})\geq u_i(b, x_{-i}) ,\,\forall b\in\mc A_i\right\}\,.$$
A (pure strategy) \emph{Nash equilibrium} is a configuration $x$ in $\mc X$ such that $x_i\in\mc B_i(x_{-i})$ for every player $i$ in $\mc V$:  
it is \emph{strict} if $\{x_i\}=\mc B_i(x_{-i})$ for every player $i$ in $\mc V$. Following \cite{Monderer.Shapley:96}, we refer to a game as: 
\begin{itemize}
\item \emph{ordinal potential} if there exists a function $ \Psi:\mc X\to \mathbb{R}$, to be referred to as an ordinal potential, such that  
\be\label{ordinal-potential}u_i(x)<u_i(y) \;\Leftrightarrow\; \Psi(x)<\Psi(y)\,,\ee
for every player $i$ in $\mc V$ and every two configurations $x$ and  $y$ in $\mc X$ such that  $x_{-i}=y_{-i}$. 
\item {\emph{exact potential} if there exists a function $ \Psi:\mc X\to \mathbb{R}$, to be referred to as an exact potential, such that  
\be\label{exact-potential}u_i(y)-u_i(x)=\Psi(y)-\Psi(x)\,,\ee
for every player $i$ in $\mc V$ and every two configurations $x$ and  $y$ in $\mc X$ such that  $x_{-i}=y_{-i}$.} 
\end{itemize}
Clearly, exact potential games are also ordinal potential. It is well known \cite{Monderer.Shapley:96} that finite ordinal potential games always admit Nash equilibria. In fact, their set of Nash equilibria can be very large. For the sake of getting a more insightful classification, we introduce the following notions: 
\begin{itemize}
\item for $l\ge0$, a $(l+1)$-tuple of configurations $(x^{(0)},x^{(1)}, \dots ,x^{(l)})$ in $\mc X^{l+1}$ is a length-$l$ \emph{best response path} from $x^{(0)}$ to $x^{(l)}$ if for every $1\le k\le l$, $x^{(k)}\ne x^{(k-1)}$ and there exists $i_k$ in $\mc V$ such that
$$x^{(k)}_{i_{k}}\in\mc B_{i_{k}}\left(x^{(k-1)}_{-i_{k}}\right)\,,\qquad x^{(k)}_{-i_{k}}=x^{(k-1)}_{-i_{k}}\,,$$ i.e., it is obtained by a sequence of single player modifications choosing best response actions; 
\item a configuration $x$ in $\mc X$ is \emph{recursive} if for every other configuration $y$ in $\mc X$ such that there exists a best response path from $x$ to $y$, there is also a best response path from $y$ to $x$.
\end{itemize}
For ordinal potential games, recursive configurations enjoy the following remarkable properties. 
\begin{lemma}\label{lemma:recursive-Nash}For a finite ordinal potential game:
\begin{enumerate}
\item[(i)] \tcb{every strict Nash equilibrium is recursive;}
\item[(ii)] \tcb{every maximizer of the potential function over the configuration space is recursive;} 
\item[(iii)] every recursive configuration is a Nash equilibrium; 
\item[(iv)]  the set of recursive Nash equilibria is invariant with respect to best response paths; 
\item[(v)] \tcb{every subset of Nash equilibria that is invariant with respect to best response paths contains only recursive Nash equilibria;}
\item[(vi)] from every configuration there is a best response path to a recursive Nash equilibrium. 
\end{enumerate}
\end{lemma}
\begin{proof}
Let $\Psi:\mc X\to\R$ be an ordinal potential function. 

\emph{(i)} If $x^*$ is a strict Nash equilibrium, then, by definition, the only best response path starting in $x^*$ is the trivial length-$0$ one. This readily implies that $x^*$ is recursive.  

\emph{(ii)}  Let $(x^{(0)},x^{(1)}, \dots ,x^{(l)})$ be a best response path starting in a maximizer of the potential function $x^{(0)}$ in $\argmax\{\Psi(x):x\in\mc X\}$. By \eqref{ordinal-potential}, we have 
$\Psi(x^{(0)})\le\Psi(x^{(1)})\le\dots\le\Psi(x^{(l)})\le\Psi(x^{(0)})$. 
This yields $\Psi(x^{(0)})=\Psi(x^{(1)})=\dots=\Psi(x^{(l)})$ and \eqref{ordinal-potential}
implies that $(x^{(l)},x^{(l-1)}, \dots ,x^{(0)})$ is also a best response path, thus proving that $x^{(0)}$ is recursive. 

\emph{(iii)} Let $x$ in $\mc X$ be a configuration that is not a Nash equilibrium. By definition, there exist $i$ in $\mc V$ and $y$ in $\mc X$ with $y_{-i}=x_{-i}$ and $u_i(x)<u_i(y)$. 
Then, \eqref{ordinal-potential} implies that $\Psi(x)<\Psi(y)$. Now, if $x$ were recursive, then there would exist a best response path $(y=x^{(0)},x^{(1)}, \dots ,x^{(l)}=x)$ from $y$ to $x$. 
But then, \eqref{ordinal-potential} again would imply that $\Psi(y)=\Psi(x^{(0)})\le \Psi(x^{(1)})\le\Psi(x^{(l)})=\Psi(x)$, thus leading to a contradiction. We have thus shown that, if $x$  is not a Nash equilibrium, then it cannot be recursive. 

\emph{(iv)} Let $x$ in $\mc X$ be recursive and assume that there exists a best response path $\gamma_1$ from $x$ to another configuration $y$ in $\mc X$.  
Now, let there be a best response path $\gamma_2$ from $y$ to a third configuration $z$ in $\mc X$. Since the concatenation of $\gamma_1$ and $\gamma_2$ is itself a best response path from $x$ to $z$ and $x$ is recursive, there should exist a best response path $\gamma_3$ from $z$ to $x$. Then, the concatenation of $\gamma_3$ and $\gamma_1$ is a best response path from $z$ to $y$. This proves that $y$ is recursive. 

\tcb{\emph{(v)} Let $\mc F$ be a subset of Nash equilibria that is invariant with respect to best response paths. Let  $(x^{(0)},x^{(1)}, \dots ,x^{(l)})$ be a best response path starting in some $x^{(0)}$ in $\mc F$, so that $x^{(k)}\in\mc F$ is thus a Nash equilibrium for all $0\le k\le l$. It then follows that, for every $0\le k<l$ there exists $i_k$ in $\mc V$ such that $x^{(k+1)}_{-i_k}=x^{(k)}_{-i_k}$ and $u_{i_k}(x^{(k+1)})=u_{i_k}(x^{(k)}_{-i_k})$, so that $(x^{(l)},x^{(l-1)}, \dots ,x^{(0)})$ is also a best response path. This implies that $x^{(0)}$ is recursive.}

\emph{(vi)} Consider the finite directed graph $\mc H=(\mc X,\mc F)$ whose node set is the set of configurations $\mc X$ and where there is a link from $x$ to $y$ is and only if $y_{-i}=x_{-i}$ and $u_i(x)\le u_i(y)$ for some $i$ in $\mc V$. Let $\mc H_1=(\mc X_1,\mc F_1),\mc H_2=(\mc X_2,\mc F_2),\ldots,\mc H_s=(\mc X_s,\mc F_s)$, be the sink connected components of $\mc H$. Then, from every $x$ in $\mc X$, some $y$ in $\mc Y=\cup_{h=1}^s\mc X_h$ is reachable. the proof is completed by observing that, by construction, all configurations in $\mc Y=\cup_{h=1}^s\mc X_h$ are recursive. 	
\qed\end{proof}
\tcb{The following example clarifies the difference between Nash equilibria, recurrent Nash equilibria, strict Nash equilibria, and maximizers of the potential function. 
\begin{example}\label{ex:2player-game}
Consider a two-player game with action set $\mc A=\{a,b,c,d\}$ for both players and identical utilities 
$$u_1(x)=u_2(x)=\left\{\ba{ll}2&\se x_1\text{ and } x_2\in\{c,d\}\\1&\se  x_1=x_2=b\\0& \text{ otherwise}\,.\ea\right.$$
(See also the table in Figure \ref{fig:2player-game}.) 
This is an identical interest game, hence an exact potential game with potential function $\Psi(x)=u_1(x)=u_2(x)$. Notice that: 
$x^{*}=(a,a)$ is a Nash equilibrium that is not recurrent; 
$x^{\bullet}=(b,b)$ is a strict Nash equilibrium but it is not a maximizer of the potential $\Psi$; 
$x^{\circ}=(c,c)$ is a non-strict recurrent Nash equilibrium and it is a maximizer of the potential $\Psi$. 
\begin{figure}
\begin{center}\includegraphics[height=3cm]{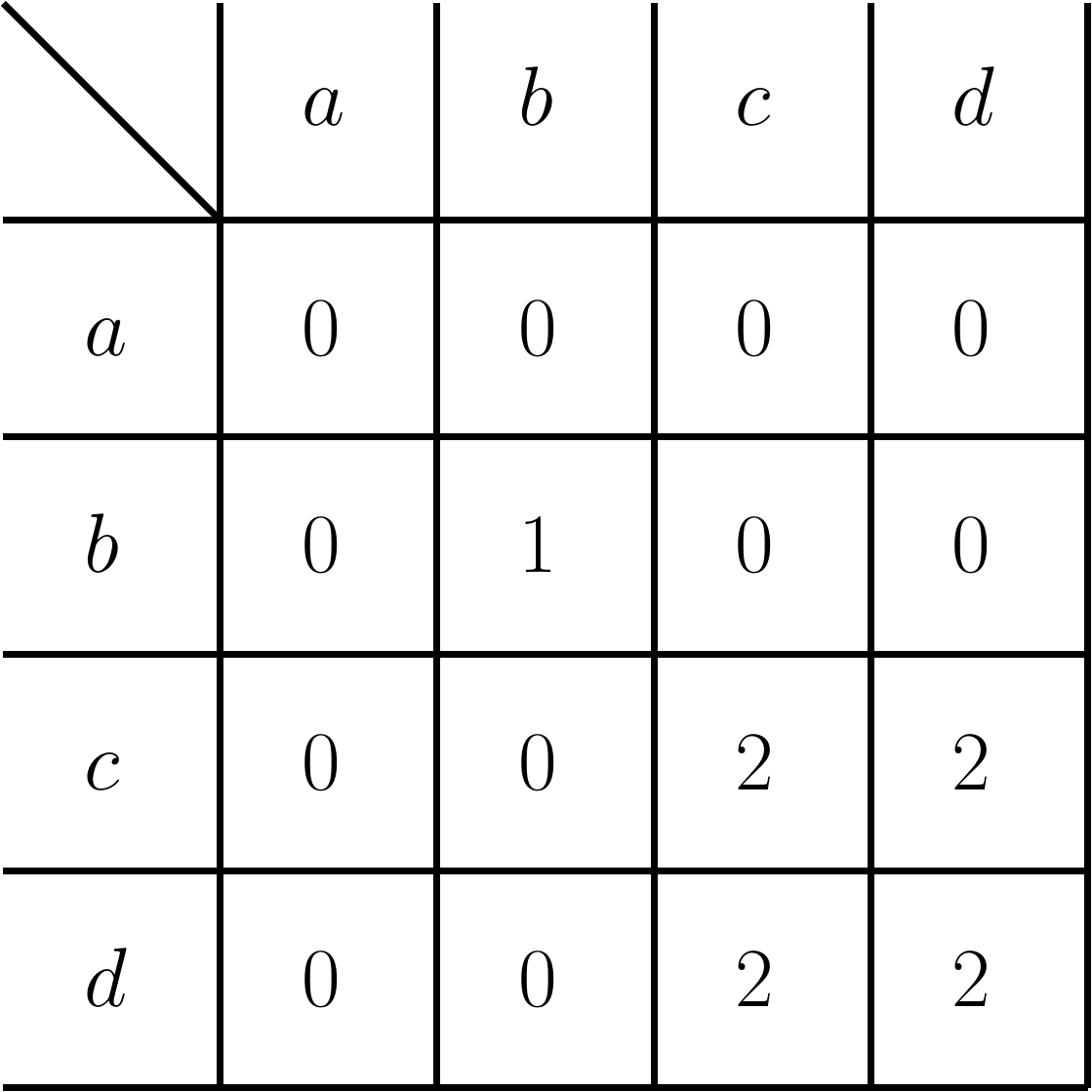}\end{center}
\caption{\label{fig:2player-game}Table representation of the 2-player identical interest game in Example \ref{ex:2player-game}.}
\end{figure}
\end{example}}
\begin{figure}\begin{center}
\includegraphics[height=2cm]{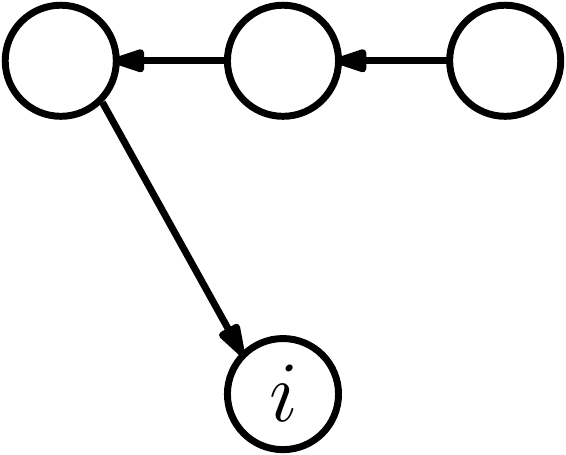}\hspace{0.3cm}
\includegraphics[height=2cm]{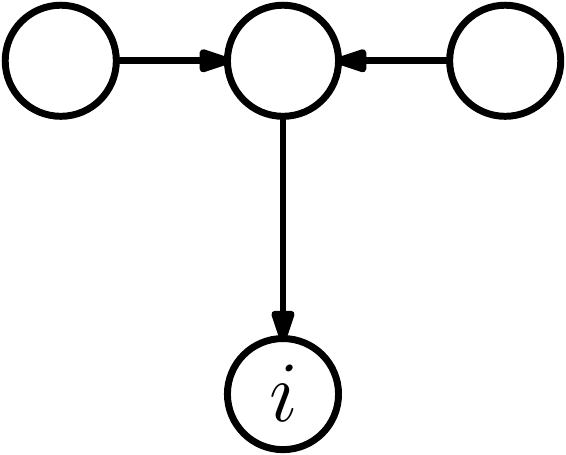}\hspace{0.3cm}
\includegraphics[height=2cm]{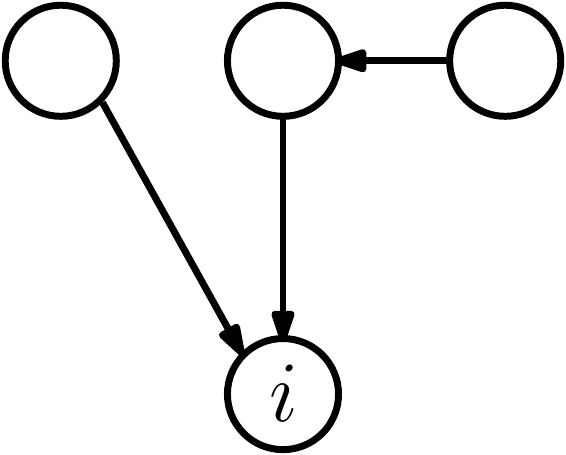}\hspace{0.3cm}
\includegraphics[height=2cm]{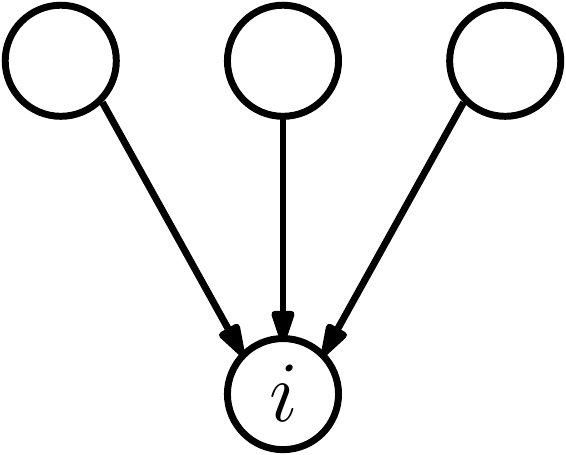}\end{center}
\tcb{\caption{\label{fig:rooted}For $|\mc V|=4$, all four possible ---up to isomorphisms--- spanning directed rooted trees in a given node $i$.  }}
\end{figure}
We will now show that centrality games are ordinal potential games. \tcb{Towards this goal, for a node $i$ in $\mc V$, consider the set $\mathbb T_i$ of \emph{spanning directed rooted trees}, i.e., acyclic graphs $\mc T=(\mc V, \mc E_{\mc T})$ with node set  $\mc V$, whereby each node has out-degree $1$ except for a node $i$ that has out-degree $0$ (\tcb{see Figure \ref{fig:rooted} for an illustration}). Then, for a configuration $x$ in $\mc X$,  define
\be\label{ni(x)def}n_i(x_{-i})=\sum_{\mc T\in\mathbb T_i}\prod_{(j,k)\in\mc E_{\mc T}}\left(\beta\1_{x_j}(k)+(1-\beta)\eta_k\right)\,,\ee
for every node $i$ in $\mc V$, and let 
\be\label{Z(x)} Z(x)=\sum_{i\in\mc V}n_i(x_{-i})\,.\ee 
Observe that the right-hand side of \eqref{ni(x)def} coincides with sum, over all possible spanning directed rooted trees $\mc T$ in $\mathbb T_i$, of the product of the weights $P_{jk}(x)=\left(\beta\1_{x_j}(k)+(1-\beta)\eta_k\right)$ of all the $n-1$ links $(j,k)$ in $\mc E_{\mc T}$. A key observation is then that, since no link $(j,k)$ in $\mc E_{\mc T}$ of any spanning directed rooted tree $\mc T$ in $\mathbb T_i$ is an outgoing link from node $i$, the right-hand side of \eqref{ni(x)def} does not depend on the action $x_i$ of node $i$, which justifies writing, as we did, the left-hand side  of \eqref{ni(x)def}  as a function of $x_{-i}$ only.} 
\tcb{We then  have the following instrumental result. 
\begin{lemma}
In a centrality game $ \Gamma(\mc V,\beta,\eta,\degprof)$, 
\be\label{strat-eq}u_i(x)=\frac{n_i(x_{-i})}{Z(x)}\,,\ee
for every player $ i$ in $\mc V $ and configuration  $x$ in  $\mc X$.
\end{lemma}
\begin{proof} By the Markov Chain Tree Theorem \cite{Anantharam.Tsoucas:1989}, the entries of the invariant distribution $\pi(x)$ of the irreducible stochastic matrix $P(x)=\beta R(x)+(1-\beta)\mathbf{1}\eta^\top$ are as in \eqref{strat-eq}. 
\qed\end{proof}}

\tcb{
\begin{proposition}\label{prop:ordinal_potential} Consider a centrality game $\Gamma=\Gamma(\mc V,\beta,\eta,\degprof)$ and let $\ov\Gamma$ be the game with the same player set $\mc V$ and configuration space $\mc X$, and utilities $$\overline u_i(x)=\log u_i(x)\,,\qquad \forall i\in\mc V\,,\quad\forall x\in\mc X\,.$$ 
Let $\Psi(x)=-\log Z(x)$. 
Then:  
\begin{enumerate}
\item[(i)] $\Psi(x)$ is an exact potential for $\ov\Gamma$;  
\item[(ii)] $\Psi(x)$ is an ordinal potential for $\Gamma$. 
\end{enumerate}
\end{proposition}}
\begin{proof} \tcb{
\emph{(i)} For every $i$ in $\mc V$ and $x$ in $\mc X$, \eqref{strat-eq} implies that
$\log u_i(x)=-\log Z(x)+\log n_i(x_{-i})$. 
For $x$ and $y$ in $\mc X$ such that $x_{-i}=y_{-i}$, the above implies that 
\be\label{point1}\ba{rcl}\ov u_i(y)-\ov u_i(x)
&=&\log u_i(y)-\log u_i(x)\\
&=&-\log Z(y)+\log Z(x)+\log n_i(y_{-i})-\log n_i(x_{-i})\\
&=&-\log Z(y)+\log Z(x)\\
&=&\ov\Psi(y)-\ov\Psi(x)
\,,\ea\ee
thus showing that $\ov\Psi(x)=-\log Z(x)$ is an exact potential for the game with utilities $\ov u_i(x)$.}

\tcb{\emph{(ii)} This follows from \emph{(i)} and the fact that $\sgn (u_i(x)-u_i(y))=\sgn(\ov u_i(x)-\ov u_i(y))$ for every configurations $x$ and $y$ in $\mc X$.
\qed}\end{proof}

\tcb{We shall denote by $\nash$ the set of all Nash equilibria of a centrality game $\Gamma(n,\beta,\eta,\degprof)$ and with the symbols
$\recnash$ and $\strictnash$, the subsets, respectively, of the recursive and of the strict Nash equilibria.
We shall also denote by $\nashZ=\argmin\{Z(x):x\in\mc X\}$ the set of maximizers of the ordinal potential function $\Psi(x)=-\log Z(x)$. 
It then follows from Lemma \ref{lemma:recursive-Nash} that   
$$\nash\supseteq \recnash\supseteq \strictnash\,,\qquad \nash\supseteq \recnash\supseteq \nashZ\,.$$ }
\tcb{The previous results have direct implications on the asymptotic behavior of some standard learning dynamics based on stochastic strategy revision processes. Consider asynchronous dynamics modeled as discrete-time Markov chains with finite state space $\mc X$ whereby, at every time step $t=0,1,2,\ldots$, conditioned on the current configuration $X(t)=x$, one player $i$ is selected uniformly at random from the player set $\mc V$ and she updates her action to a value $X_i(t+1)=a$ sampled from a conditional distribution $p_i(a|x_{-i})$ while all other players keep playing the same action $X_{-i}(t)=x_{-i}$. In particular, we shall refer to the case when $p_i(a|x_{-i})=p^{(0)}_i(a|x_{-i})=\1_{\mc B_i(x_{-i})}(a)/|\mc B_i(x_{-i})|$ (so that the active player choses her new action uniformly at random from the her best response set $\mc B_i(x_{-i})$) as the (asynchronous) best response dynamics (c.f.~\cite{Blume:1995}). On the other hand, we shall refer to the case when $p_i(a|x_{-i})=p_i^{(\gamma)}(a|x_{-i})=(u_i(a,x_{-i}))^{1/\gamma}/(\sum_{b\in\mc A_i}(u_i(b,x_{-i}))^{1/\gamma})$ for some $\gamma>0$ as the noisy best response dynamics with noise level $\gamma$: this can be recognized as the log-linear learning dynamics \cite{Blume:1993,Marden.Shamma:2012} for the game $\ov\Gamma$ with utilities $\ov u_i(x)=\log u_i(x)$. Observe that $p_i^{(\gamma)}(a|x_{-i})\to p_i^{(0)}(a|x_{-i})$ as $\gamma\downarrow0$. }
\tcb{\begin{corollary}\label{coro:dynamics} Consider a centrality game $\Gamma(\mc V,\beta,\eta,\degprof)$. 
Then, 
\begin{enumerate}
\item[(i)] for every initial configuration, the best response dynamics gets absorbed in finite time in the set of recursive Nash equilibria $\recnash$; 
\item[(ii)] for every noise level $\gamma\ge0$, the noisy best response dynamics is reversible with respect to its unique stationary distribution 
$$\mu_x^{(\gamma)}=\frac{Z(x)^{-1/\gamma}}{\sum_yZ(y)^{-1/\gamma}}\,,$$
and 
$$\mu_x^{(\gamma)}=\lim_{\gamma\downarrow0}\mu_x^{(\gamma)}=\frac1{|\nashZ|}\1_{\nashZ}(x)\,.$$
\end{enumerate}
\end{corollary}
\begin{proof}The result follows directly from Proposition \ref{prop:ordinal_potential} and \cite{Blume:97}.\qed\end{proof}.}

\section{Equilibrium Analysis for General Out-Degree Profiles}\label{sec:mainResults}
In this section, we present our main results on the structure of Nash equilibria in centrality games $ \Gamma(\mc V,\beta,\eta,\degprof) $ with general out-degree profile $\degprof$. 
\tcb{We recall our standing notation: given a configuration $x$, we denote by $\mc G(x)$ the corresponding graph and use a similar functional dependence for every other graph-theoretic concept. E.g., we shall write $\mc N^{-l}_i(x)$, $\mc N^{-\infty}_i(x)$ and often, since such sets only depend on $x_{-i}$, also  $\mc N^{-l}_i(x_{-i})$. Similarly, we shall denote by $\tau_j^i(x_{-i})$ and $\tau_{\eta}^i(x_{-i})$ the expected hitting times of a node $i$ in $\mc V$ for the Markov chain with transition probability matrix $P(x)=\beta R(x)+(1-\beta)\mathbf{1}\eta^\top$, since as already observed in Section \ref{sec:graph}, they do not depend on $x_i$. }

\subsection{Locality of Best Response}
Observe that Lemma \ref{lemma:tau} allows us to rewrite the utility of a player $i$ in $\mc V$ as
\be\label{eq:best-times}
u_i(x)
=\left(1+(1-\beta)\tau_{\eta}^i(x_{-i})+\frac{\beta}{\deg_i}\sum_{j\in x_i}\tau_j^i(x_{-i})\right)^{-1}\,,\qquad x\in\mc X\,, 
\ee
leading to the following result. 
\begin{proposition}\label{cor:best_response}
In a centrality game $ \Gamma(\mc V,\beta,\eta,\degprof)$, for every $ i$ in $\mc V $ and  $ x_{-i}$ in  $\mc X_{-i} $,
\be\label{br-character} x_i\in  \mathcal{B}_i(x_{-i}) \qquad\Longleftrightarrow\qquad \tau_{k}^i(x_{-i})\geq \max_{j\in x_i}\tau_{j}^i(x_{-i}),\quad \forall k\notin x_i\,.\ee
\end{proposition}
\begin{proof} The only term in the right hand side of \eqref{eq:best-times} that depends on the action $x_i$ of player $i$ is the set over which the summation runs. It follows that 
$$
\mathcal{B}_i(x_{-i})=\underset{x_i\in\mathcal{A}_i}{ \text{argmin}}\sum_{j\in x_i}\tau_j^i(x_{-i})\,.
$$
The minimization above is easily solved by choosing any $\deg_i$-tuple of nodes in $\mc V\setminus\{i\}$ that have the smallest hitting times to $i$, thus proving the claim. 
\qed\end{proof}

Proposition \ref{cor:best_response} reduces the computation of the best response actions of a player $i$ to finding the $\deg_i$ nodes that have the smallest expected hitting times \tcb{of node $i$}. 
The main difficulty in applying it directly stems from the fact that, as already observed at the end of Section \ref{sec:graph}, such expected hitting times depend not only on the action profile $x_{-i}$ of the rest of the players, which determines the graph except for node $i$'s out-neighborhood $x_i$, but also on the discount factor $\beta$ and on the intrinsic centrality $\eta$. 
We now present a remarkable result asserting that certain fundamental inequalities however always hold, independently from $\beta$ and $\eta$. 
\begin{proposition}\label{prop:best_repsonse} Let $\mc G=(\mc V,\mc E)$ be a graph, $\beta$ a scalar in $(0,1)$, and $\eta$ a probability vector in $\R_+^{\mc V}$. Then, for every $i$ in $\mc V$, the expected hitting times solution of \eqref{eq:system_tau2} satisfy the following properties:
\begin{enumerate}
\item[(i)] the relative order of $\{\tau_j^i:\,j\in\mc V\}$ does not depend on $\eta$;  
\item[(ii)] the relative order of $\{\tau_j^i:\,j\in\mc  N^{-\infty}_i\}$ only depends on the subgraph $\mc G_{\mc N^{-\infty}_i}$; 
\item[(iii)] for every $h,j$ in $\mc V\setminus\mc N^{-\infty}_i $ and $k$ in $\mc N^{-\infty}_i $, 
\begin{equation}\label{eq:tau-infty}\tau^i_h=\tau^i_j>\tau^i_k;\end{equation}
\item[(iv)] for every $k$  in $\mc N^{-\infty}_i $ and cut-set $\mc C \subseteq \mc V\setminus \lbrace k,i\rbrace $ between $k$ and  $i$,
\be\label{eq:min_WB}\tau^i_k> \min_{j\in \mc C}\,\tau^i_j\,.\ee
\tcb{\item[(v)] for every three nodes $h$, $i$, and $j$ in $\mc V$, such that $h\ne i\ne j$ and $\deg_h=\deg_j$, 
 $$\mc N_j\setminus\{h\}=\mc N_h\setminus\{j\}\qquad\Longrightarrow\qquad \tau^i_h=\tau^i_j\,.$$
}
\end{enumerate}
\end{proposition}
\begin{proof} 
We start with a simple but crucial observation. For a node $i$ in $\mc V$, consider the recursive relations \eqref{eq:system_tau2} characterizing the expected hitting times $\tau_j^i$ as $j$ varies in $\mc V$. Define
\begin{equation}\label{transformation}\tilde\tau_j^i=\frac{\tau_j^i}{1+(1-\beta)\tau_{\eta}^i}\,,\qquad i,j\in\mc V\,,
\end{equation}
and notice that these quantities satisfy the relations 
\begin{equation}\label{eq:system_tau2bis}\tilde\tau_i^i=0\,,\qquad 
\tilde\tau_j^i=1+\frac{\beta}{\deg_j}\sum\limits_{k\in\mc N_j}\tilde\tau_k^i\,\qquad j\ne i\, .
\end{equation}
This yields that the values $\tilde\tau_j^i$ coincide with the hitting times on node $i$ for the case when $\eta_j=\delta^i_j$. 
As the transformation \eqref{transformation} preserves the relative order of the hitting times, this proves point \emph{(i)}. 

Notice that when $\eta=\delta^i$ the equations in \eqref{eq:system_tau2} corresponding to nodes $j$ in $\mc N^{-\infty}_i$ are completely decoupled from the remaining equations, so that their solutions (hence, their relative order) only depend on the subgraph {induced by} $\mc N^{-\infty}_i$. Together with point \emph{(i)}, this proves point \emph{(ii)}.

\tcb{Thanks again to point \emph{(i)}, it is sufficient to prove points \emph{(iii)}, \emph{(iv)}, and \emph{(v)} in the special case $\eta_j=\delta^i_j$, as their statements concern only on the relative order of the hitting times.} 
In this case, for every $j$ in $\mc V\setminus\mc N^{-\infty}_i$, we have that $\mc N_j\subseteq\mc V\setminus\mc N^{-\infty}_i$, so that
\tcb{the equations in \eqref{eq:system_tau2} corresponding to nodes $j$ in $\mc V\setminus\mc N^{-\infty}_i$ are decoupled from those corresponding to nodes $j$ in $\mc N^{-\infty}_i$ and it can be directly verified by substitution that the unique solution $\tilde\tau^i$ of \eqref{eq:system_tau2bis} is such that} \be\label{tildetau1}\tilde\tau_j^i=1/(1-\beta)\,,\qquad\forall j\in\mc V\setminus\mc N^{-\infty}_i\,.\ee
\tcb{Let now
$\mc M=\argmax\{\tilde\tau_j^i:\,j\in\mc N^{-\infty}_i\}$, and, for $j$ in $\mc M$, let $\alpha^i_j=|\mc N_j\cap\mc N_i^{-\infty}|/\deg_j$. 
Then, \eqref{eq:system_tau2bis}, the fact that $\tilde\tau_k^i\le\tilde\tau_j^i$ for all $k$ in $\mc N_j\cap\mc N_i^{-\infty}$, and \eqref{tildetau1} imply that 
$$\tilde\tau_j^i=\displaystyle 1+\frac{\beta}{\deg_j}\sum\limits_{k\in\mc N_j}\tilde\tau_k^i
=1+\frac{\beta}{\deg_j}\sum\limits_{k\in\mc N_j\cap\mc N_i^{-\infty}}\tilde\tau_k^i+\frac{\beta}{\deg_j}\sum\limits_{k\in\mc N_j\setminus\mc N_i^{-\infty}}\tilde\tau_k^i
\leq 1+\beta\alpha^i_j\tilde\tau_j^i+\beta(1-\alpha^i_j)\frac{1}{1-\beta}\,,$$
which implies that \be\label{tildetau}\tilde\tau_j^i\leq 1/(1-\beta)\,,\qquad \forall j\in\mc N_i^{-\infty}\,.\ee To prove \emph{(iii)}, we need to show that the inequality in \eqref{tildetau} is strict. Assume by contradiction that $\tilde\tau_j^i= 1/(1-\beta)$ for some $j$ in $\mc N_i^{-\infty}$. Then, for every $j$ in $\mc M$,  \eqref{tildetau} implies that $\tilde\tau_j^i= 1/(1-\beta)$ and $\mc N_j\cap\mc N_i^{-\infty}\subseteq \mc M$. Since $j\in\mc N^{-\infty}_i$, an iteration of this argument implies that also $i\in\mc M$, so that $\tilde\tau^i_i=1/(1-\beta)$, which contradicts the first identity in \eqref{eq:system_tau2bis}. Thus point \emph{(iii)} is proven. }

Let $\mc W$ be the set of nodes in $\mc N^{-\infty}_i \setminus\mc C$ such that every path connecting them to $i$ hits the cut set $\mc C$.  Suppose by contradiction that 
$\min\{\tilde\tau_j^i:\,j\in \mc W\}\leq \min\{\tilde\tau_j^i:\,j\in \mc C\}\,,$
and let $j$ in $\mc W$ be any minimum point. Notice that $\mc N_j\subseteq \mc W\cup\mc C\cup (\mc V\setminus\mc N^{-\infty}_i)$ Then, by construction,
$$\tilde\tau_j^i=\displaystyle 1+\frac{\beta}{\deg_j}\sum\limits_{k\in\mc N_j}\tilde\tau_k^i\geq 1+\beta \tilde\tau_j^i\;\Rightarrow\; \tilde\tau_j^i\geq\frac{1}{1-\beta}$$
This contradicts point \emph{(iii)} and proves point \emph{(iv)}. 

\tcb{Finally, to prove point \emph{(v)}, observe that for three nodes $h$, $i$, and $j$ in $\mc V$, such that $h\ne i\ne j$ and $\deg_h=\deg_j$, 
we have $\mc N_j\setminus\{h\}=\mc N_h\setminus\{j\}$ if and only if either $\mc N_j=\mc N_h$ do not contain neither $h$ nor $j$, or 
$h\in\mc N_j$, $j\in\mc N_h$, and $\mc N_j\setminus\{h\}=\mc N_h\setminus\{j\}$. In the former case, \eqref{transformation} and \eqref{eq:system_tau2bis} imply that 
$$\tilde\tau^i_h=1+\frac{\beta}{\deg_h}\sum_{k\in\mc N_h}\tilde\tau^i_k=1+\frac{\beta}{\deg_j}\sum_{k\in\mc N_j}\tilde\tau^i_k=\tilde\tau^i_j\,,$$
On the other hand, in the latter case,  \eqref{transformation} and \eqref{eq:system_tau2bis} imply that 
$$\tilde\tau^i_h-\tilde\tau^i_j=\frac{\beta}{\deg_h}\sum_{k\in\mc N_h}\tilde\tau^i_k-\frac{\beta}{\deg_j}\sum_{k\in\mc N_j}\tilde\tau^i_k=
\frac{\beta}{\deg_h}(\tilde\tau^i_h-\tilde\tau^i_j) \,,$$
so that $(1-\beta/\deg_h)(\tilde\tau^i_h-\tilde\tau^i_j)=0$, i.e., $\tilde\tau^i_h=\tilde\tau^i_j$. }
\qed\end{proof}

A number of important properties of the structure of Nash equilibria of centrality games are a direct consequence of Propositions \ref{cor:best_response} and \ref{prop:best_repsonse}. 
Our first main result concerns the structure of the best response sets and shows how the optimal wiring strategy of a node is independent from the intrinsic centrality $\eta$ and satisfies a locality property. 
Precisely, given the choices of all the other players,  the $ \deg_i $ nodes to which a node $i$ needs to link in order to maximize her utility are nodes from which node $i$ itself can be reached in at most $\deg_i$ hops, provided that there is a sufficient number of such nodes. 

\begin{theorem}[\textbf{Independence from $\eta$ and locality of the best response}]\label{thm:best_response}
Let $ \Gamma(\mc V,\beta,\eta,\degprof)$ be a centrality game. Then, for every player $i$ in $\mc V$ and action profile $x_{-i}$ in $\mc X_{-i}$:
\begin{enumerate}
\item[(i)]the best response set $\mathcal{B}_i(x_{-i})$ does not depend on the intrinsic centrality $\eta$;
\item[(ii)] if $\vert \mc N^{-\infty}_i(x_{-i}) \vert {>} \deg_i$, then
\begin{equation}
x_i\in\mathcal{B}_i(x_{-i})\,\, \Rightarrow\,\, x_i\subseteq\mc N^{-\deg_i}_i(x_{-i}){\setminus\{i\}}\,,
\end{equation}
and, from every $j$ in $x_i$ there exists a path $(r_0, r_{1},\dots,r_{l-1},r_l)$ from $r_0=j$ to $r_l=i$ in $ \mc G(x) $ such that $\{r_k:\,\le k<l\}\subseteq x_i$ and 
\be\label{tau>}\tau_{j}^i(x_{-i})=\tau_{r_{0}}^i(x_{-i})>\tau_{r_{1}}^i(x_{-i})>\cdots >\tau_{r_{l-1}}^i(x_{-i})>\tau_{r_{l}}^i(x_{-i})=\tau_{i}^i(x_{-i})=0\,.\ee
\item[(iii)] if $\vert \mc N^{-\infty}_i(x_{-i}) \vert {\leq} \deg_i $, then %{$\mc  N^{-\infty}_i(x_{-i})=\mc N^{-\deg_i}_i(x_{-i})$ and} 
\begin{equation}
x_i\in\mathcal{B}_i(x_{-i})\,\, \Leftrightarrow\,\, x_i\supseteq {\mc N^{-\deg_i}_i(x_{-i})\setminus\{i\}}\,.
\end{equation}
\end{enumerate}
\end{theorem}
\begin{proof}
\emph{(i)} This is a direct consequence of Propositions \ref{cor:best_response} and \ref{prop:best_repsonse}\emph{(i)}.

\emph{(ii)} Let $x_i$ in $\mathcal{B}_i(x_{-i})$ be a best response action. Then, Proposition \ref{cor:best_response} and Proposition \ref{prop:best_repsonse}\emph{(iii)} imply that $x_i\subseteq \mc N_i^{-\infty}(x_{-i})$. Given any $j$ in $x_{i}$, 
suppose by contradiction that there is no path 
$(j=r_0, r_{1},\dots, r_l=i)$ from $j$ to $i$ in $ \mc G(x) $ such that $r_k\in x_i$ for all $0\le k<l$. Then, the set $\mc C=\mc V\setminus (x_i\cup\{i,j\})$ would be a cut between $j$ and $i$ and Proposition \ref{prop:best_repsonse}\emph{(iv)} would imply that 
$$\max\{\tau^i_k(x_{-i}):\,k\in x_i\}\ge\tau_{j}^j(x_{-i})>\min\{\tau^i_k(x_{-i}):\,k\in \mc C\}\,,$$ thus violating \eqref{br-character}. 
Hence, there exists at least one path $(j=r_0, r_{1},\dots, r_l=i)$ from $j$ to $i$ in $ \mc G(x) $ such that $r_k\in x_i$ for $0\le k<l$. For the shortest such path, we have $l\leq \deg_i$, thus implying that $j\in \mc N_i^{-\deg_i}(x_{-i})$. By the arbitrariness of $j$ in $x_i$, we then get that $x_i\subseteq\mc N_i^{-\deg_i}(x_{-i})$. 

To complete the proof of point \emph{(ii)}, first observe that 
the last inequality in \eqref{tau>} is straightforward since $\tau_{r_{l-1}}^i(x_{-i})\ge1>0=\tau_{i}^i(x_{-i})=\tau_{r_l}^i(x_{-i})$.  On the other hand, for $k= l-2,l-3,\ldots,0$, since $x_{r_k}=\mc N_{r_k}$ is a cut between $r_k$ and $i$, Proposition \ref{prop:best_repsonse}\emph{(iv)} implies that $\tau_{r_k}^i(x_{-i})>\min\{\tau_h^i(x_{-i}):\,h\in x_{r_k}\}$. As $x_i\in\mathcal{B}_i(x_{-i})$, there must exist $r_{k+1}$ in $x_{-i}\cap x_{r_k}$  such that  $\tau_{r_k}^i(x_{-i})>\tau_{r_{k+1}}^i(x_{-i})$. In this way, we can recursively construct a path from $j$ to $i$ satisfying  \eqref{tau>}.

\emph{(iii)} The claim follows from Propositions \ref{cor:best_response} and \ref{prop:best_repsonse}\emph{(iii)}, by observing that  $\mc N_i^{-\infty}(x_{-i})=\mc N_i^{-\deg_i}(x_{-i})$.
\qed
\end{proof}

Notice that Theorem \ref{thm:best_response}\emph{(ii)} implies that, if $\vert \mc  N^{-\infty}_i(x_{-i}) \vert {>} \deg_i$, i.e, if node $i$ is reachable by at least $\deg_i$ other nodes, then  in any best response action $x_i$ in $\mc B_i(x_{-i})$ one out-link of node $i$ is necessarily towards a node in its in-neighborhood, a second out-link is towards another  node in its in-neighborhood or to a node that is in the in-neighborhood of the previous node, and so on. 

An immediate consequence of Theorem  \ref{thm:best_response}\emph{(i)} is the following. 
\begin{corollary}\label{coro:independence} 
In a centrality game $ \Gamma(\mc V,\beta,\eta,\degprof)$, {the set of Nash equilibria $\nash$ and the subsets of the strict and of the recursive ones, respectively,  $\strictnash$ and  $\recnash$,} are all independent from $\eta$.
\end{corollary}
\begin{remark}While Corollary \ref{coro:independence} ensures their independence from the intrinsic centrality $\eta$, {the sets $\strictnash$,  $\recnash$, and $\nash$}  {may} depend on the discount factor $\beta$. On the other hand, the set of potential maximizing equilibria $\nashZ$ may also depend on $\eta$. 
\end{remark}

\subsection{Examples and Elementary Properties of Nash Equilibria}
In this subsection, for the sake of simplicity of notation, we often identify a configuration $x$ and the corresponding graph $\mc G(x)$ and often attribute game-theoretic properties directly to the latter (e.g., we will say that a certain graph is a Nash equilibrium).

We start by presenting a number of graphs that turn out to be Nash equilibria for a centrality game $ \Gamma(\mc V,\beta,\eta,\degprof)$ for every possible value of the discount factor $\beta$, as can be proved from the following symmetry result. 
\begin{corollary}\label{coro:symmetric}
{Consider a centrality game $ \Gamma(\mc V,\beta,\eta,\degprof)$ and a configuration $x$ such that $\mc G(x)=(\mc V,\mc E)$ is an undirected graph. If for every $i$ in $\mc V$ and $j$ and $k$ in $\mc N_i(x)$, there exists an automorphism $f$ of $\mc G(x)$ such that $f(i)=i$ and $f(j)=k$, then $x$ is a strict Nash equilibrium for every value of $\beta$ and $\eta$. }
\end{corollary}
\begin{proof}
{The assumption implies that $\tau_j^i(x_{-i})=\tau_k^i(x_{-i})$ for every $i$ in $\mc V$, $j$ and $k$ in $\mc N_i$. For every $h$ in $\mc V\setminus\mc N_i(X)$, since $\mc N_i(x)$ is a cut between $h$ and $i$, Proposition \ref{prop:best_repsonse} implies that $\tau^i_h(x_{-i})> \tau^i_j(x_{-i})$, thus proving that the unique best response of node $i$ is indeed $x_{i}$.}
\qed\end{proof}
\tcb{\begin{example}\label{ex:star} 
The complete bipartite graph $K_{l,m}$ is an undirected graph 
$(\mc V,\mc E)$ with node set $\mc V=\mc I\cup\mc J$, where $\mc I\cap\mc J=\emptyset$, $|\mc I|=l$, and $|\mc J|=m$, and link set  $\mc E=\{(i,j),(j,i):\,i\in\mc  I,\,j\in\mc J\}$. As a special case, when  $l=1$  and $m=n-1$, we recover the star graph $S_n$ displayed in Figure \ref{fig:complete-star-ring}(b). For $l,m\ge1$, $K_{l,m}$ satisfies the   symmetry assumption of Corollary \ref{coro:symmetric}, so that it is a strict Nash equilibrium.
\end{example}}\medskip
\begin{example}\label{ex:ring} 
Let $R_n$ be the ring graph, as displayed in Figure \ref{fig:complete-star-ring}(c). Clearly, $R_n$ satisfies the symmetry assumption of Corollary \ref{coro:symmetric}, so that it is a strict Nash equilibrium.
\end{example}\medskip
\begin{example}\label{ex:platonic} All platonic graphs, \tcb{i.e., graphs that have one of the Platonic solids as their skeleton} \cite[pp.~263 and 266]{Read.Wilson:98}, satisfy the symmetry assumption of Corollary \ref{coro:symmetric}, hence {they are all strict Nash equilibria}. 
\end{example}\medskip

While all the examples  above are undirected graphs, general strongly connected Nash equilibria may as well contain directed links, as it will become clear from the examples reported in Section \ref{sec:m=2}. 
The following result provides a lower bound on the number of undirected links \tcb{(i.e., pairs of nodes $\{i,j\}$ such that both directed  links $(i,j)$ and $(j,i)$ are links)} and 3-cycles in {any} strongly connected Nash equilibrium graph. 
\begin{corollary}\label{prop:degrees}
Let $x$ in $\nash$ be a Nash equilibrium of a centrality game $ \Gamma(\mc V,\beta,\eta,\degprof)$. Assume that $\mc G(x)=(\mc V, \mc E)$ is strongly connected.
Let $n=|\mc V|$ be the number of nodes,  $d_{*}=\min\{d_i :\,i\in\mc V\}$ be the minimum out-degree, $c_2$, and $c_3$ be, respectively, the number of undirected links and  of $3$-cycles in $\mc G(x)$. Then, 
\be\label{inequalities}2c_2\geq n\,,\qquad d_{*}c_3+2c_2\geq nd_{*}\,.\ee
\end{corollary}
\begin{proof}
For a node $i$ in $\mc V$, let $\deg_i^{\leftrightarrow}=|\{j\in\mc V:\,(i,j)\in\mc E,(j,i)\in\mc V\}|\le d_i$ be the number of its outgoing links forming undirected links. 
Since $\mc G(x)$ is strongly connected, $|\mc N_i^{-\infty}(x)|=n> \deg_i$, so that Theorem \ref{thm:best_response}\emph{(ii)}  implies that $\deg_i^{\leftrightarrow}\ge1$. 
Thus, $2c_2=\sum_i\deg_i^{\leftrightarrow}\ge n$, proving the first inequality in \eqref{inequalities}. 
If $\deg_i^{\leftrightarrow}<\deg_i$, by Theorem \ref{thm:best_response}\emph{(ii)} there exist $j\ne k $ in $\mc V$ such that
$(i,j), (j,i), (i,k), (k,j)\in\mc E$, while $(k,i)\not\in\mc E$. Hence, 
$c_3\geq \sum_{i}\min\{\deg_i-\deg_i^{\leftrightarrow},1\}\geq  \sum_{i}(\deg_i-\deg_i^{\leftrightarrow})/{\deg_i}\ge n-{\sum_i\deg_i^{\leftrightarrow}}/{\deg_*}= n-{2c_2}/{\deg_*}$, yielding 
the second inequality in \eqref{inequalities}.
\qed\end{proof}

A useful consequence of Proposition \ref{prop:best_repsonse}  is that the disjoint union of strongly connected graphs that are Nash equilibria remains a Nash equilibrium. Precisely, the following result holds true.
\begin{corollary}\label{cor:union} 
Consider  two disjoint sets $\mc V_1$ and $\mc V_2$, for $h=1,2$  the centrality games $ \Gamma(\mc V_h,\beta,\eta^{(h)},\degprof^{h}) $ and for each of them a (strict) Nash equilibrium  {$x^h$}. 
 Consider now the centrality game $ \Gamma(\mc V,\beta,\eta,\degprof) $ where  $\mc V=\mc V_1\cup\mc V_2$, $\eta_i=\eta^{(h)}_i$ and $\degprof_i=\degprof^{(h)}_i$ for all $i$ in $\mc V_h$ and $h=1,2$. Then, the configuration $x$ such that $x_i=x^h_i$ for all $i$ in $\mc V_h$ and $h=1,2$ is a (strict) Nash equilibrium.
 \end{corollary}
\begin{proof} Notice that $\mc G(x)$ is simply the disjoint union of the two graphs $\mc G(x^h)$ for $h=1,2$.
Fix $h$ in $\{1,2\}$ and $i$ in $\mc V_h$. Since, $\mc N_i^{-\infty}(x)=\mc V_h$ and $\deg^{h} _i\leq |\mc G(x^h)|-1$, it follows from Theorem \ref{thm:best_response}\emph{(ii)} that all best response sets for node $i$  in the centrality game $ \Gamma(\mc V,\beta,\eta,\degprof) $ are subsets of $\mc V_h\setminus\{i\}$. Moreover, it follows from  Proposition \ref{prop:best_repsonse}\emph{(ii)} that the set $\{\tau_j^i(x):\,j\in\mc V_h\}$ has the same order as the set $\{\tau_j^i(x^h):\,j\in\mc V_h\}$. This implies that node $i$ is currently playing a best response action in the union graph and such best response action is unique if it was unique for the centrality game $ \Gamma(\mc V_h,\beta,\eta^{(h)},\degprof^{h}) $, thus proving the claim. 
\qed\end{proof}
\begin{example} Any disjoint union of complete graphs, star graphs, and ring graphs is a strict Nash equilibrium for every value of the discount factor $\beta$ and of the intrinsic centrality $\eta$.
\end{example}

\subsection{Connectivity Structure of Nash Equilibria}
We now investigate in more generality the structure of the Nash equilibria of the centrality games.
Theorem \ref{thm:best_response} has important consequences on the connectivity structure of Nash equilibria. We start with a preliminary result.
\begin{lemma}\label{lemma:nash} Let $x$ in $\nash$ be a Nash equilibrium of a centrality game $ \Gamma(\mc V,\beta,\eta,\degprof)$ and let $\mc G_{\mc K}(x)$ be a connected component of $\mc G(x)$.
Then: 
\begin{enumerate}
\item[(i)] if $\deg_i< |\mc K|$ for every $i$ in $\mc K$, then $\mc G_{\mc K}(x)$ is a sink; 
\item[(ii)] if $\deg_i\geq  |\mc K|$ for some $i$ in $\mc K$, then $\mc G_{\mc K}(x)$ is a source.
\end{enumerate}
\end{lemma}
\begin{proof}Fix some $i$ in $\mc K$. Since $\mc G_{\mc K}(x)$ is a connected component, 
\begin{equation}\label{double-inclusion} \mc N_i\cap \mc N^{-\infty}_i\subseteq \mc K\subseteq\mc N^{-\infty}_i\,\end{equation}
so that in particular $|\mc K|\le |\mc N^{-\infty}_i|$. 

\emph{(i)} If  $\deg_i< |\mc K|$, then  $\deg_i{<} |\mc N^{-\infty}_i|$, so that Theorem \ref{thm:best_response}\emph{(ii)} implies that $\mc N_i \subseteq\mc N^{-\deg_i}_i$. This, by \eqref{double-inclusion}, yields
$\mc N_i\subseteq \mc K$. If this is true for every $i$ in $\mc K$, then $\mc G_{\mc K}(x)$ is necessarily a sink.

\emph{(ii)} If $\deg_i\geq |\mc K|$, then $\mc N_i \not\subseteq \mc K\setminus\{i\}$,  so that $\mc G_{\mc K}(x)$ cannot be a sink. Condition \eqref{double-inclusion} now implies that $\mc N_i\not\subseteq\mc  N^{-\infty}_i$,
so that  $d_i{\geq}|\mc N_i^{-\infty}|$ by Theorem \ref{thm:best_response}\emph{(ii)} and, consequently, $\mc N^{-\infty}_i{\setminus\{i\}}\subseteq \mc N_i$ by Theorem \ref{thm:best_response}\emph{(iii)}. Finally, using again  \eqref{double-inclusion}, we obtain that $\mc N^{-\infty}_i=\mc K$.  This implies that $\mc G_{\mc K}(x)$ is necessarily a source.
\qed\end{proof}
\tcb{\begin{remark}\label{rem:source-hom} In the special case of homogeneous out-degree profiles, i.e., $\degprof =d\mathbf 1$, Lemma \ref{lemma:nash} implies that a connected component $\mc G_{\mc K}(x)$ that is a source is necessarily such that $d\geq |\mc K|$. This in turn implies that $\mc G_{\mc K}(x)$ is a $|\mc K|$-clique and every node within it has $d-|\mc K|+1$ out-links towards other connected components that are sinks.
\end{remark}\medskip}

We conclude this section with the following result providing necessary conditions on the connectivity properties for a  graph to be a Nash equilibrium of a centrality game. \tcb{Before stating it, we remind the reader that, by convention, isolated components of a graph are classified as sinks but not as sources.}
\begin{theorem}[\textbf{Connectivity of Nash equilibria}]\label{thm:condensation_graph_generalm}
Let $x$ in $\nash$ be a Nash equilibrium of a centrality game $ \Gamma(\mc V,\beta,\eta,\degprof)$. Then, 
\begin{enumerate}
\item[(i)] every connected component of $\mc G(x)$ is either a sink or a source. 
\end{enumerate}
Moreover: 
\begin{enumerate}
\item[(ii)] if $x$ is recursive, then at most one of  the connected components of $\mc G(x)$ is a source;
\item[(iii)] if $x$ is strict and \tcb{$\max_i\deg_i<n-1 $}, then all the connected components of $\mc G(x)$ are isolated.
\end{enumerate}
\end{theorem}
\begin{proof}
\emph{(i)}  This follows directly from Lemma \ref{lemma:nash}.

\emph{(ii)} \tcb{By contradiction, let $\mc G_{\mc I}(x)$ and $\mc G_{\mc K}(x)$ be two distinct connected components of $\mc G(x)$ that are sources. Observe that, since $\mc G_{\mc I}(x)$ is not a sink, there must exist some $i$ in $\mc I$ such that $x_i\nsubseteq\mc I$. Then, by point \emph{(i)}, there exist a connected component $\mc G_{\mc J}(x)$ that is a sink and node $j$ in $x_i\cap\mc J$. 
In particular, $j\not\in\mc I\cup \mc K$.
It follows from Theorem \ref{thm:best_response}\emph{(iii)} that, for every $k$ in $\mc K$, $y_i=(x_{i}\setminus \{j\})\cup \{k\}$ is a best response action for player $i$. 
The graph $\mc G(y)$ possesses a connected component $\mc G_{\mc K}(y)$ that is neither a source nor a since. Thus, by item \emph{(i)}, $y$ is not a Nash equilibrium, hence Lemma \ref{lemma:recursive-Nash}\emph{(iii)} implies that $y$ is not recursive. Since $(x,y)$ is a length-$1$ best response path and $y$ is not recursive, the Nash equilibrium $x$ is not recursive.}

\tcb{\emph{(iii)} By contradiction, let  $\mc G_{\mc I}(x)$ be  a connected component of $\mc G(x)$ that is a source. Since $\mc G_{\mc I}(x)$ is not a sink, there exists $i$ in $\mc I$ such that $x_i\nsubseteq\mc I$. Moreover, since $\mc G_{\mc I}(x)$ is a source, we have that $\mc N_i^{-\infty}(x_{-i})=\mc I$. Hence, $|\mc I|=|\mc N_i^{-\infty}(x_{-i})|\le \deg_i$, for otherwise Theorem \ref{thm:best_response}\emph{(ii)} would imply that $x_i\subseteq\mc I$. Then, Theorem \ref{thm:best_response}\emph{(iii)} implies that every $y_i\subseteq\mc V\setminus\{i\}$ such that $|y_i|=\deg_i$ and $\mc N_i^{-\infty}(x_{-i})\setminus\{i\}\subseteq y_i$ is a possible best response action for player $i$. There are $\binom{n-|\mc I|}{\deg_i+1-|\mc I|}$ such subsets. By assumption,  $n>\deg_i+1$, so that $\binom{n-|\mc V'|}{\deg_i+1-|\mc I|}>1$, so that $x$ cannot be a strict Nash equilibrium.} \qed
\end{proof}
\begin{figure}\centering
\subfigure[\ ]{\includegraphics[height=2.5cm]{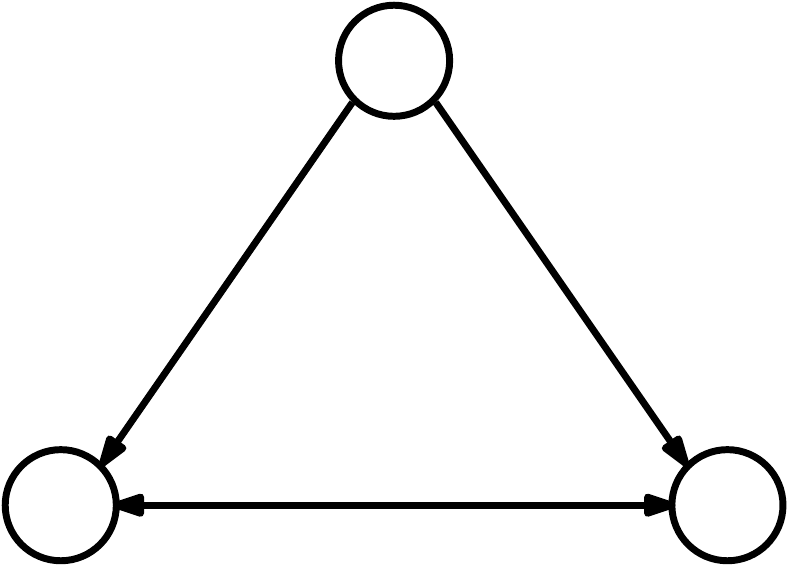}}\hspace{1cm} 
\subfigure[\ ]{\includegraphics[height=4cm]{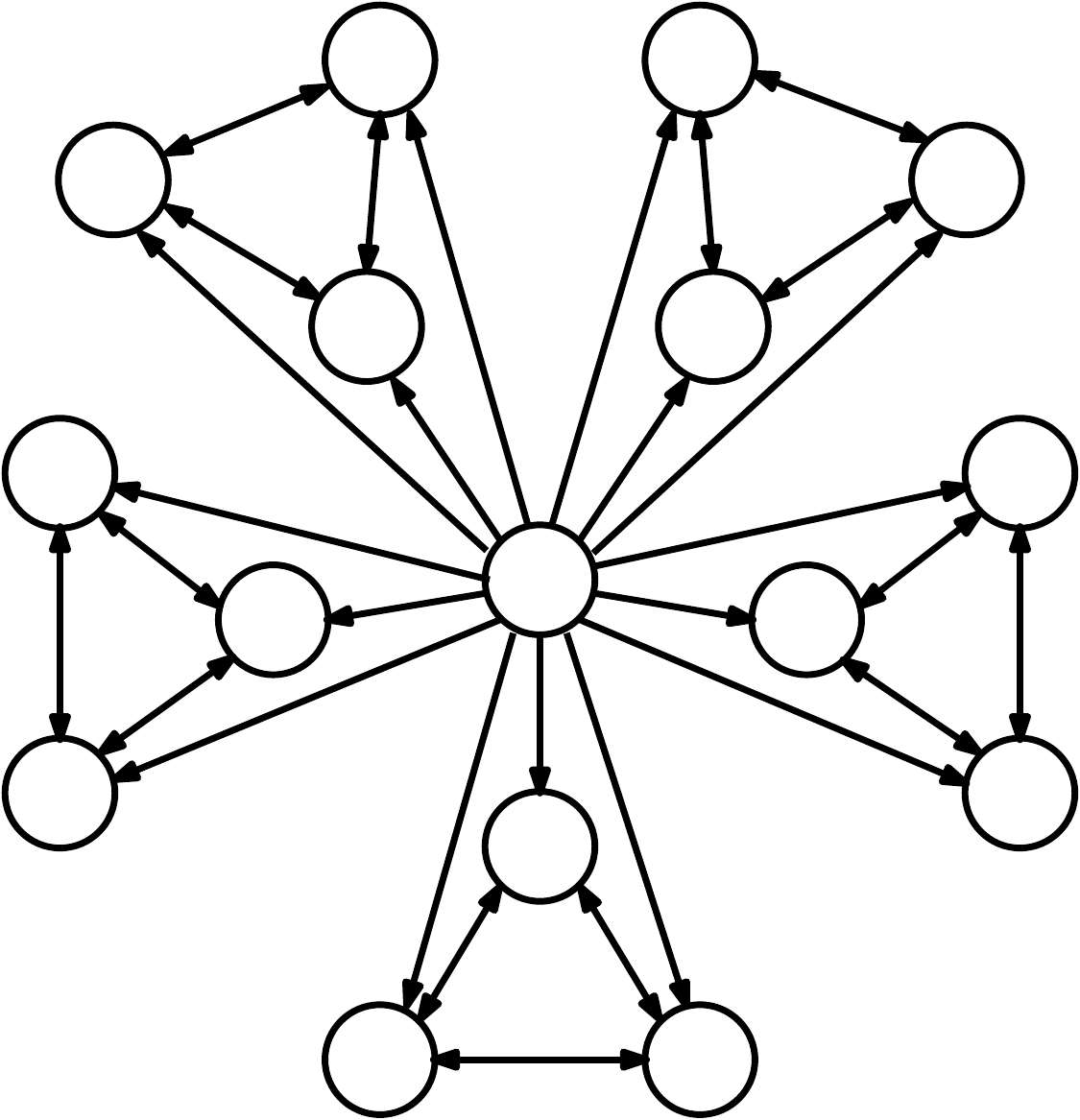}} 
\tcb{\caption{\label{fig:hub-spoke} Graphs of Example \ref{ex:hub-spoke} in the special cases: (a) $n=3$ and $d=1$; (b) $n=16$ and $d=2$. }}
\end{figure}
\tcb{\begin{remark} The assumption $\max_i\deg_i<n-1$ in Theorem \ref{thm:condensation_graph_generalm}(iii) cannot be removed. Indeed, e.g., for $n=3$ and $\degprof=(2,1,1)$, the configuration $x=(\{2,3\},\{3\},\{2\})$ is a strict Nash equilibrium, yet $\mc G(x)$ admits a connected component $\mc G_{\{1\}}(x)$ that is a source (see Figure \ref{fig:hub-spoke}(a)).  \end{remark}}

\subsection{\tcb{Analysis of Potential Maximizing Equilibria}}\label{sec:potential-maximizers}
\tcb{In this subsection, we focus on the subset $\nashZ$ of Nash equilibria where the potential $\Psi(x)=-\log Z(x)$ achieves its maximum value over the configuration space $\mc X$. In particular, we characterize the structure of such potential maximizing equilibria for values of the discount factor $\beta$ close to $1$.} 

\tcb{For a centrality game $ \Gamma(\mc V,\beta,\eta,\degprof)$, define the function $m:\mc X\to\N$ by
$$m(x)=\max_{i\in\mc V}|\mc N_i^{-\infty}(x_{-i})|\,,$$
associating to every configuration $x$ the maximum over all players $i$ in $\mc V$ of the number of nodes from which node $i$ is reachable in $\mc G(x)$. 
The key fact is the following result that determines the asymptotic behavior of the function $Z(x)$ as $\beta$  approaches $1$.} 

\tcb{\begin{proposition} \label{prop:betato1}   In a centrality game $ \Gamma(\mc V,\beta,\eta,\degprof)$, for every configuration $x$ in $\mc X$,
\be\label{Z-asymp}Z(x)\asymp(1-\beta)^{n-m(x)}\,,\ee
as $\beta\to1$. 
\end{proposition}
\begin{proof} For every configuration $x$ in $\mc X$ and spanning directed rooted tree $\mc T=(\mc V,\mc E_{\mc T})$ in $\mathbb T_i$, let
$$w_{\mc T}(x_{-i})=\prod_{(j,k)\in\mc E_{\mc T}}\left(\beta\1_{x_j}(k)+(1-\beta)\eta_k\right)\,,$$
be the weight of $\mc T$ in $x$. 
Let also $\mc E_{\mc T}^x=\mc E_{\mc T}{\cap}\mc E(x)$ be the set of links of $\mc T$ that are also links of $\mc G(x)$, 
and let $\ov{\mc E}_{\mc T}^x=\mc E_{\mc T}\setminus\mc E(x)$ be the set of  links of $\mc T$ that are not  links of $\mc G(x)$. 
Then,  define 
$$B_{\mc T}(\beta)=\prod_{(j,k)\in\mc E_{\mc T}^x}\left(1-(1-\beta)(1-\eta_k)\right)\,,\qquad 
C_{\mc T}=\prod_{(j,k)\in\ov{\mc E}^x_{\mc T}}\eta_k\,,$$
and observe that the weight of $\mc T$ in $x$ satisfies 
$$w_{\mc T}(x_{-i}) = 
\prod_{(j,k)\in\mc E^x_{\mc T}}\left(1-(1-\beta)(1-\eta_k)\right)
\prod_{(j,k)\in\ov{\mc E}^x_{\mc T}}(1-\beta)\eta_k
=B_{\mc T}(\beta)C_{\mc T}(1-\beta)^{|\ov{\mc E}^x_{\mc T}|}\,.$$
Since $B_{\mc T}(\beta)\stackrel{\beta\to1}{\longrightarrow}1$, it follows that 
$$w_{\mc T}(x_{-i})=B_{\mc T}(\beta)C_{\mc T}(1-\beta)^{|\ov{\mc E}^x_{\mc T}|}
= B_{\mc T}(\beta)C_{\mc T}(1-\beta)^{n-1-|\mc E^x_{\mc T}|}
\sim C_{\mc T}(1-\beta)^{n-1-|\mc E^x_{\mc T}|}\,,$$
as $\beta\to1$.
Now, notice that, for a given configuration $x$ in $\mc X$, the quantity $|\mc E^x_{\mc T}|$ is maximized by those spanning directed rooted trees $\mc T$ in $\mathbb T_i$ whose restriction to $\mc N_i^{-\infty}(x_{-i})$ is a subgraph of $\mc G(x)$, so that 
$\max_{\mc T\in\mathbb T_i}|\mc E^x_{\mc T}|=|\mc N_i^{-\infty}(x_{-i})|-1\,.$
It follows that 
$$n_i(x_{-i})=\sum_{\mc T\in\mathbb T_i}w_{\mc T}(x_{-i}) \sim \sum_{\mc T\in\mathbb T_i} C_{\mc T}(1-\beta)^{n-1-|\mc E^x_{\mc T}|}
\sim H_i^x(1-\beta)^{n-|\mc N_i^{-\infty}(x_{-i})|}\, $$ 
as $\beta\to1$, where $H_i^x=\sum_{\mc T\in\mathbb T_i^x}C_{\mc T}$ and $\mathbb T_i^x=\argmax\{|\mc E^x_{\mc T}|:\mc T\in\mathbb T_i\}$.
Hence,  \be\label{Zxsim}Z(x)=\sum_{i\in\mc V}n_i(x_{-i})\sim\sum_{i\in\mc V}H_i^x(1-\beta)^{n-|\mc N_i^{-\infty}(x_{-i})|}\sim K_x(1-\beta)^{n-m(x)}\,,\ee
as $\beta\to1$, where $K_x=\sum_{i\in\mc V_x}H_i^x$ and $\mc V_x=\argmax\{|\mc N_i^{-\infty}(x_{-i})|:\,i\in\mc V\}$. 
The result then follows from \eqref{Zxsim}.
\qed\end{proof}
Proposition \ref{prop:betato1}   allows one to prove the following result, stating that, for large enough values of the discount factor $\beta$, all configurations maximizing the potential of a centrality game 
are minimizers of the positive-integer valued function $m(x)$. 
\begin{theorem} \label{theo:betato1} In a centrality game $ \Gamma(\mc V,\beta,\eta,\degprof)$, 
there exists $\ov\beta<1$ such that 
\be\label{minZ}\nashZ\subseteq\argmin_{x\in\mc X}m(x)\,,\ee
for every $\beta$ in $(\ov\beta,1)$. 
\end{theorem}
\begin{proof} {It follows from Proposition \ref{prop:betato1}   and the fact that $\mc X$ is a finite set that there exists two positive numbers $0<c_1<c_2$ such that 
\be\label{asymp1}c_1\leq \frac{Z(x)}{(1-\beta)^{n-m(x)}}\leq c_2
\ee
for every $x$ in $\mc X$ and $\beta$ in $(0,1)$. Put $\ov\beta =1-c_1/c_2$ and take $\beta>\ov\beta$. Let $x^*$ in $\nashZ$ be a potential maximizer and assume by contradiction that $x^*\not\in \argmin\{m(x):x\in\mc X\}$. This implies that there exists $x^{**}$ in $\mc X$ such that $m(x^{**})\leq m(x^*)-1$. From relation \eqref{asymp1} we then obtain
$$Z(x^*)\geq c_1(1-\beta)^{n-m(x^*)}
\geq c_1(1-\beta)^{n-m(x^{**})-1}
\geq \ds\frac{c_1}{c_2(1-\beta)}Z(x^{**})
\geq\ds \frac{c_1}{c_2(1-\beta)}Z(x^{*})$$
This yields $\beta\leq 1-c_1/c_2=\ov\beta$ contradicting the assumption $\beta>\ov\beta$. Hence, the result follows.}
\qed\end{proof} 
Observe that there exists a simple lower bound on $m^*=\min\{m(x):x\in\mc X\}$, as stated below. 
\begin{lemma}\label{lemma:m*}
In a centrality game $ \Gamma(\mc V,\beta,\eta,\degprof)$, 
$$m^*\ge1+{\lceil}\ov d{\rceil}\,,\qquad \ov d=\frac1n\sum_{i\in\mc V}\deg_i\,.$$
\end{lemma}
\begin{proof}
Let $x$ in $\mc X$ be any configuration. 
Since the total in-degree of $\mc G(x)$ is equal to its total out-degree, we have that 
$$\sum_{i\in\mc V}|\mc N_i^{-1}(x_{-i})|=\sum_{i\in\mc V}\deg_i{+n}=n(\ov d{+1})\,.$$ 
Then, by the pigeonhole principle, there exists at least a node $i$ in $\mc V$ such that 
$|\mc N_i^{-1}(x_{-i})|\ge\lceil\ov d\rceil+1$. Since ${\mc N_i^{-\infty}(x_{-i})}\supseteq\mc N_i^{-1}(x_{-i})$,  it follows that 
$$m(x)=\max_{i\in\mc V}|\mc N_i^{-\infty}(x_{-i})|=\max_{i\in\mc V}(1+|\mc N_i^{-1}(x_{-i}))|\ge1+\lceil\ov d\rceil\,.$$
The claim then follows from the arbitrariness of the configuration $x$ in $\mc X$. 
\qed
\end{proof}}

\tcb{\begin{example}\label{ex:hub-spoke} 
Let $n$ and $d$ be positive integers such that $n-1$ is a multiple of $d+1$.
Consider an out-degree-profile such that $n-1$ nodes have the same out-degree $\deg_i=d$ for $i=1,2,\ldots,n-1$, and a hub node $n$ has out-degree $\deg_n=n-1$. 
Observe that $\ov d=(n-1)(d+1)/n$ so that $\lceil\ov d\rceil=d+1$. 
By Lemma \ref{lemma:m*} we have that 
$$m^*\ge1+\lceil\ov d\rceil=2+d\,.$$
Then, observe that $m^*=\min\{x\in\mc X:m(x)={d+2}\}$ is achieved by all those configurations $x$ in $\mc X$ such as $\mc G(x)$ is the union of $(n-1)/{d+1}$ $(d+1)$-cliques plus the hub node pointing all other nodes (see Figure \ref{fig:hub-spoke} for an illustration in the special cases $n=3$  and $d=1$, and  $n=16$ and $d=2$, respectively). It then follows from Theorem \ref{theo:betato1} that for for values of the discount factor $\beta$ sufficiently close to $1$, all  potential maximizers $x$ in $\nashZ$ are of this form. 
\end{example}}

\section{Equilibrium Analysis for Homogeneous Out-Degree Profiles} \label{sec:homogeneous}
In this section, we refine the equilibrium analysis in centrality games in the special case of homogeneous out-degree profiles. 
First, we analyze the structure of the potential maximizers $x^*$ in $\nashZ$ when $\degprof=d\cdot\mathbf 1$ for arbitrary positive integer $\deg$. 
Then, we focus on two special cases of homogeneous out-degree profiles: $\degprof=\mathbf 1$ and $\degprof=2\cdot\mathbf 1$, respectively. 
For these, we are able to  fully characterize the sets of Nash equilibria $\nash$, recursive Nash equilibria $ \recnash$, and strict Nash equilibria $\strictnash$.

\subsection{Potential Maximizing Equilibria for Homogeneous Out-Degree Profiles}\label{sec:m=k}
\begin{figure}
\subfigure[\ $\mc G(x)$]{\includegraphics[height=2.3cm]{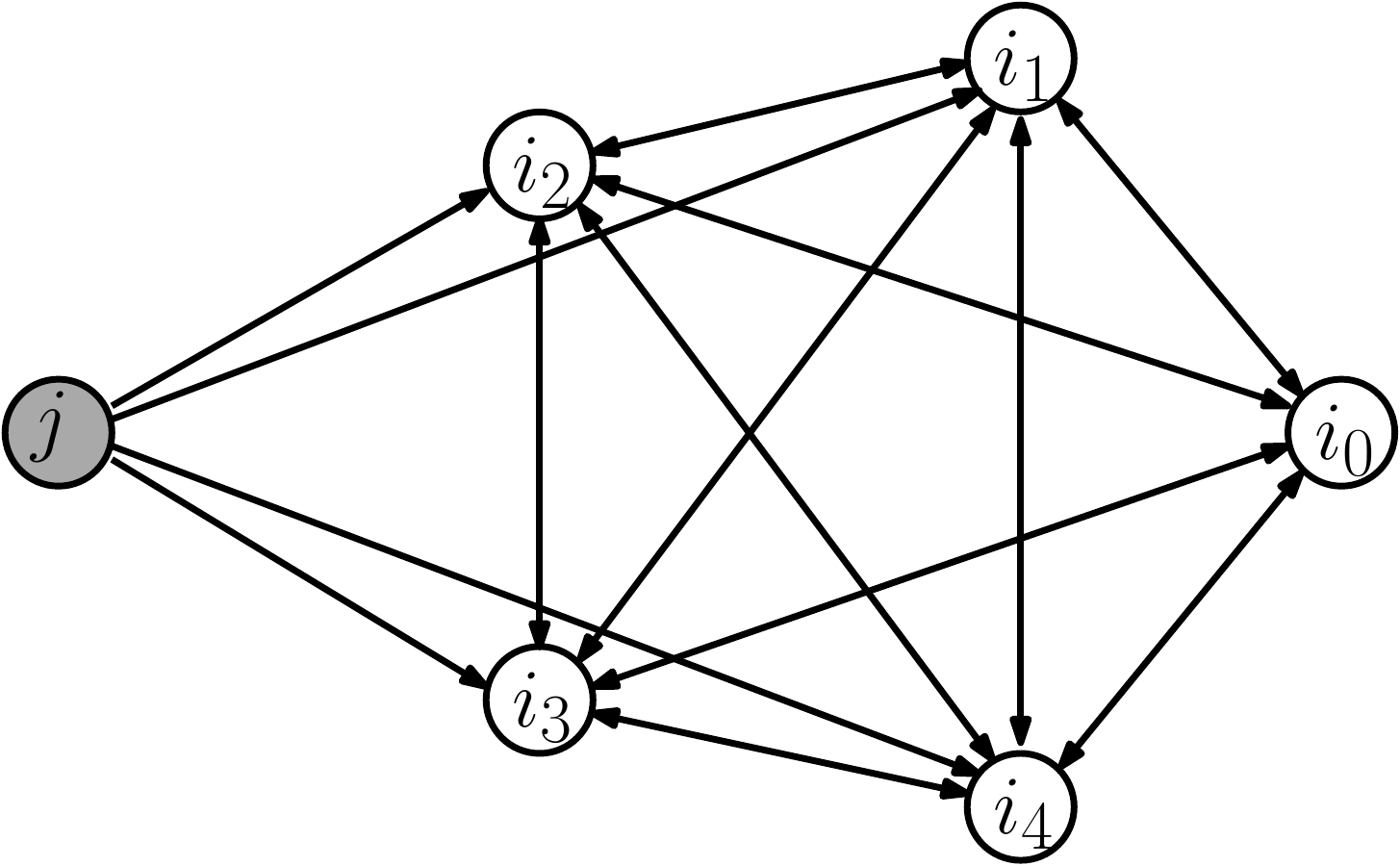} }\hspace{.2cm}
\subfigure[\ $\mc G(y)$]{\includegraphics[height=2.3cm]{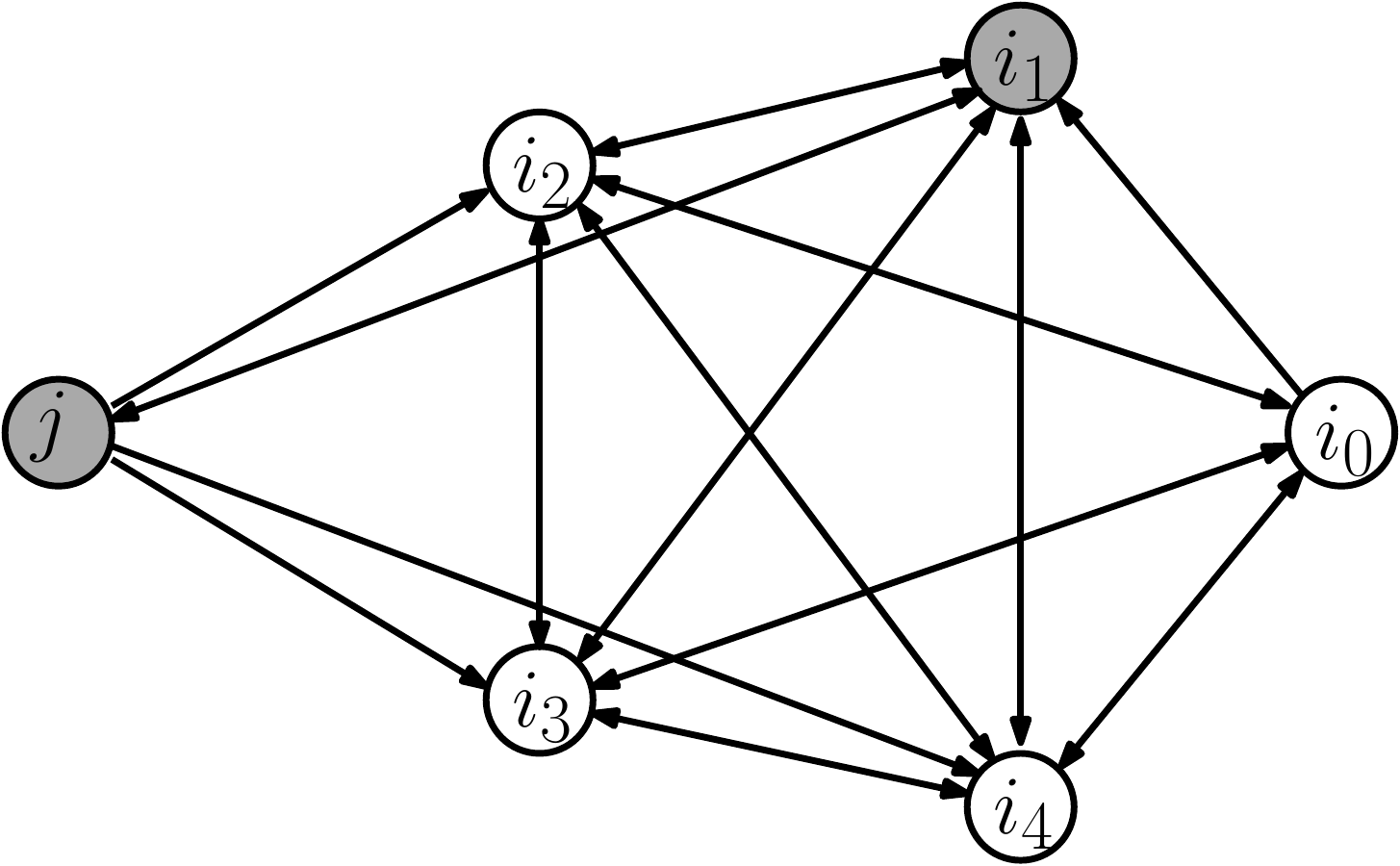} }\hspace{.2cm}
\subfigure[\ $\mc G(z)$]{\includegraphics[height=2.3cm]{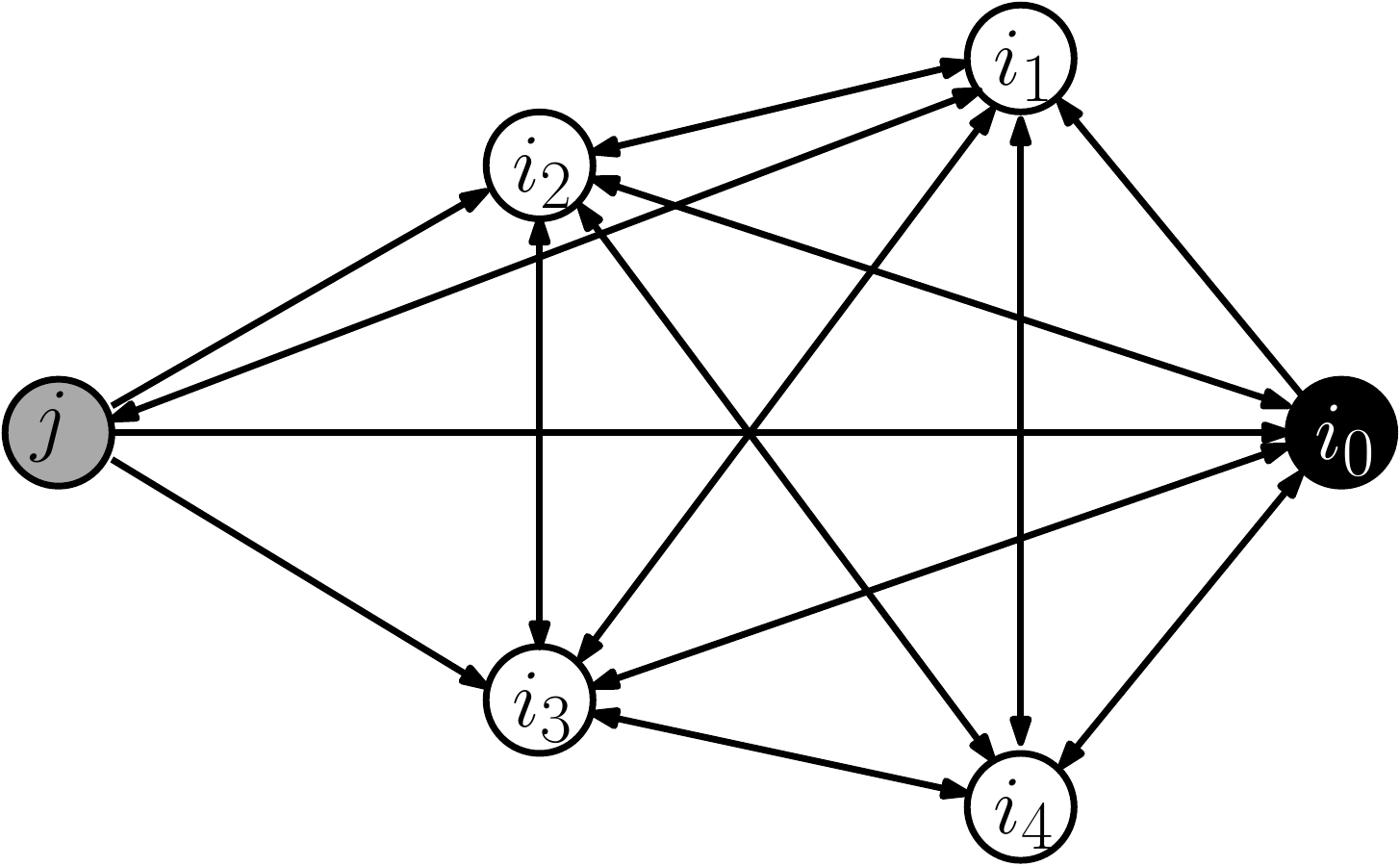} }
\tcb{\caption{\label{fig:noNash} Graphs associated to the best response path $(x,y,z)$ in the proof of Lemma \ref{lemma:notaNash} in the special case $d=4$.  
Configuration $z$ is not a Nash equilibrium since node $i_0$ strictly prefers linking to $j$ instead of $i_1$. }}
\end{figure}

We start with the following result, proving that certain configurations can never be recursive. 
\tcb{\begin{lemma} \label{lemma:notaNash}
Let $x$ in $\nash$ be a Nash equilibrium of a centrality game $ \Gamma(\mc V,\beta,\eta,\degprof)$.  If $\mc G(x)$ contains a $(d+1)$-clique $\mc G_{\mc K}(x)$ that is a sink component and a source node $j$ in $\mc V\setminus\mc K$ with $\deg_j=d$ and $x_j\subseteq\mc K$, then  $x$ is not recursive. 
\end{lemma}}
\begin{proof}
\tcb{We are going to construct a length-$2$ best response path from $x$ to a configuration $z$ in $\mc X$ that is not a Nash equilibrium (see Figure \ref{fig:noNash} for an illustration in the special case $d=4$). {By Lemma \ref{lemma:recursive-Nash}\emph{(iii)-(iv)},} this will imply that $x$ is not recursive.} 
\tcb{We start by labeling the nodes in the clique as $\mc K=\{i_0,i_1,\ldots,i_d\}$ in such a way that $x_j=\{i_1,\ldots,i_d\}$. 
We then consider a configuration $y$ such that $y_{-i_1}=x_{-i_1}$ and $y_{i_1}={(}x_{i_1}\setminus\{i_0\}{)}\cup\{j\}=\{i_2,\ldots,i_d,j\}$. 
Observe that since $x_{i_1}\in\mc B_{i_1}(x_{-{i_1}})$ (because $x$ is a Nash equilibrium) and $x_{j}=\{i_1,\ldots,i_d\}=x_{i_0}$, Proposition \ref{prop:best_repsonse}\emph{(v)} and Proposition \ref{cor:best_response} imply that also $y_{i_1}\in\mc B_{i_1}(x_{-{i_1}})$. 
Let then $z$ be a configuration such that $z_{-j}=y_{-j}$ and $z_{j}={(}y_{j}\setminus\{i_d\}{)}\cup\{i_0\}$. 
Observe that, since $y_{i_0}\setminus\{i_d\}=\{i_1,\ldots,i_{d-1}\}=y_{i_d}\setminus\{i_0\}$ and $y_{j}\in\mc B_{j}(y_{-j})$, Proposition \ref{prop:best_repsonse}\emph{(v)} implies that also $z_{j}\in\mc B_{j}(y_{-j})$. 
This proves that $(x,y,z)$ is a {best response} path. }
\tcb{To prove that $z$ is not a Nash equilibrium we show that $z_{i_0}\notin\mc B_{i_0}(z_{-i_0})$. Indeed, %let $a=z_{i_0}\setminus\{i_1\}\cup\{j\}$. Then, 
since $z_{i_1}=\{j\}\cup\{i_2,\ldots,i_d\}$ and $z_j=\{i_0, i_1,\ldots, i_{d-1}\}$, from \eqref{eq:system_tau2} we get 
{$$\ba{rcl}
\ds\tau_{i_1}^{i_0}(z_{-i_0})-\ds\tau_{j}^{i_0}(z_{-i_0})
&=&\ds\frac{\beta}d\sum_{h\in z_{i_1}}\tau_{h}^{i_0}(z_{-i_0})-\ds\frac{\beta}d\sum_{h\in z_{j}}\tau_{h}^{i_0}(z_{-i_0})\\ 
&=&\ds\frac{\beta}d\left(\tau_{j}^{i_0}(z_{-i_0})-\tau_{i_1}^{i_0}(z_{-i_0})\right)+\frac{\beta}d\tau_{i_d}^{i_0}(z_{-i_0})\,,
\ea$$
This yields }
$$\tau_{i_1}^{i_0}(z_{-i_0})-\tau_{j}^{i_0}(z_{-i_0})=\frac{\beta}{d-\beta}\tau_{i_d}^{i_0}(z_{-i_0})>0\,.$$
It then follows from Proposition \ref{cor:best_response} that $z_{i_0}\notin\mc B_{i_0}(z_{-i_0})$, so that $z$ is not a Nash equilibrium. By Lemma \ref{lemma:recursive-Nash}\emph{(iii)} that $z$ is not recursive. Finally, Lemma \ref{lemma:recursive-Nash}\emph{(iv)} implies that $x$ is not recursive.} \qed\end{proof}
\tcb{\begin{theorem}\label{theo:d=k}
Consider a centrality game $ \Gamma(\mc V,\beta,\eta, \degprof)$ with homogeneous out-degree profile $\degprof=d\cdot\mathbf 1$ with $1\le d<n=|\mc V|$. Then, there exists $\ov\beta<1$ such that, for every $\ov\beta<\beta<1$ and $x^*$ in $\nashZ$: 
\begin{enumerate} 
\item[(i)] if $n$ is a multiple of $d+1$, then $\mc G(x^*)$ is the union of isolated $(d+1)$-cliques; 
\item[(ii)] if $n>(d+1)d$, then $\mc G(x^*)$ is the union of isolated connected components each of order either $d+1$ or $d+2$. 
\end{enumerate} 
\end{theorem}}
\begin{proof} \tcb{We first establish some inequalities concerning the minimum value of the function $m(x)$.
Write $n=(d+1)q+r$ for some integers $q\ge1$ and $0\le r\le {d}$.  Consider any configuration $\ov x$ in $\mc X$ such that $\mc G(\ov x)$ is the union of $q$ disjoint $(d+1)$-cliques, {while the remaining $r$ nodes are split into} $c=\min\{r,q\}$ source components, each one of order at most $\lceil r/q\rceil$ and with all outgoing links pointing to the same clique, and such that distinct source components point to distinct cliques. Notice that constructing such configuration $\ov x$  is possible as $c\lceil r/q\rceil=\min\{r\lceil r/q\rceil,q\lceil r/q\rceil\}\geq r$. 
By construction, we have $m(\ov x)\le d+1+\lceil r/q\rceil$, so that 
 \be\label{s<=r/q-1} \min\{m(x):x\in\mc X\}\le m(\ov x)\le d+1+\lceil r/q\rceil\,.\ee  
On the other hand, observe that, in every configuration $x$ in $\mc X$, every node $i$ in a sink component $\mc G_{\mc K}(x)$ of $\mc G(x)$ points to nodes within $\mc K$ itself, so that $|\mc K|\ge\max\{d_i:i\in\mc K\}+1$. Since $\degprof=d\cdot\mathbf 1$, it follows that every sink component of $\mc G(x)$ has order $|\mc K|\ge d+1$, with equality if and only if $\mc G_{\mc K}$ is a $(d+1)$-clique.}
\tcb{Now, let $x$ in $\recnash$ be a recursive Nash equilibrium. Theorem \ref{thm:condensation_graph_generalm}\emph{(i)-(ii)} implies that all connected components of $\mc G(x)$ are sinks, except for possibly one source. Let $\mc G_{\mc S}(x)$ be such source component and let $s(x)=|\mc S|$ be its order, with $s(x)=0$ when there is no source component in $\mc G(x)$. Observe that every node $i$ in a sink component $\mc G_{\mc K}(x)$ that is linked by the source component $\mc G_{\mc S}(x)$ is reachable from all nodes in $\mc K\cup\mc S$ so that $|\mc N^{-\infty}_i(x_{-i})|{=}|\mc K|+|\mc S|\ge d+1+s(x)$. Hence, $ \forall x\in\recnash $, 
\be\label{m(x)>d+s}
 d+1+s(x)\le\min\{|\mc N^{-\infty}_i(x_{-i})|:i\in\mc K\}\le\max\{|\mc N^{-\infty}_i(x_{-i})|:i\in\mc V\}=m(x)\,.\ee 
Let $0<\ov\beta<1$ be as in Theorem \ref{theo:betato1} and consider a discount factor $\beta$ in $(\ov\beta,1)$. Let  $x^*$ in $\nashZ$ be a maximizer of the potential function, so that $m(x^*)=\min\{m(x):x\in\mc X\}$  by Theorem \ref{theo:betato1}. By Lemma \ref{lemma:recursive-Nash}\emph{(ii)}, $x^*\in\recnash$, so that \eqref{m(x)>d+s} and  \eqref{s<=r/q-1} imply that $\forall x^*\in\nashZ$,
\be\label{s<=r/h} d+1+s(x^*)\le m(x^*)=\min\{m(x):x\in\mc X\}\le d+1+\lceil r/q\rceil\, .\ee} 

\tcb{We are now ready to prove the two statements.
On the one hand, when $n$ is a multiple of $d+1$, so that $r=0$, \eqref{s<=r/h} implies that $s(x^*)=0$, so that no source component is present, and $m(x^*)=d+1$, so that every sink component has size $d+1$. As already observed, this can only happen if every sink is indeed an isolated $(d+1)$-clique, thus proving \emph{(i)}.
On the other hand, notice that we only need to prove \emph{(ii)} when $n>(d+1)d$ is not a multiple of $d+1$: in this case, necessarily, $\lceil r/q\rceil=1$ so that
\eqref{s<=r/h} implies that $s(x^*)\le1$ and $d+1\leq m(x^*)\leq d+2$ for every $x^*$ in $\nashZ$. Notice now that it cannot be that $m(x^*)= d+1$ as otherwise $\mc G(x)$ should be the union of $q$ disjoint $(d+1)$-cliques, which is impossible when $n$ is not a multiple of $(d+1)$. Therefore the only possibility is that $m(x^*)=d+2$. We now prove that also necessarily $s(x^*)=0$.
Assume by contradiction that $s(x^*)=1$ and let $j$ be the unique source node. If $j$ links to a sink component $\mc G_{\mc K}(x^*)$, we must have that $|\mc K|+1\leq m(x^*)=d+2$. Since, as already observed, $|\mc K|\ge d+1$ for every sink component, this implies that $|\mc K|=d+1$. Therefore, $\mc G_{\mc K}(x^*)$ is a $(d+1)$-clique. By Theorem \ref{thm:best_response}\emph{(iii)}, there exists a length-$1$ best response path, consisting in the rewiring of all the out-links of node $j$ towards $\mc K$, leading to a configuration $x^{**}$ where  $x_j^{**}\subseteq\mc K$. % all out-links from $j$ lead to the component $\mc G_{\mc K}(x^*)$.
Lemma \ref{lemma:notaNash} implies that $x^{**}$ is not recursive. By Lemma \ref{lemma:recursive-Nash}\emph{(iii)}, also $x^*\notin\recnash$. Then, Lemma \ref{lemma:recursive-Nash}\emph{(ii)} implies that $x^*\notin\nashZ$, leading to a contradiction.
Hence, necessarily, $s(x^*)=0$, so that $\mc G(x^*)$ is the union of isolated connected components. Since $m(x^*)=d+2$, the order of such components can be either $d+1$ or $d+2$. }
\qed\end{proof}

\begin{remark} It is worth noting that, if $n\le d(d+1)$ is not a multiple of $d+1$, then we cannot rule out the possibility that, for some potential maximizer $x^*$ in $\nashZ$, the associated graph $\mc G(x^*)$ contains a connected component that is a source. E.g., for $n=5$ and $d=2$,  a configuration with associated graph as the one displayed in Figure \ref{fig:trasitionButterfly}(b) is recursive and minimizes $m(x)$. \end{remark}

\subsection{The Case $\degprof=\mathbf 1$}\label{sec:m=1}
In this subsection, we consider centrality games with homogeneous out-degree profiles $\degprof=\mathbf 1$, i.e., when every node has a single out-link. 
In this case, the best response presented in Theorem \ref{thm:best_response} takes the following special form.

\begin{proposition}\label{prop:best_response1}
Consider a centrality game $ \Gamma(\mc V,\beta,\eta, \mathbf1) $. Then, 
\be\label{BR1}\mathcal{B}_i(x_{-i})=\left\{\begin{array}{lcl}\mc N_i^{-1}(x_{-i})\setminus\{i\} &\se&\mc N_i^{-1}(x_{-i})\neq \{i\}\\[7pt]
\mc V \setminus\{ i\}&\se&\mc N_i^{-1}(x_{-i})= \{i\}\end{array}\right.\ee 
for every player $i$ in $\mc V$ and action profile $x_{-i}$ in $\mathcal{X}_{-i}$. 
\end{proposition}

\begin{proof}
Since $\degprof=\mathbf1$, for every $j$ in $\mc N^{-1}_i(x_{-i})\setminus\{i\}$  we have that $x_j=\{i\}$, so that \eqref{eq:system_tau2} implies that $ \tau^i_j=1+(1-\beta)\sum_{v\in \mc V}\eta_v\tau^i_v$. Hence, if $\mc N_i^{-1}(x_{-i})\neq \{i\}$, then Theorem \ref{thm:best_response}\emph{(ii)} implies the result.
If $ \mc N_i^{-1}(x_{-i})= \{i\} $, then the result follows directly from Theorem \ref{thm:best_response}\emph{(iii)}.\qed
\end{proof}

We now introduce the family of graphs
$\mathcal K_2^{r}$  obtained by adding to a finite set of disjoint $2$-cliques, $r$ extra nodes, each of which having exactly one out-link pointing towards an arbitrary node belonging to any of the $2$-cliques. 
The following result provides a complete characterization of the sets of Nash equilibria, recursive Nash equilibra, and strict Nash equilibria, respectively, for centrality games with homogeneous out-degree profile $\degprof=\mathbf 1$. 

\begin{theorem}[\textbf{Nash equilibria for $\degprof =\mathbf 1$}]\label{thm:nash_m=1}  
For a centrality game $\Gamma(\mc V,\beta,\eta, \mathbf 1) $ and a configuration $x$:   
\begin{enumerate}
\item[(i)]  $x$ is a Nash equilibrium if and only if  $ \mc G(x)\in\bigcup_{r\geq 0}\mc K_2^{r}$; 
\item[(ii)] $x$ is a strict Nash equilibrium if and only if  $ \mc G(x)\in \mc K_2^{0}$. 
\item[(iii)] $x$ is a recursive Nash equilibrium if and only if  $ \mc G(x)\in\mc K_2^{0}\cup \mc K_2^{1}$; 
\end{enumerate}
\end{theorem}
\begin{proof}
\emph{(i)} If $x$ in $\mc X$ is such that $\mc G(x)\in\mc K^r_2$, then both nodes that belong to a $2$-clique as well nodes that link to a $2$-clique are playing a best response action according to  \eqref{BR1}, so that $x$ is a Nash equilibrium. Conversely, if $x$ is a Nash equilibrium, then for every node $i$ in $\mc V$, Proposition \ref{prop:best_response} guarantees that either there is another node $j$ such that both $(j,i)\in\mc E$ and  $(i,j)\in\mc E$, or $\mc N_i^{-1}=\{i\}$. In the former case, $\mc G_{\{i,j\}}(x)$ is a connected component that is a sink. In the latter case, $\mc G_{\{i\}}(x)$ is a connected component that is a source. Hence, $\mc G(x)$ is necessarily of type $ \mc K_2^{r} $.

\emph{(ii)} It follows from \tcb{Theorem}  \ref{thm:condensation_graph_generalm}\emph{(iii)} {that if $x$ is a strict Nash equilibrium, then}
$\mc G(x)\in\mc K^0_2$. On the other hand, if $x$ in $\mc X$ is  such that $\mc G(x)\in \mc K^0_2$, then it follows from \eqref{BR1} that every node has just one incoming link and is thus playing its unique best response action. This implies that $x\in \strictnash$.

\emph{(iii)} \tcb{It follows from Theorem  \ref{thm:condensation_graph_generalm}\emph{(ii)} that every recursive Nash equilibrium $x$ is such that all connected components of $\mc G(x)$ are sinks except for possibly one source. Since $\degprof=\mathbf1$, necessarily the sinks have order $2$ and the source has order $1$, so that} 
$\mc G(x)\in\mc K^0_2\cup\mc K^1_2$. Since $\mc G(x)\in\mc K^0_2$ implies that $x\in\strictnash$ by point \emph{(ii)}, it remains to show that $\mc G(x)\in\mc K^1_2$ implies that $x\in\recnash$. 
For that, let $x$ in $\mc X$ be such that $\mc G(x)\in \mc K^1_2$. Denote by $s$ the \tcb{unique source} node and let $i,j$ be the nodes in the $2$-clique such that $(s,i)\in\mc E(x)$. From \eqref{BR1}, all nodes in $\mc V\setminus\{s,i\}$ are currently playing their unique best response and no transition is thus possible. Also, from \eqref{BR1}, we deduce that $\mc B_s(x)=\mc V\setminus \{s\}$ and $\mc B_{i}(x)=\{j,s\}$.  
Therefore, \tcb{every best response path starting from $x$ makes a first step $y$ such that either 
$y_{-s}=x_{-s}$ and  $y_s\in\mc V\setminus\{i,s\}$, or $y_{-i}=x_{-i}$ and $y_{i}=s$. In both cases $\mc G(y)\in\mc K_2^1$. We now apply Lemma \ref{lemma:recursive-Nash}\emph{(v)} to the set of configurations $x$ such that $\mc G(x)\in\mc K_2^1$ and conclude that they are all recursive.}
\qed\end{proof}

 \begin{remark} Notice that, if the number of players $n$ is even, then Theorem \ref{thm:nash_m=1} implies that the set of recursive Nash equilibria of the centrality game $\Gamma(\mc V,\beta,\eta,\mathbf 1)$ coincides with the set of its strict Nash equilibra as they are both given by $\mc K_2^0$. Instead, if $n$ is odd, the centrality game $\Gamma(\mc V,\beta,\eta,\mathbf 1)$  admits no strict Nash equilibria, while the set of recursive Nash equilibra coincide with $\mc K_2^1$. 
 \end{remark}
 \subsection{The Case $\degprof=2\cdot\mathbf1$}\label{sec:m=2}
In this subsection, we focus on the special case of the centrality game $ \Gamma(\mc V,\beta, \eta,\degprof)$ with homogeneous out-degree profile $\degprof=2\cdot\mathbf1$, i.e.,  when every node has to place exactly two out-links. 
We shall first present a full classification of recursive and strict Nash equilibria and then provide a complete classification of the Nash equilibria of the centrality game. 
We start with following result. 
\begin{proposition}\label{prop:best_response}
Consider the centrality game $ \Gamma(\mc V,\beta,\eta,\degprof)$ with $\degprof=2\cdot\mathbf 1$.  Then,  
\be\label{BR2}\mathcal{B}_i(x_{-i})=\left\{\begin{array}{lcl}
\bigl\lbrace \lbrace j,k\rbrace\! : j,k\!\in\! \mc V,\, i\ne j\neq k\ne i \bigr\rbrace \quad &\text{ if }& \mc N_i^{-\infty}(x_{-i})=\{i\}\\[7pt]
\bigl\lbrace \lbrace j,k\rbrace \!:  k\in \mc V,\,  i\ne k\ne j  \bigr\rbrace  \qquad &\text{ if }&   \mc N_i^{-\infty}(x_{-i})=\{i,j\}\,,\end{array}\right.\ee
for every player $i$ in $\mc V$ and strategy profile $x_{-i}$ in $\mc X_{-i}$. 
Moreover, if $ |\mc N_i^{-\infty}(x_{-i})|> 2 $, then 
\begin{align}\label{BR2b} 
\mathcal{B}_i(x_{-i})\subseteq
&\left\{ \lbrace j,k\rbrace \!: j,k\in\mc  N^{-1}_i(x_{-i}),\, i\ne j\ne k\ne i\right\} \cup \\ &\cup \left\{ \lbrace j,k\rbrace \!: j\!\in\!\mc N^{-1}_i(x_{-i}),\, k\!\in\!\mc N^{-1}_j(x_{-j}),\,i\ne j\ne k\ne i\right\}\,. \notag
\end{align}
\end{proposition}
\begin{proof}
Relation \eqref{BR2} follows from Theorem \ref{thm:best_response}\emph{(iii)}. \tcb{On the other hand, \eqref{BR2b} follows from Theorem \ref{thm:best_response}\emph{(ii)}: if node $i$ is reachable in configuration $x$ by at least two nodes besides itself, then either its best response is a pair of nodes $j\ne k$ that both point directly towards it, or it consists of a node $j$ that points directly towards it and of a node $k$ that points directly towards $j$.}
\qed
\end{proof}
\begin{figure}\centering
\subfigure[\ $B_5$]{\includegraphics[height=2.3cm]{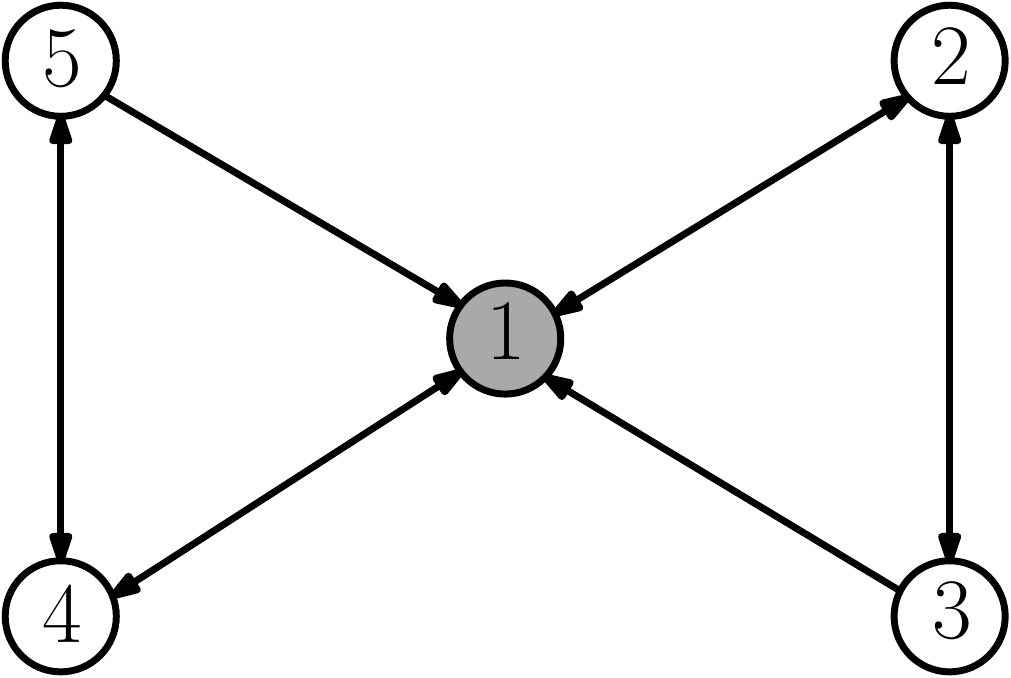} }\hspace{.4cm}
\subfigure[\ $B_5'$]{\includegraphics[height=2.3cm]{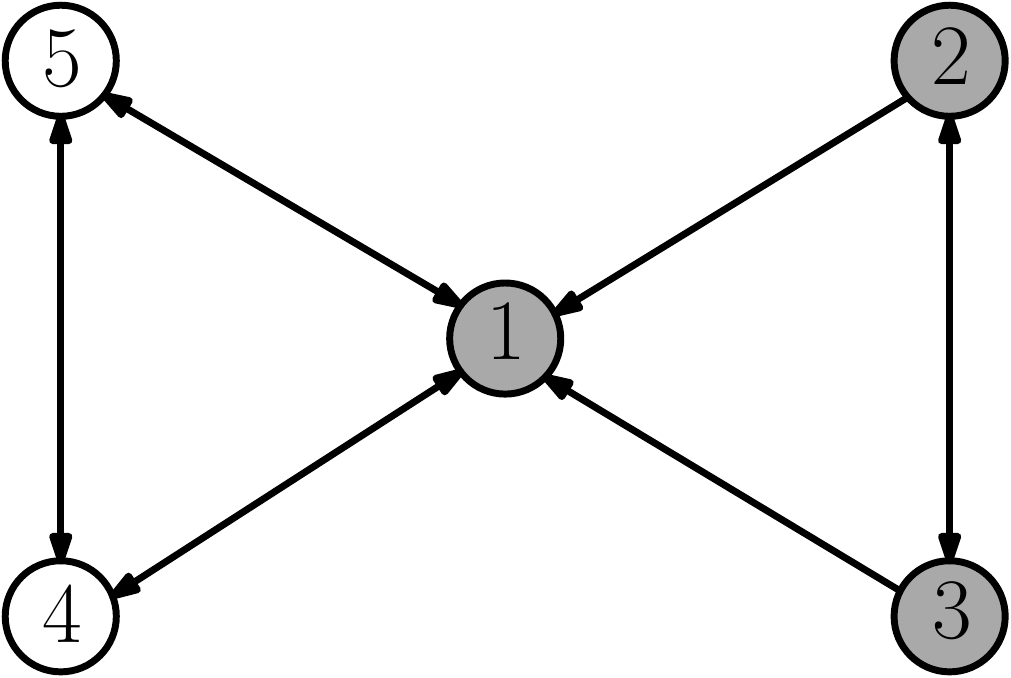} }\hspace{.4cm}
\subfigure[\ $B_5''$]{\includegraphics[height=2.3cm]{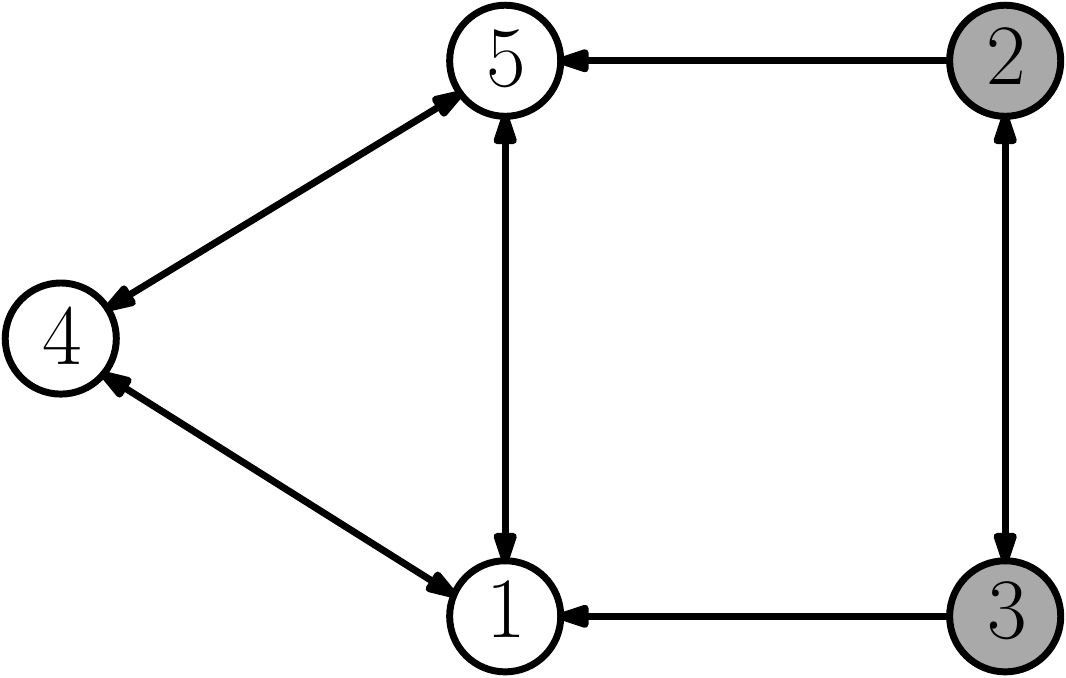} }
\tcb{\caption{\label{fig:trasitionButterfly} {The Butterfly graphs}. Grey nodes have more than one best response action. }}
\end{figure}
  
As discussed in Example \ref{ex:ring}, every ring graph $R_n$ is a Nash equilibrium of the centrality game $ \Gamma(\mc V,\beta, \eta, 2\cdot\mathbf 1) $. Notice that ring graphs are the only connected undirected graphs where all nodes have degree $2$. A remarkable fact is that there exists a recursive strongly connected Nash equilibrium that is not undirected. This  is displayed in Figure \ref{fig:trasitionButterfly} (a) and will be referred to as the \textit{butterfly graph} $B_5$. Figures \ref{fig:trasitionButterfly} (b) and \ref{fig:trasitionButterfly} (c) display two more graphs that turn out to be best response evolutions of $B_5$, as for the result below.
\begin{lemma}\label{lemma:B5} Consider the centrality game $ \Gamma(\mc V,\beta,\eta,2\cdot\mathbf 1) $ and let $x$ in $\mc X$ be such that $\mc G(x)=B_5$. Then: 
\begin{enumerate}
\item[(i)] $x$ is a non-strict Nash equilibrium; 
\item[(ii)] $x$ is a recursive Nash equilibrium and the 
configurations reachable from $x$ by a best response path are exclusively configurations $y$ such that $\mc G(y)$ is isomorphic to some of the graphs $B_5$,  $B_5'$, and $B_5''$ of Figure \ref{fig:trasitionButterfly}.
\end{enumerate}
\end{lemma}
\begin{proof} 
\emph{(i)} {We analyze the best response set of every node in configuration $x$, starting with node $1$.
Proposition \ref{prop:best_repsonse}\emph{(v)} applied to the two triples $\{1, 4,5\}$ and $\{1, 2,3\}$ implies that $\tau_4^1(x)=\tau_5^1(x)$ and $\tau_2^1(x)=\tau_3^1(x)$. By symmetry, we then conclude that all four hitting times are equal to each other. This implies that every pair of nodes is a best response for node $1$ in configuration $x$.}
Moving to node $4$, we can see that it has two in-neighbors: node $5$ and  node $1$. From system (\ref{eq:system_tau2}) and Proposition \ref{prop:best_repsonse}\emph{(iv)} (observing that $\{1\}$ is a cut set between node $2$ and node $4$) we obtain that $\tau_1^4(x)-\tau_5^4(x)=\frac{\beta}{2}(\tau_2^{4}(x)-\tau_1^{4}(x))>0$. Therefore, by Proposition 
\ref{cor:best_response}, the unique best response for node $4$ is $\{5,1\}$, namely the action currently played. Finally, node $5$ has only one in-neighbor, node $4$, and node $4$ has only one in-neighbor different from node $5$, that is node $1$. Therefore, from Theorem \ref{thm:best_response}\emph{(ii)}, we deduce that the unique best response of node $5$ is $\{4,1\}$. By symmetry, also nodes $2$ and $3$ are playing their unique best response. This says that $x$  is a Nash equilibrium.
It is not strict because the best response of node $1$ contains six different possible pairs. 

\emph{(ii)} We first notice that four out of the six best response actions of player $1$ lead to configurations $x'$ such that $\mc G(x')$ is isomorphic to $B_5$, while two of them lead to configurations $y$ such that $\mc G(y)$ is equal (or isomorphic) to $B_5'$. This graph consists of a sink that is a $3$-clique and of a source that is a $2$-clique with both nodes of it out-linking to one of the nodes of the $3$-clique (see Figure \ref{fig:trasitionButterfly} (b)).
We now show that {such configurations $y$ are Nash equilibria. For simplicity, we assume that $\mc G(y)$ coincides with $B'_5$ as in Figure \ref{fig:trasitionButterfly}} (b). Regarding node $5$, {Proposition \ref{prop:best_repsonse}\emph{(iv)-(v)} (using a cut argument with cut set $\{1\}$) 
yield} $\tau^5_2(y)=\tau^5_3(y)>\tau^5_1(y)=\tau^5_4(y)$. Hence, the only best response for node $5$ is $\{1,4\}$. By symmetry,  the only best response for node $4$ is $\{1,5\}$. Finally, because of \eqref{BR2}, the best response of node $2$ is any pair $\{3, s\}$ with $s$ in $\{1,4,5\}$ and the best response of node $3$ is any pair $\{2, s\}$ with $s$ in $\{1,4,5\}$. This implies that {$y$} is a Nash equilibrium. The inverse transition is a best response for $1$ in $y$ and leads back to $x$. The only other possible transitions from $y$ are through a modification of one of the out-links of either node $2$ or node $3$ and lead to 
configurations $z$ such that $\mc G(z)$ is isomorphic to $B_5''$ (see Figure \ref{fig:trasitionButterfly} (c)). 
We show that also {such configurations $z$ are Nash equilibria (for simplicity we assume that $\mc G(z)$ coincides with $B''_5$ as in Figure \ref{fig:trasitionButterfly} (c))}. {An argument completely analogous to the one used in configuration $y$ yields that node $5$ is playing its unique best response. }
{Regarding node $1$, notice that $\tau_4^1(z)=\tau_5^1(z)$ thanks again to Proposition \ref{prop:best_repsonse}\emph{(v)}. Moreover, from system \eqref{eq:system_tau2}, we obtain that 
$$\tau_5^1(z)-\tau_2^1(z)=\frac{\beta}{2}(\tau_4^1(z)-\tau_3^1(z)), \tau_4^1(z)-\tau_3^1(z)=\frac{\beta}{2}(\tau_5^1(z)-\tau_2^1(z)-\tau_4^1(z))=-\frac{\beta}{2}\tau_2^1(z)\,,$$
and this implies that both $\tau_2^1(z)$ and $\tau_3^1(z)$ are both strictly greater than $\tau_4^1(z)=\tau_5^1(z)$.}
As a consequence, also node $1$ (and, by symmetry, node $4$) is currently playing its unique best response. Finally, 
\eqref{BR2} shows that node $2$ and node $3$ are still playing a best response {in configuration $z$. This proves that $z$ is a Nash equilibrium. Finally notice that in configuration $z$, only nodes $2$ and $3$ can modify their action and that any modification will lead to a configuration whose graph is either isomorphic to $B'_5$ or to $B''_5$. We have thus proven that all the configurations whose graph is isomorphic to one of the three graphs $B_5$, $B'_5$, or $B''_5$ are Nash equilibria and that such set is closed by best response paths. By Lemma \ref{lemma:recursive-Nash}\emph{(v)}, this implies that all such Nash equilibria are recursive. }
\qed
\end{proof}

\begin{figure}\centering
\subfigure[\ $T_{(s,j),i}$ ]{\includegraphics[height=2.3cm]{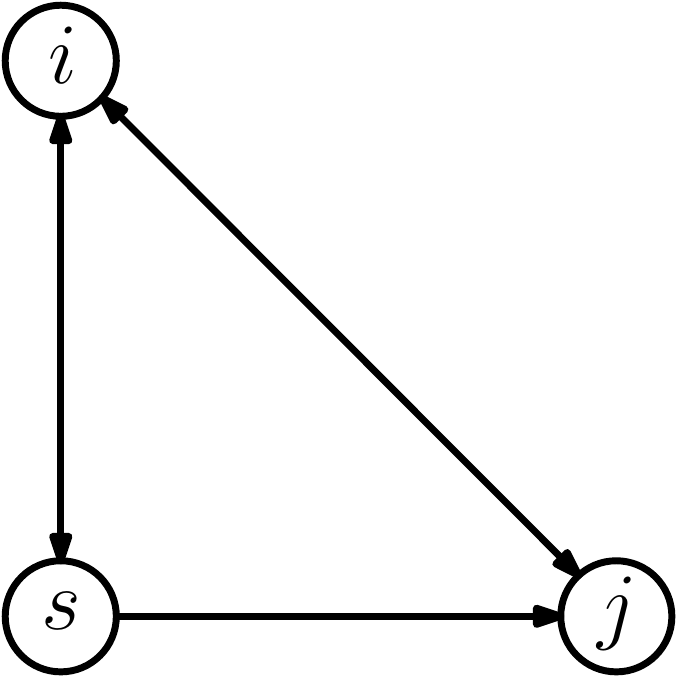} }\hspace{.4cm}
\subfigure[\ ]{\includegraphics[height=2.3cm]{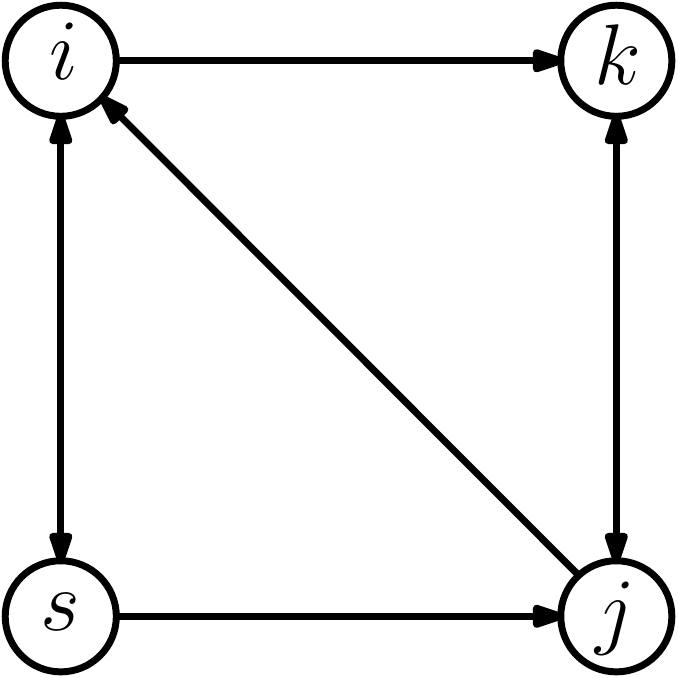} }\hspace{.4cm}
\subfigure[\ ]{\includegraphics[height=2.3cm]{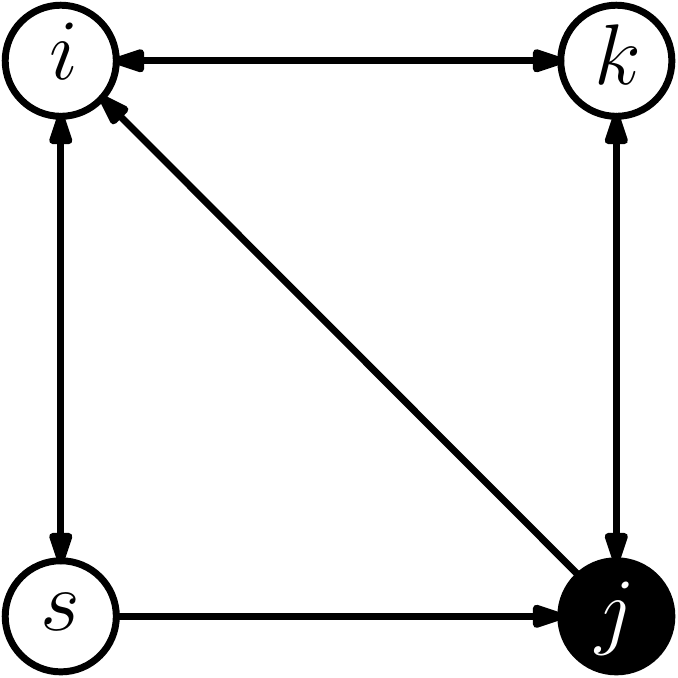} }\hspace{.4cm}
\subfigure[\ ]{\includegraphics[height=2.3cm]{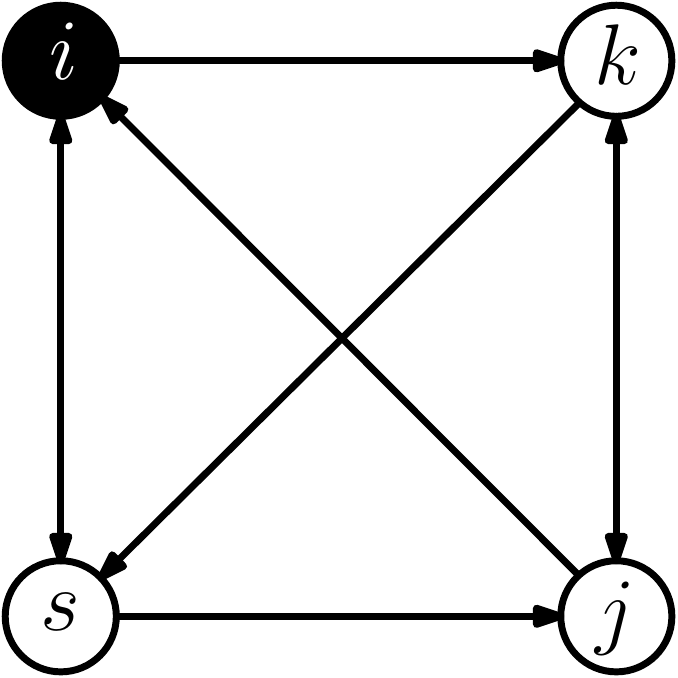} }
{\caption{\label{fig:nash_2} \label{fig:cycle_graph} The directed graph $T_{(s,j),i}$ and other explanatory graphs for the proof of Lemma \ref{lem:triangle}. }}
\end{figure}

We now go more in depth with our analysis of Nash equilibria proving a necessary condition on the connected components of  Nash equilibria of the centrality game $\Gamma(\mc V,\beta, \eta,2\cdot\mathbf 1)$. In particular,  we show that ring graphs $R_n$ and butterfly graphs $B_5$ are, in this context, the only possible strongly connected Nash equilibria.
Towards this goal, we first introduce  the graph $T_{(s,j),i}$ as the directed graph on the node set $ \lbrace i,j,s\rbrace $ having one directed link $(s,j) $ and all the other links undirected (see Figure \ref{fig:cycle_graph} (a)). Notice that node $s$ and node $i$ have out-degree $2$ and thus they have no choice but to connect to the remaining two nodes. Instead, node $j$ has out-degree $1$ and its best response is either linking to $i$ or to node $s$.  If node $j$ moves its out-link from $i$ to $s$,  we obtain the isomorphic graph $T_{(j,i),s}$.
This says that $T_{(s,j),i}$ is a non-strict, though recursive, Nash equilibrium of the centrality game $\Gamma(\mc V,\beta,\eta,\degprof)$ with $\mc V=\{s,i,j\}$ and $\degprof=(2,2,1)$. The following result illustrates the role played by the graph $T_{(s,j),i}$ in a centrality game $\Gamma(\mc V,\beta,\eta,\degprof)$ with $\degprof=2\cdot\mathbf1$. 

 \begin{lemma}\label{lem:triangle}
Let $ x$ in $\nash$ be a Nash equilibrium for a centrality game $\Gamma(\mc V,\beta,\eta,\degprof)$ with $\degprof=2\cdot\mathbf1$. Let $\mc G_{\mc K}(x)=(\mc K,\mc E')$ be a connected component of $ \mc G(x)$ that is a sink. If there exists a link $(s,j)$ in $\mc E'$ such  that $(j,s)\not\in\mc E'$, then $\mc G_{\mc K}(x)$ contains a subgraph of type $T_{(s,j),i}$.
\end{lemma}
\begin{proof}
Since {$\mc G_{\mc K}(x)=(\mc K,\mc E')$} is a sink connected component of $\mc G(x)$ and every node has out-degree $2$, it follows that ${|\mc K|}\geq 3$. In particular, $|\mc N_s^{-\infty}(x)|\geq{|\mc K|\geq 3}$ so that Theorem \ref{thm:best_response}\emph{(ii)} implies that $x_s\subseteq\mc N_s^{-2}(x)$. By assumption, $j\notin\mc N^{-1}_s(x)$, hence there exists a node $i$ in $\mc K$ such that $(j,i)\in\mc E'$ and $(i,s)\in\mc E'$. Moreover, the second claim in Theorem \ref{thm:best_response}\emph{(ii)} implies that $(s,i)\in \mc E'$. We are left to prove that also $(i,j)\in\mc E'$.
By contradiction, suppose this is not the case. The argument just used for the pair of nodes $s$ and $j$, can now be applied to the pair of nodes $j$ and $i$ and deduce the existence of a fourth node $ k$ in $\mc V'\setminus\{i,j\}$ such that $(i,k), (k,j), (j,k)\in\mc E'$. Notice that, as a consequence, $k\neq s$. The graph depicted in Figure \ref{fig:nash_2} (b) would thus be a subgraph of {$\mc G_{\mc K}(x)$}.  Now there are two possibilities: either $(k,i)\in\mc E'$ or $(k,i)\notin\mc E'$ . In the former case, we are in the situation of Figure \ref{fig:nash_2} (c) and we claim that $ j $ is not playing a best response. Indeed, subtracting the expected hitting times from nodes $i$ and $s$ to node $j$ and using  \eqref{eq:system_tau2}, we obtain that $ (1+(\beta/2))(\tau^j_i{(x)}-\tau_s^j{(x)})=(\beta/2)\tau^j_k{(x)}>0 $ and so 
$\lbrace s,k\rbrace$ gives $ j $ a strictly better utility than $x_j=\lbrace i,k\rbrace$. Consider finally the case when $(k,i)\not\in\mc E'$. Arguing for the pair of nodes $i$ and $k$ as we did before we deduce that $(k,s)\in\mc E'$, i.e., we are in the situation of Figure \ref{fig:nash_2} (d). Using again \eqref{eq:system_tau2}, we obtain $ (1+(\beta/2))(\tau_k^i{(x)}-\tau_j^i{(x)})=(\beta/2)\tau^i_s{(x)}>0 $ and so $ i $ would not be playing best response, for $\lbrace j,s\rbrace$ would give $i$ a strictly better utility than $x_i=\lbrace s,k\rbrace$.
This completes the proof.  \qed\end{proof}

We can now prove the following result. 
\begin{proposition}\label{thm:condensation_graph}
Consider a centrality game $\Gamma(\mc V,\beta,\eta,\degprof)$ with $\degprof=2\cdot\mathbf1$.  Let $x$ in $\nash$ be a Nash equilibrium and let $\mc G_{\mc K}(x)=(\mc K, \mc E')$ be a connected component of $\mc G(x)$. Then: 
\begin{enumerate}
\item[(i)] if $ \mc G_{\mc K}(x)$ is a source component, then it is either a singleton or a $2$-clique;
\item[(ii)] if $ \mc G_{\mc K}(x)$ is a sink component, then it is either a ring graph $R_k$ or the Butterfly graph $B_5$.
\end{enumerate}
\end{proposition}

\begin{figure}\centering
\subfigure[\ ]{\includegraphics[height=2.3cm]{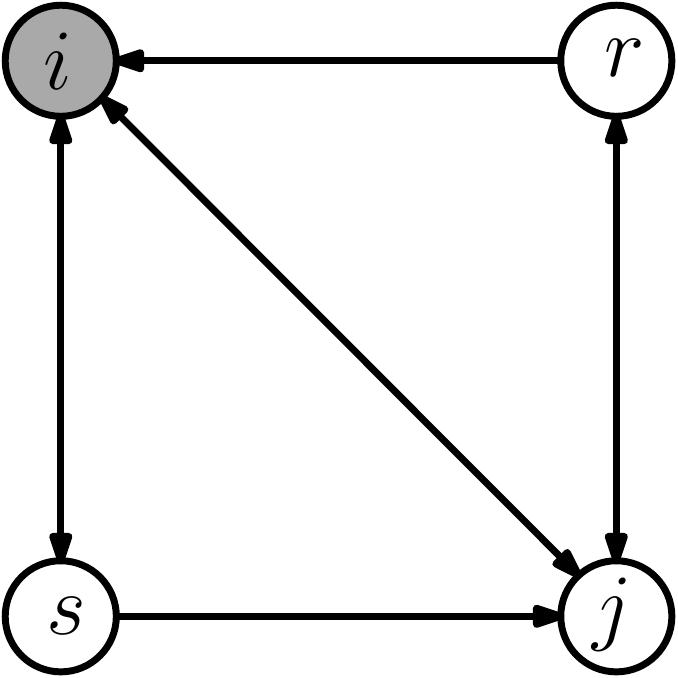} }\hspace{.4cm}
\subfigure[\ ]{\includegraphics[height=2.3cm]{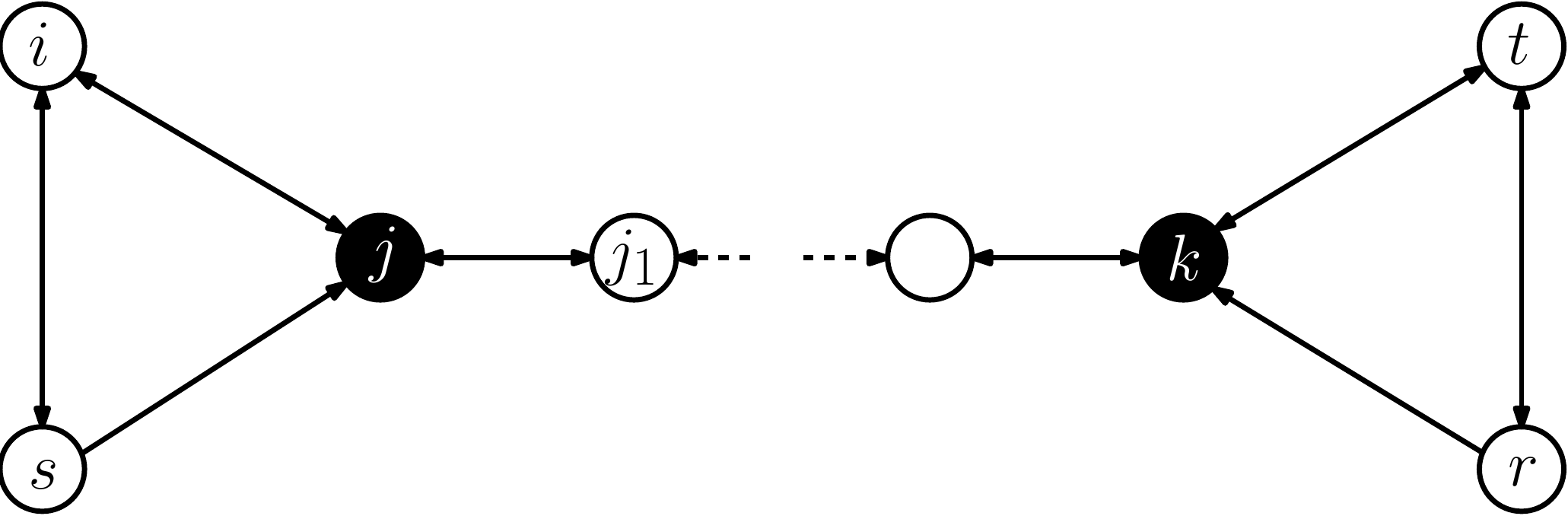} }\
{\caption{\label{fig:long}  Explanatory graph for the proof of Proposition \ref{thm:condensation_graph}. Gray nodes have multiple best response and black nodes are not playing best response. }}
\end{figure}

\begin{proof}
\emph{(i)} \tcb{This follows directly from Remark \ref{rem:source-hom}.}

\emph{(ii)}  \tcb{Since $\mc G_{\mc K}(x)$ is a connected component of a graph $\mc G(x)$ that is a sink and all its nodes have out-degree exactly $2$, if  $\mc G_{\mc K}(x)$ is undirected then it must be a ring graph.  Otherwise, if $\mc G_{\mc K}(x)$ is not undirected, then there must exist at least two directed links in $\mc E'$, say $(s,j)$ and $(r,k)$. Lemma \ref{lem:triangle} implies the existence of two subgraphs $T_{(s,j),i}$ and $T_{(r,k),t}$. Suppose first that these two subgraphs intersect (in one or two nodes). Direct degree considerations show that the only way that  $T_{(s,j),i}$ and $T_{(r,k),t}$ can intersect is if either (a) $k=i$ $t=j$ or (b) $j=k$ and $\{s,i\}\cap\{r,t\}= \emptyset$. In case (a) we obtain that $\mc G_{\mc K}(x)$ coincides with the graph in Figure \ref{fig:long}(a) and,
thanks to Proposition \ref{prop:best_repsonse}\emph{(v)}, $\tau_j^i(x)=\tau_{r}^i(x)$. This implies that if $\{j,s\}$ is a best response for node $i$, then also $\{s, r\}$ is such. If $i$ chooses this action, however, we get a configuration $y$ whose component $\mc G_{\mc K}(y)$ coincides with the graph in Figure \ref{fig:nash_2}(c) that in the proof of Lemma \ref{lem:triangle} was shown not to correspond to a Nash equilibrium. In case (b), the graph obtained corresponds to the Butterfly graph $B_5$ in Figure \ref{fig:trasitionButterfly}(a). Since in graph $B_5$ every node has out-degree equal to $2$, it necessarily coincides with the sink component $\mc G_{\mc K}(x)$. Suppose instead that 
$T_{(s,j),i}$ does not intersect any other of such triangles. Since in $T_{(s,j),i}$ node $j$ has out-degree $1$, there must exist a fourth node, say $j_1$, such that $(j,j_1), (j_1,j)\in \mc E'$. A recursive argument based on the finiteness of the graph now shows that 
there must exist a sequence of distinct nodes $j=j_0,j_1,\dots , j_l$, with $l\geq 2$, such that $\{j_a,j_{a+1}\}$ are $2$-cliques in $\mc G_{\mc K}(x)$ for $a=1,\dots ,l-1$ and from $j_l$ there is a directed link to some node $r\in \{s, j,i, j_1,\dots j_{l-1}\}$. Lemma \ref{lem:triangle} and a direct out-degree argument shows that the only possibility is that $r=j_{l-2}$.  This implies that $l\geq 3$ and
 we obtain
 the graph depicted in Figure \ref{fig:long}(b). Since again every node has out-degree equal to $2$, it necessarily coincides with the sink component $\mc G_{\mc K}(x)$.
A direct computation show that node $j$  is not playing best response in this configuration and, by symmetry, neither is $k$. Indeed, a direct computation from  \eqref{eq:system_tau2} implies that $ (1-(\beta^{2}/4))(\tau^j_{j_2}-\tau^j_{s})=(\beta^{2}/4)\tau^j_{j_4}>0 $ and thus  $y_j=\lbrace i,s\rbrace$ gives $ j $ a strictly better utility than $x_j=\lbrace i,j_2\rbrace$. This completes the proof.}
\qed\end{proof}

In particular, Proposition \ref{thm:condensation_graph}\emph{(ii)} provides a complete classification of the graphs corresponding to Nash equilibria that are strongly connected or consist of isolated connected components. We are now ready to classify all strict and recursive Nash equilibria of the centrality game  $\Gamma(\mc V,\beta, \eta,2\cdot\mathbf 1)$ in terms of the  following graph families:
\begin{itemize}
\item $\mc R$ is the family of graphs obtained by taking an arbitrary disjoint union of ring graphs $R_k$ with $k\geq 3$; 
\item $\mc K_{3,2}$ 
is the family of graphs that consist of a disjoint union of $3$-ring graphs $R_3$ and of a unique $2$-clique source component with two outgoing links each pointing to any of the nodes belonging to the ring graphs; 
\item $\mc K_{3,\mc B}$ is the family of graphs that are a disjoint union of $3$-ring graphs $R_3$ and of a unique butterfly graph $B_5$. 
\end{itemize}

\begin{figure}\centering
\subfigure[\ ]{\includegraphics[height=3cm]{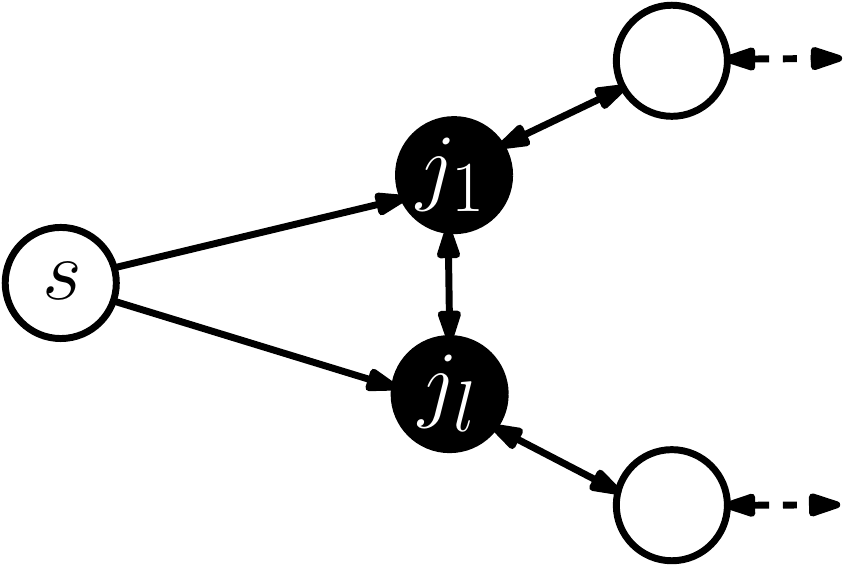} }\hspace{.8cm}
\subfigure[\ ]{\includegraphics[height=3cm]{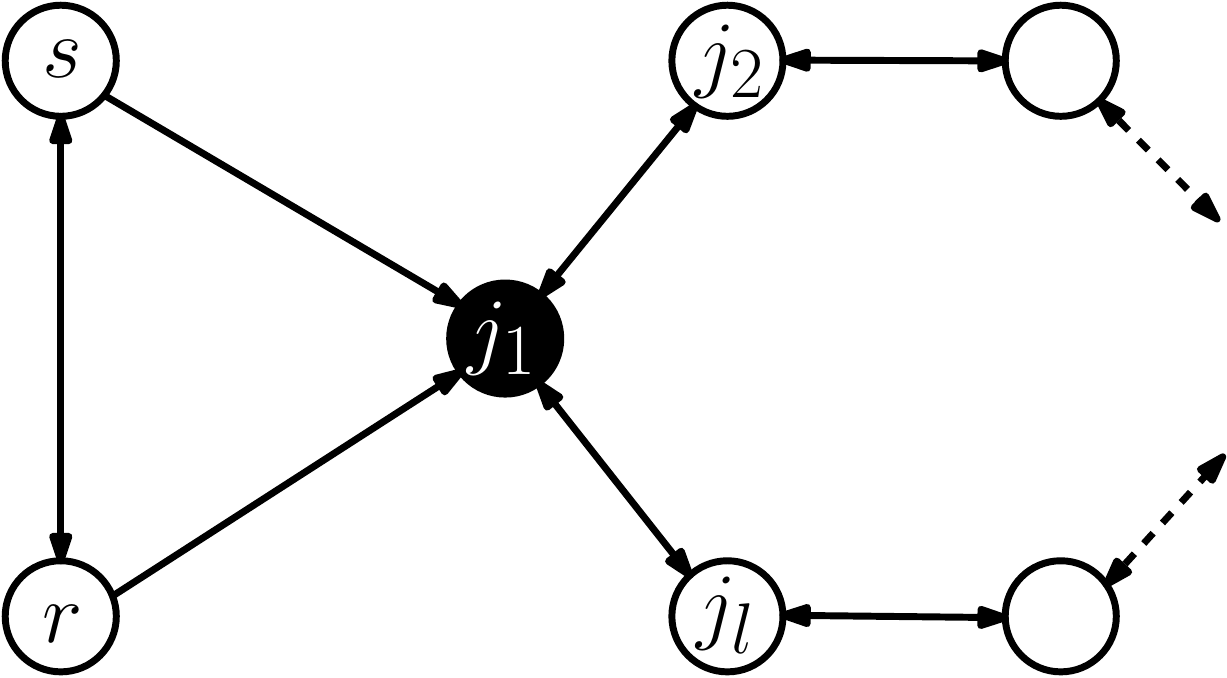} }\
{\caption{\label{fig:singleton_ring_NE}  (a) Singleton source linking to two adjacent nodes in a ring graph. (b) 2-clique source linking to a single node in a ring graph. Black nodes are not playing a best response. }}
\end{figure}

\begin{theorem}[\textbf{Strict and recursive Nash equilibria with $\degprof=\mathbf{2} $}]\label{thm:trapping_setsM2}
For a centrality game $ \Gamma(\mc V,\beta, \eta,2\cdot\mathbf 1) $ and a configuration $x$ in $\mc X$: 
\begin{enumerate}
\item[(i)] $x$ is a strict Nash equilibrium if and only if  $ \mc G(x)\in\mc R$;
\item[(ii)] $x$ is a recursive Nash equilibrium if and only if  $ \mc G(x)\in\mc R\cup \mc K_{3,2}\cup \mc K_{3,\mc B}$. 
\end{enumerate}
\end{theorem}
\begin{proof}
\emph{(i)} If $\mc G(x)$ is a disjoint union of ring graphs, then $x$ is a strict Nash equilibria thanks to Corollary \ref{cor:union}. On the other hand, it follows from Proposition \ref{thm:condensation_graph}, Lemma \ref{lemma:B5}, and Theorem \ref{thm:condensation_graph_generalm}\emph{(iii)} that there cannot be other strict Nash equilibria.

\emph{(ii)} We first show that if $\mc G(x)\in \mc K_{3,2}$, then $x$  is a Nash equilibrium. \tcb{Let $\mc G_{\mc K}(x)$, with $\mc K=\{r,s\}$, be the unique $2$-clique source component. If both remaining out-links of $r$ and $s$ are directed to one or two nodes of the same $3$-ring component $\mc G_{\mc T}(x)$, the subgraph induced by $\mc K\cup \mc T$ is isomorphic to either $B_5'$ or $B_5''$ (see Figure \ref{fig:trasitionButterfly}). The graph $\mc G(x)$ is in this case the disjoint union of $3$-rings and of a graph isomorphic to either $B_5'$ or $B_5''$. Consequently, $x$ is Nash because of Lemma \ref{lemma:B5} and Corollary \ref{cor:union}.
Suppose instead that the remaining out-links of $r$ and $s$ are directed to nodes of two different disjoint $3$-rings $\mc G_{\mc T_1}(x)$ and $\mc G_{\mc T_2}(x)$.
Suppose that $\mc T_1=\{i,j,k\}$ and that $r$ links to $k$. Then $i$ (and analogously $j$) is playing its unique best response by noticing that $\tau_j^i(x)=\tau_k^i(x)$ because of Proposition \ref{prop:best_repsonse}\emph{(v)} and the fact that $k$ is a cut set with respect to the remaining nodes in the graph (see Proposition \ref{prop:best_repsonse}\emph{(iv)}). Regarding node $k$, we have that $\tau_j^k(x)=\tau_i^k(x)$ because of Proposition \ref{prop:best_repsonse}\emph{(v)} and from we obtain that
$\tau^k_r(x)-\tau_j^k(x)=\frac{\beta}{2}(\tau^k_s(x)-\tau_i^k(x))>\frac{\beta}{2}(\tau^k_r(x)-\tau_j^k(x))$ that implies $\tau^k_s>\tau^k_r(x)>\tau_j^k(x)$. Since $\mc N^{-2}_k(x)=\{k, i, j, r, s\}$, it follows from Proposition \ref{prop:best_response} that $k$ is playing its unique best response. Since $\mc N^{-2}_r(x)=\{r, s\}$ also $r$ is playing a best response and by symmetry, also $s$ and the nodes in the other $3$-ring are playing a best response. $\mc G(x)$ is thus a Nash equilibrium because again of Corollary \ref{cor:union}.}
Similarly, any graph in $\mc K_{3,\mc B}$
is a Nash equilibrium because of Lemma \ref{lemma:B5} and  Corollary \ref{cor:union}. Previous considerations and Lemma \ref{lemma:B5}\emph{(ii)} also show that, from any $\mc G$ in $\mc K_{3,2}\cup\mc K_{3,\mc B}$, the graphs reachable in a best response path are all graphs in $\mc K_{3,2}\cup\mc K_{3,\mc B}$. By Lemma \ref{lemma:recursive-Nash}\emph{(v)}, this implies that all configurations $x$ such that $\mc G(x)\in \mc K_{3,2}\cup\mc K_{3,\mc B}$ are recursive.
We are left with proving that if $x$ is recursive, then $\mc G(x)\in  \mc R\cup \mc K_{3,2}\cup\mc K_{3,\mc B}$. Suppose that $\mc G(x)\not\in\mc R$. By Proposition \ref{thm:condensation_graph}, there are three possibilities to analyze: (a) $\mc G(x)$ contains a butterfly $B_5$ as a sink component; (b) $\mc G(x)$ contains a $2$-clique $K_2$ as a source component;
(c) $\mc G(x)$ contains a singleton as a source component.
\tcb{In case (a), Lemma \ref{lemma:B5}\emph{(ii)} implies that there exists best response transitions that generate a source component from the $B_5$. For this reason, Theorem \ref{thm:condensation_graph_generalm}\emph{(ii)} forbids the presence of source components in $\mc G(x)$ and also of other sink components isomorphic to a $B_5$.}
Moreover, Lemma \ref{lemma:B5}\emph{(ii)} also yields that there exists a best response path from $x$ to a configuration $y$ with associated graph $\mc G(y)$ whose connected components are all sinks consisting of ring graphs except for a $2$-clique $\{r,s\}$ that is the unique source with both $r$ and $s$ linking to a node $j_1$ in a \tcb{ring of maximal length $l$. If $l>3$, we can argue as follows. Let $j_1$ be the node in the ring to which both $r$ and $s$ point to, and let $j_2$ and $j_l$ be its adjacent nodes in the ring (see Figure \ref{fig:singleton_ring_NE}(b)).} 
By system (\ref{eq:system_tau2}), $\tau_{s}^{j_1}(y)- \tau_{j_2}^{j_1}(y)= (\beta/2) (\tau_{s}^{j_1}(y)- \tau_{j_3}^{j_1}(y))$. Since $\tau_{j_3}^{j_1}(y)>\min\{\tau_{j_2}^{j_1}(y),\tau_{j_l}^{j_1}(y)\}=\tau_{j_2}^{j_1}(y)$ (\tcb{where the inequality follows from Proposition \ref{prop:best_repsonse}(iv), since $\mc C=\{j_2,j_l\}$ is a cut between $j_3$ and $j_1$, and the equality by symmetry}), it follows that $\tau_{j_2}^{j_1}(y)>\tau_s^{j_1}(y)$. This says that $j_1$ is not playing a best response action.
Consequently $y$ is not a Nash equilibrium and thus $x$ is not a recursive Nash equilibrium. Therefore, if $\mc G(x)$  contains a butterfly  graph $B_5$, then $\mc G(x)\in \mc K_{3,\mc B}$.  A completely analogous argument, in case (b), shows that if $\mc G(x)$ contains a $2$-clique $K_2$ as a source component, then $\mc G(x)\in \mc K_{3,2}$.
Finally,  suppose we are in case (c):
$\mc G(x)$ contains a singleton source node $s$. From $x$, there exists a best response path leading to a configuration $y$ where node $s$ is linking to two adjacent nodes \tcb{$j_1$ and $j_l$} in the same ring of maximal length $l\geq 3$, as in Figure \ref{fig:singleton_ring_NE}(a). \tcb{By labeling the nodes in the length-$l$ ring as $j_1,j_2,\ldots,j_l$ so that $j_h$ and $j_k$ are adjacent if and only if $|h-k|=1$ modulo $l$,  and using  \eqref{eq:system_tau2}, we get that}
\be\label{singleton}\tau_{s}^{j_1}(y)- \tau_{j_2}^{j_1}(y) = \frac{\beta}{2} (\tau_{j_l}^{j_1}(y)- \tau_{j_3}^{j_1}(y))\,.\ee
If $l>3$, a cut and a symmetry argument again implies that $\tau_{j_l}^{j_1}(y)- \tau_{j_3}^{j_1}(y)<0$. This again shows that \tcb{$j_1$} is not playing a best response. Finally, if $l=3$, the nodes $j_3$ and $j_l$ coincide. In this case, (\ref{singleton}) and Proposition \ref{prop:best_repsonse} \emph{(v)} yield $\tau_{s}^{j_1}(y)= \tau_{j_2}^{j_1}(y)= \tau_{j_3}^{j_1}(y)$ and, thus, the pair $\{j_2, s\}$ is a best response for node \tcb{$j_1$}. This choice leads to a configuration $z$ whose graph $\mc G(z)$ is isomorphic to the one in Figure \ref{fig:nash_2} (c) that can not be a Nash because of Lemma \ref{lem:triangle}. Therefore, $x$ is not recursive and this completes the proof.
\qed\end{proof}

\begin{remark} In both the homogeneous cases analyzed, the set of strict Nash equilibria $\strictnash$ and that of recursive Nash equilibria $\recnash$ are both independent from the discount factor $\beta$. As we will see later on, however, in the case of two links the set of Nash equilibria $\nash$ generally depends on the value of $\beta$.
\end{remark}\medskip
\begin{remark}
\tcb{Notice that the disjoint union of two recursive Nash equilibria may not be a recursive Nash equilibrium in general. E.g. consider two disjoint butterfly graphs $B_5$. Individually, such configurations are recursive according to Lemma \ref{lemma:B5}. However, Theorem \ref{thm:trapping_setsM2}(ii) implies that a recursive equilibrium cannot admit the simultaneous presence of two disjoint subgraphs isomorphic to $B_5$.}
\end{remark}\medskip

\section{Numerical Results}\label{sec:num_results}
In this section, we present some numerical results that corroborate our theoretical findings and highlight further features of the equilibrium structure of the considered centrality games.

\tcb{Our numerical experiments consider centrality games $\Gamma=\Gamma(\mc V,\beta,\eta,\degprof)$ and simulate the asynchronous best response dynamics introduced in Section \ref{sec:CMG-sub}: an initial configuration $X(0)$ is chosen at random from $\mc X$ and subsequent configurations are recursively generated at times $t=0,1,2,\ldots$  by sampling a node $i$ uniformly at random from $\mc V$ and moving to a new configuration $X(t+1)$ such that $X_{-i}(t+1)=X_{-i}(t)$, while $X_{i}(t+1)$ is chosen uniformly from the best response set $\mc B_i(X_{-i}(t))$. By Corollary \ref{coro:dynamics}\emph{(i)}, with probability one $X(t)$ gets absorbed in finite time in the set $\recnash $ of recursive Nash equilibria. In our simulations, we stop the dynamics at some time $T$ and for the final configuration $x=X(T)$ we compute the number of connected components  $c(x)$.}

\tcb{At every time $t$, given the current configuration $X(t)=x$, best response actions for the sampled node $i$ are computed from their characterization \eqref{br-character} in terms of expected hitting times. We make use of Theorem \ref{thm:best_response} in order to reduce the computational burden and restrict the search of nodes $j$ with the lowest expected hitting times on $i$ to the set $\mc N_i^{-d_i}(x_{-i})$ of nodes from which node $i$ can be reached in at most $\deg_i$ hops. Building on Proposition \ref{prop:best_repsonse}\emph{(i)-(ii)}, we compute the expected hitting times $\tilde\tau^i_{j}$ for the case $\eta=\delta^i$: using \eqref{tildetau1}, this requires solving the linear system \eqref{eq:system_tau2bis} in the unknowns $(\tilde\tau^i_j)_{j\in\mc N_i^{-\infty}(x_{-i})}$. The dimension $|\mc N_i^{-\infty}(x_{-i})|$ of this system depends of course on the configuration $x$, but may result much smaller than the network order $n$: in particular, this occurs when $m(x)=\max_{i\in\mc V}|\mc N_i^{-\infty}(x_{-i})|$ is small. As illustrated in Section \ref{sec:potential-maximizers}, for values of the discount factor $\beta$ close to $1$, this corresponds to high values of the potential function $\Psi(x)$, which is non-decreasing along the best-response dynamics. 
}

\subsection{The Homogeneous Case}
\tcb{The first set of simulations is carried on for centrality games  $\Gamma=\Gamma(\mc V,\beta,\eta,\degprof)$ with homogeneous degree profiles, namely $\degprof=d\cdot\mathbf 1$.} 

\tcb{Notice that, for such centrality games, Theorem \ref{theo:d=k} characterizes the class of potential maximizing equilibria (that are a subclass of  recursive Nash equilibria) when the discount factor $\beta$ is sufficiently close to $1$ and $n$ is sufficiently larger than $\deg$ (precisely, $n\geq d(d+1)$). All such potential maximizing equilibria are composed of isolated components of size $(d+1)$ and $(d+2)$ and thus, for such equilibria $x$, it holds that \be\label{comp-hom}\frac{n}{d+2}\leq c(x)\leq \frac{n}{d+1}\,.\ee
Notice that any recursive equilibrium is composed of a number of sink connected components that must have size at least $d+1$ and of at most a source connected component of size at most $\deg$. This implies that, for every recursive equilibrium $x$ (not necessarily a potential maximizer), we have the bound
$$c(x)\leq 1+\frac{n-1}{d+1}\,,$$
that for large $n$ is asymptotically equivalent to the upper bound in \eqref{comp-hom}. However, there exist recursive equilibria for which instead the left inequality in \eqref{comp-hom} is not satisfied: e.g., configurations $x$ such that $\mc G(x)$ is the ring graph $R_n$ are strongly connected strict Nash equilibria for $d=2$.
}
\begin{figure}
\centering
\includegraphics[width=0.49\textwidth]{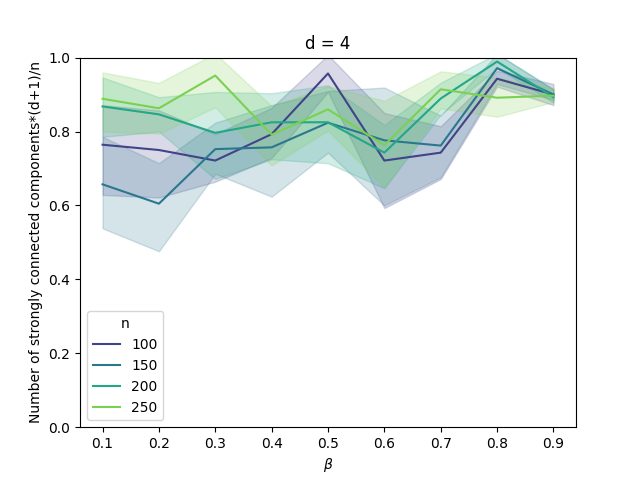}
\includegraphics[width=0.49\textwidth]{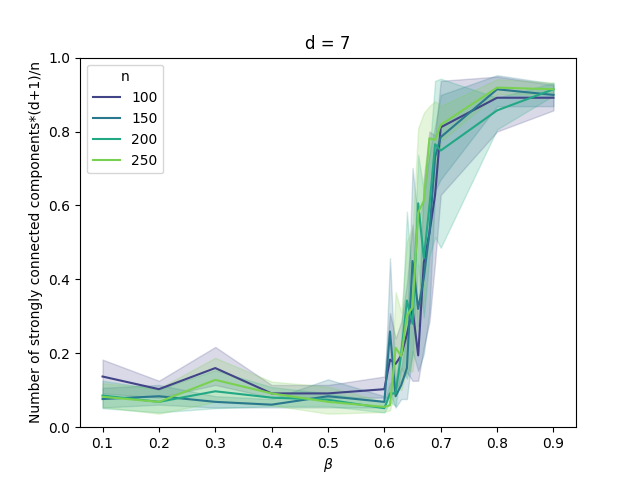}
\caption{Simulations of the best response dynamics after $T=100000$ time steps, for different values of $n$, $\beta$ and $d$. The solid lines refer to the average normalized number of connected components over seven initially generated networks against $\beta$ for fixed $n,d$, while shaded areas indicate the corresponding variances. }
\label{fig:reg_conn_comp_varying_m}
\end{figure}
\tcb{For $n=100,150,200,250$,  $ d=4,7 $, and $\beta\in [0.1, 0.9]$, we randomly generate seven networks with $n$ nodes and out-degree $d$ by choosing, independently for each node $ i $, the set of its $ d $ out-neighbors uniformly at random in $\mc V\setminus \{i \}$.
We then numerically simulate the best response dynamics  for $T=100000$ time steps and compute  $c(x)$  in the final configuration $X(T)=x$.}
\tcb{Figure \ref{fig:reg_conn_comp_varying_m} shows the results of the simulations for (a) $d=4$ and (b) $d=7$ as $\beta$ varies, for the different values of $n$. We plot the normalized index $C(x)=c(x)(d+1)/n\in (0,1]$ 
against $\beta$ for the various values of $n$. 
Solid lines are the averages while shaded areas indicate the corresponding variances. We notice that for $d=4$, $C(x)$ is always above $0.6$ suggesting that the recursive Nash equilibria obtained are largely disconnected for every value of $\beta$.
Instead, for $d=7$ a transition phase phenomenon is suggested to happen around $\beta=0.65$ (sampling has been consequently refined around this point). Above this value, we obtain again maximally disconnected networks close to the potential maximizers investigated in Theorem \ref{theo:d=k}, while below such value  more connected networks emerge.}

\subsection{Power-Law Out-Degree Distributions}
\tcb{ A second set of experiments has been carried on for centrality games  $\Gamma=\Gamma(\mc V,\beta,\eta,\degprof)$ whose out-degree profile has been randomly sampled from a truncated power-law distribution: 
\begin{equation}\label{eq:power_law}
 \mathbb{P}(\deg_i=d) =\left\{\ba{lcl}\dfrac{d^{-\alpha}}{{\sum_{k=1}^{n-1}k^{-\alpha}}}&\text{ if }&1\le d< n\\[15pt]0&\text{ if }& d\ge n\,,\ea\right.
\end{equation}
where $\alpha>0$ is a parameter. 
A peculiar feature of graphs with such power law distributions (particularly when $\alpha$ is small) is the presence of \emph{hubs}, meant as nodes with very large out-degree. }
 \begin{figure}
\centering
\subfigure[ ]{\includegraphics[width=0.46\textwidth]{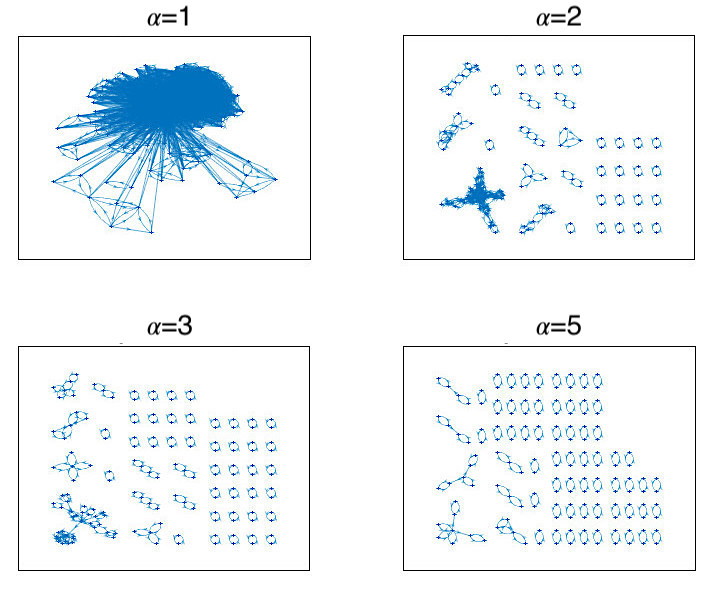}}\hspace{0.4cm}
\subfigure[ ]{\includegraphics[width=0.49\textwidth]{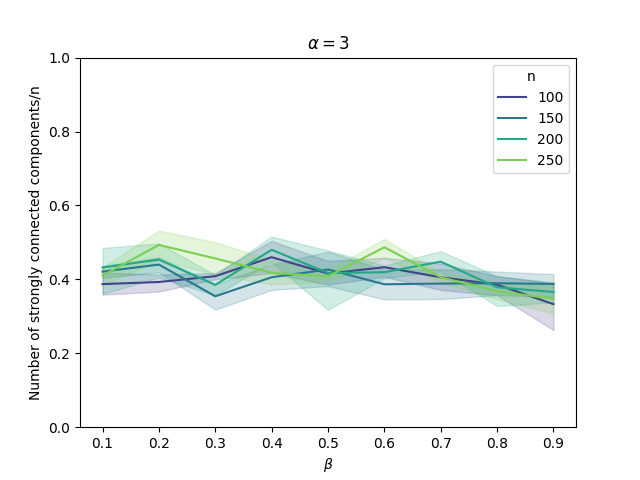}}
\caption{Simulations of the best response dynamics after $T=100000$ time steps on networks generated by the power law (\ref{eq:power_law}). 
(a) Final configurations reached with $n=150$, $\beta=0.5$, and various values of $\alpha$. (b) 
Normalized number of connected components in final configurations against $\beta$ for $\alpha=3$ and various $n$ (solid lines indicate averages while shaded areas indicate the corresponding variances).
} 
\label{fig:VariousExponents}\label{fig:powerlaw_numcomp}
\end{figure}
 \tcb{ We have carried on several experiments with various values of $n$, $\beta$, and $\alpha$.
In all cases, the initial configuration has been randomly generated by first sampling the out-degree distribution $\degprof$ according to \eqref{eq:power_law} and then choosing, independently for each node $ i $, the set of its $ \degprof_i$ out-neighbors uniformly at random in $\mc V\setminus \{i \}$. As in the previous case, the best response dynamics has been simulated for $T=100000$ time steps.}
\tcb{Figure \ref{fig:powerlaw_numcomp}(a) shows four final configurations reached by the asynchronous best response dynamics started from a random initial configuration, with $n=150$, $\beta=0.5$, and $\alpha\in \{1,2,3,5 \}$. We can notice the presence of a hub for $\alpha=1$, while higher values of $\alpha$ yield more and more isolated components as maximal out-degree gets smaller and smaller.}

\tcb{In consideration of the fact that typically the power-law distribution is considered for $\alpha >2$ (otherwise the average degree is unbounded in $n$), we have carried on more extensive results in the style we had followed for the homogeneous case, for the value $\alpha=3$. In Figure \ref{fig:powerlaw_numcomp}(b), we have plotted average (over seven instances) and variances of the normalized index $c(x)/n$ for the final configuration $x$ against $\beta$ for $n \in \{100,150,200,250\}$ and $\alpha=3$. These plots suggest that the final network remains largely disconnected with the number of connected components that seems to increase linearly in $n$, in agreement with Figure \ref{fig:powerlaw_numcomp}(a). We can also notice that the parameter $\beta$ does not appear to have a significant effect on the structure of the emerging equilibria.
}

\section{Conclusion}\label{conclusions}
\tcb{In this paper, we have proposed and analyzed a family of network formation games in which every node $i$, equipped with a fixed number of out-links $\deg_i$, is free to choose how to direct them in order to maximize its PageRank centrality. We have first showed that the considered model is a potential game. Our results show that best responses are essentially local: a player $i$ tends to link to nodes from which node $i$ can be reached in at most $\deg_i$ steps. This fact yields fundamental information on  the structure of networks that are Nash equilibria: connected components can only be sources or sinks and at most one source can show up in the class of recursive Nash equilibria, where best response dynamics is known to get absorbed in finite time. This implies that typically equilibria are largely disconnected, with several undirected links and small cycles. For the special case of homogeneous out-degree profiles with $\deg_i=d$ for every node $i$, the analysis of the potential function allows to reach further insight on the subclass of recursive Nash equilibria that maximize the potential (when the discount factor of the centrality is sufficiently high): they are all composed of isolated components of size $d+1$ and $d+2$. For the case of $d=1,2$ we have a complete classification of the recursive Nash equilibria. }

\tcb{In the final section, we carry on some initial numerical studies. Besides corroborating our theoretical results, our numerical results suggest the possible presence of phenomena not yet investigated, such as for instance the possible presence of threshold type behaviors with respect to the discount parameter of the PageRank centrality. This is left for future investigation.}

\tcb{Fragmentation and lack of connectivity seem to be the norm in this network formation process and we might question how realistic this can be. Indeed, in some real-world networks this may occur: notable examples are the citation graphs, the World Wide Web, or other social networks like the sentimental relation graph reported in \cite[Figure 2.7]{Easley.Kleinberg:2010}. Nevertheless, there are contexts where being part of a larger community may bring an advantage to the individuals. Some preliminary work in this direction is reported in \cite{Como.ea:22} where a community is engaged in an inferential task whose performance depends on the size of the community.
A challenging problem (and new at the best of our knowledge) is to combine the two mechanisms and consider games where players have to trade off between trying to be central in their community and, at the same time, being part of a community large enough to well perform some collective task.}

\section*{Acknowledgments.}
This research was carried on within the framework of the MIUR-funded {\it Progetto di Eccellenza} of the {\it Dipartimento di Scienze Matematiche G.L.~Lagrange}, Politecnico di Torino, CUP: E11G18000350001. It received partial support from the MIUR-funded project PRIN 2017 ``Advanced Network Control of Future Smart Grids'' and by the {\it Compagnia di San Paolo}.  

\bibliographystyle{plain}
\bibliography{bibl_centrality_game}

\begin{thebibliography}{10}

\bibitem{Alon10}
N.~Alon, E.~D. Demaine, M.~Hajiaghayi, , and T.~Leighton.
\newblock Basic network creation games.
\newblock In {\em Proceedings of the 22nd ACM symposium on Parallelism in
  algorithms and architectures}, volume SPAA '10, pages 106 -- 113. ACM, New
  York, 2010.

\bibitem{Anantharam.Tsoucas:1989}
V.~Anantharam and P.~Tsoucas.
\newblock A proof of the {M}arkov chain tree theorem.
\newblock {\em Statistics \& Probability Letters}, 8:189--192, 1989.

\bibitem{Avrachenkov06}
K.~Avrachenkov and N.~Litvak.
\newblock The effect of new links on {G}oogle {P}age{R}ank.
\newblock {\em Stochastic Models}, 22(2):319--331, 2006.

\bibitem{Bala00}
V.~Bala and S.~Goyal.
\newblock A noncooperative model of network formation.
\newblock {\em Econometrica}, 68(5):1181 -- 1229, 2000.

\bibitem{Belhaj16}
M.~Belhaj, S.~Bervoets, and F.~Fr{\'e}d{\'e}ric~Dero{\"\i}an.
\newblock Efficient networks in games with local complementarities.
\newblock {\em Theoretical Economics}, 11:357 -- 380, 2016.

\bibitem{Blume:1993}
L.~Blume.
\newblock The statistical mechanics of strategic interaction.
\newblock {\em Games and Economic Behavior}, 5:387--424, 1993.

\bibitem{Blume:1995}
L.~Blume.
\newblock The statistical mechanics of best response strategy revision.
\newblock {\em Games and Economic Behavior}, 11(2):111--145, 1995.

\bibitem{Blume:97}
L.~Blume.
\newblock {\em The Economy As An Evolving Complex System II}, chapter
  Population Games.
\newblock CRC Press, 1997.

\bibitem{Boldi2014}
P.~Boldi and S.~Vigna.
\newblock Axioms for centrality.
\newblock {\em Internet Mathematics}, 3-4(10):222--262, 2014.

\bibitem{PB:87}
P.~Bonacich.
\newblock Power and centrality: {A} family of measures.
\newblock {\em American Journal of Sociology}, 92(5):1170--1182, 1987.

\bibitem{SB-LP:98}
S.~Brin and L.~Page.
\newblock The anatomy of a large-scale hypertextual {W}eb search engine.
\newblock {\em Computer Networks}, 30:107--117, 1998.

\bibitem{Castaldo.ea:2020}
M.~Castaldo, C.~Catalano, G.~Como, and F.~Fagnani.
\newblock On a centrality maximization game.
\newblock {\em IFAC}, 53(2):2844--2849, 2020.

\bibitem{Chen2009}
W.~Chen, S.-H. Teng, Y.~Wang, and Y.~Zhou.
\newblock On the $\alpha$-sensitivity of {N}ash equilibria in
  {P}age{R}ank-based network reputation games.
\newblock In {\em Frontiers in Algorithmics}, pages 63--73, 2009.

\bibitem{Chien2004}
S.~Chien, C.~Dwork, R.~Kumar, D.~R. Simon, and D.~Sivakumar.
\newblock Link evolution: Analysis and algorithms.
\newblock {\em Internet Mathematics}, 1(3):277--304, 2004.

\bibitem{scarsini}
R.~Cominetti, M.~Quattropani, and M.~Scarsini.
\newblock The buck-passing game.
\newblock {\em Mathematics of Operations Research}, 47(3):1731--1756, 2022.

\bibitem{Como.Fagnani:2015}
G.~Como and F.~Fagnani.
\newblock Robustness of large-scale stochastic matrices to localized
  perturbations.
\newblock {\em IEEE Transactions on Network Science and Engineering},
  2(2):53--64, 2015.

\bibitem{Como.Fagnani:2016}
G.~Como and F.~Fagnani.
\newblock From local averaging to emergent global behaviors: The fundamental
  role of network interconnections.
\newblock {\em Systems and Control Letters}, 95:70--76, 2016.

\bibitem{Como.ea:22}
G.~Como, F.~Fagnani, and A.~Proskurnikov.
\newblock Reaching optimal distributed estimation through myopic
  self-confidence adaptation.
\newblock {\em IFAC PapersOnLine}, 2022.

\bibitem{Corbo06}
J.~Corbo, A.~Calvo-Armengol, and D.~Parkes.
\newblock A study of {N}ash equilibrium in contribution games for peer-to-peer
  networks.
\newblock {\em Operating Systems Review}, 40(3):61 -- 66, 2006.

\bibitem{Jungers10}
B.~C. Cs{\'a}ji, R.~M. Jungers, and V.~D. Blondel.
\newblock {P}age{R}ank optimization in polynomial time by stochastic shortest
  path reformulation.
\newblock In {\em Algorithmic Learning Theory}, pages 89--103, 2010.

\bibitem{dekerchove08}
C.~de~Kerchove, N.~Ninove, and P.~van Dooren.
\newblock Maximizing {P}age{R}ank via outlinks.
\newblock {\em Linear Algebra and its Applications}, 429(5):1254 -- 1276, 2008.

\bibitem{Demaine07}
E.~D. Demaine, M.~Hajiaghayi, H.~Mahini, and M.~Zadimoghaddam.
\newblock The price of anarchy in network creation games.
\newblock In {\em Proceedings of the twenty-sixth annual ACM symposium on
  Principles of distributed computing}, volume PODC '07, pages 292 -- 298. ACM,
  New York, 2007.

\bibitem{Easley.Kleinberg:2010}
D.~Easley and J.~Kleinberg.
\newblock {\em Networks, Crowds, and Markets: Reasoning About a Highly
  Connected World}.
\newblock Cambridge University Press, 2010.

\bibitem{Ehsani11}
S.~Ehsani, S.S. Fadaee, M.A. Fazli, A.~Mehrabian, S.S. Sadeghabad, M.A. Safari,
  and M.~Saghafian.
\newblock On a bounded budget network creation game.
\newblock In {\em Proceedings of the twenty-third annual ACM symposium on
  Parallelism in algorithms and architectures}, volume SPAA '11, pages 207 --
  214. ACM, New York, 2011.

\bibitem{Fabrikant03}
A.~Fabrikant, A.~Luthra, E.~Maneva, C.~H. Papadimitriou, and S.~Shenker.
\newblock On a network creation game.
\newblock In {\em Proceedings of the twenty-second annual symposium on
  Principles of distributed computing}, volume PODC '03, pages 347 -- 351. ACM,
  New York, 2003.

\bibitem{NEF-ECJ:90}
N.~E. Friedkin and E.~C. Johnsen.
\newblock Social influence and opinions.
\newblock {\em Journal of Mathematical Sociology}, 15(3-4):193--206, 1990.

\bibitem{Friedkin:91}
N.E. Friedkin.
\newblock Theoretical foundations for centrality measures.
\newblock {\em American Journal of Sociology}, 96:1478--1504, 1991.

\bibitem{Hopcroft_2008}
J.~Hopcroft and D.~Sheldon.
\newblock {Network Reputation Games}.
\newblock Technical report, Cornell University, 2008.

\bibitem{Ishii.Tempo:2014}
H.~Ishii and R.~Tempo.
\newblock The {P}age{R}ank problem, multiagent consensus, and web aggregation:
  A systems and control viewpoint.
\newblock {\em IEEE Control Systems Magazine}, 34(3):34--53, 2014.

\bibitem{JacksonW96}
M.~O. Jackson and A.~Wolinsky.
\newblock A strategic model of social and economic networks.
\newblock {\em Journal of Economic Theory}, 71(1):44 -- 74, 1996.

\bibitem{JacksonBook2008}
M.O. Jackson, editor.
\newblock {\em Social and Economic Networks}.
\newblock Princeton Univ. Press, Princeton and Oxford, 2008.

\bibitem{katz:53}
L.~Katz.
\newblock A new status index derived from sociometric analysis.
\newblock {\em Psychometrika}, 18:39--43, 1953.

\bibitem{Konig.ea:2014}
M.~D. K\"onig, C.~J. Tessone, and Y.~Zenou.
\newblock Nestedness in networks: A theoretical model and some applications.
\newblock {\em Theoretical Economics}, 9(3):695--752, 2014.

\bibitem{Laoutaris08}
N.~Laoutaris, L.~J. Poplawski, R.~Rajaraman, R.~Sundaram, and S.-H. Teng.
\newblock Bounded budget connection (bbc) games or how to make friends and
  influence people, on a budget.
\newblock In {\em Proceedings of the twenty-seventh ACM symposium on Principles
  of distributed computing}, volume PODC '08, pages 165 -- 174. ACM, New York,
  2008.

\bibitem{Marden.Shamma:2012}
J.~R. Marden and J.~S. Shamma.
\newblock Revisiting log-linear learning: Asynchrony, completeness and
  payoff-based implementation.
\newblock {\em Games and Economic Behavior}, 75(2):788--808, 788--808.

\bibitem{Monderer.Shapley:96}
D.~Monderer and L.~S. Shapley.
\newblock Potential games.
\newblock {\em Games and Economic Behavior}, 14(1):124 -- 143, 1996.

\bibitem{Newman:03}
M.E.J. Newman.
\newblock The structure and function of complex networks.
\newblock {\em SIAM Review}, 45(2):167--256, 2003.

\bibitem{norris_1997}
J.~R. Norris.
\newblock {\em Markov Chains}.
\newblock Cambridge Series in Statistical and Probabilistic Mathematics.
  Cambridge University Press, 1997.

\bibitem{proskurnikov:2017tutorial}
Anton~V Proskurnikov and Roberto Tempo.
\newblock A tutorial on modeling and analysis of dynamic social networks. part
  i.
\newblock {\em Annual Reviews in Control}, 43:65--79, 2017.

\bibitem{Read.Wilson:98}
R.~C. Read and R.~J. Wilson.
\newblock {\em An Atlas of Graphs}.
\newblock Oxford University Press, 1998.

\end{thebibliography}

\end{document}